\documentclass[ reprint,aps, prb, longbibliography, superscriptaddress]{revtex4-2}
\usepackage{graphicx}
\usepackage{amsmath,bm}
\usepackage{amssymb}
\usepackage{mathrsfs}
\usepackage{braket}
\usepackage{amsfonts}
\usepackage{dsfont}
\usepackage{comment,color}
\usepackage{lipsum}
\usepackage{hyperref}
\usepackage{booktabs}
\usepackage{bbm}
\usepackage[T1]{fontenc}
\usepackage{float}  
\usepackage{pifont}
\hypersetup{colorlinks=true,linkcolor=blue,anchorcolor=blue,citecolor=blue,filecolor=blue,urlcolor=blue,bookmarksnumbered=true,pdfview=FitB}

\usepackage[dvipsnames]{xcolor}

\begin{document}

\title{Pure Spin Bulk Photovoltaic Effect in an Altermagnetic Higher-Order Topological Insulator}
\author{Sibgat Ulah}
\affiliation{Department of Physics, Indian Institute of Technology Delhi, Hauz Khas, New Delhi, India 110016}
\author{Ankan Bhattacharyya}
\affiliation{Department of Physics, Indian Institute of Technology Hyderabad, Kandi, Sangareddy, Telangana, India 502285}
\author{Manisha Thakurathi}
\affiliation{Department of Physics, Indian Institute of Technology Hyderabad, Kandi, Sangareddy, Telangana, India 502285}
\date{\today}
\begin{abstract}
We investigate the bulk photovoltaic effect (BPVE) in a $\mathcal{PT}$-symmetric two-dimensional heterostructure consisting of a topological insulator coupled to a $d$-wave altermagnet. To describe the symmetry-enforced degenerate bands of this system, we develop a non-Abelian formulation of the spin BPVE, extending the conventional theory from isolated nondegenerate bands to $\mathcal{PT}$-degenerate manifolds. We show that the orientation of the N\'eel vector controls both the topological phase and the character of the nonlinear optical response. When the N\'eel vector lies in the $xy$-plane, the heterostructure realizes a second-order topological insulator (SOTI) protected by $\mathcal{C}_{4z}\mathcal{T}$ symmetry. The system also retains $\mathcal{C}_{2z}$ symmetry, which completely suppresses second-order charge photocurrents while allowing finite spin photocurrents. As a result, the BPVE becomes an intrinsically pure spin photovoltaic effect, generating a dc spin current without an accompanying charge current. We find that linearly polarized light drives a spin shift current, whereas circularly polarized light generates a spin injection current. Both responses undergo a sign reversal whenever the local Dirac mass changes sign. As the N\'eel vector is rotated toward the $z$-axis, the SOTI phase transforms into a first-order topological insulating phase, leading to the coexistence of charge and spin photocurrents. Our results establish $\mathcal{PT}$-symmetric altermagnetic topological-insulator heterostructures as a versatile platform for generating and controlling pure spin photocurrents and reveal nonlinear spin transport as a sensitive probe of topology and magnetic symmetry.
\end{abstract}
\maketitle
\section{Introduction}
In recent years, the growing demand for advanced optoelectronic technologies and sustainable energy-harvesting strategies has stimulated considerable interest in nonlinear light-matter interactions \cite{Shi, zeng}. Among these phenomena, the bulk photovoltaic effect (BPVE) has emerged as a promising mechanism for direct photocurrent generation. The BPVE refers to the generation of a second-order nonlinear dc current in a material illuminated by monochromatic light in the absence of an external bias \cite{PhysRevB.53.10751, PhysRevB.61.5337,PhysRevB.82.235204}. Conventional photovoltaic devices based on $p$-$n$ junctions rely on built-in electric fields at spatial interfaces to separate photoexcited electron-hole pairs \cite{10.1063/1.1721711}. However, BPVE manifests itself as a bulk photocurrent generated in a non-centrosymmetric crystal under uniform illumination, even in the absence of junctions or externally applied bias fields \cite{Dai,tan2016shift,PhysRevB.95.224430,PhysRevB.95.035134,PhysRevB.100.224411,PhysRevLett.90.216601,PhysRevLett.122.197702,PhysRevLett.86.4358,spin_polarised_solar_battery,PhysRevB.95.224430,PhysRevLett.110.057201}. Beyond its potential technological applications, the BPVE has attracted renewed attention owing to its intimate connection with the quantum geometry of electronic bands. Recent studies have revealed that nonlinear optical responses are governed by geometric quantities such as Berry curvature, quantum metric, and related geometric tensors, providing a powerful probe of the underlying electronic wave functions \cite{Ahn2021}. As a result, the BPVE has become an important platform for exploring the interplay between topology, symmetry, and quantum geometry in condensed-matter systems \cite{Jiang_2025}.

Two principal second-order dc photocurrent mechanisms are the shift and injection currents. Although both contribute to the bulk photovoltaic response, they originate from different microscopic processes. The shift current is an intrinsic, relaxation time-independent contribution arising from the real-space displacement of an electron during an optical interband transition and is quantified by the shift vector. In contrast, the injection current is a ballistic response resulting from an asymmetric population of photoexcited carriers in momentum space, leading to a net carrier velocity. Unlike the shift current, its magnitude is governed by the carrier relaxation time and is therefore sensitive to scattering processes \cite{Dai,PhysRevLett.90.136603}. For charge photocurrents, broken inversion symmetry is a necessary prerequisite for the bulk photovoltaic effect \cite{Cook2017}. Furthermore, the symmetry of the system determines which second-order photocurrent responses can be generated under linearly polarized light (LPL) or circularly polarized light (CPL) \cite{PhysRevX.11.011001}. In time-reversal-symmetric ($\mathcal{T}$) systems, the allowed charge photocurrents are the linear shift current and the circular injection current \cite{PhysRevX.11.011001,PhysRevB.61.5337}. In contrast, for $\mathcal{PT}$-symmetric systems, the symmetry-allowed charge photocurrents are the circular shift (gyration) current and the linear injection current \cite{PhysRevX.11.011001,PhysRevB.61.5337}. We note that additional crystalline symmetries, such as mirror symmetries, can further restrict the nonlinear conductivity tensor and may completely suppress the charge photocurrent, even in the absence of inversion symmetry \cite{Xu2021}. In such situations, a pure spin photocurrent may remain symmetry-allowed even when the corresponding charge photocurrent vanishes. 
The symmetry classification of spin photocurrents is complementary to that of charge photocurrents. In $\mathcal{PT}$-symmetric systems, linearly polarized light can give rise to a spin shift current, whereas circularly polarized light can generate a spin injection current \cite{Xu2021}. Conversely,  the spin photocurrent consists of a circular shift-current response and a linear injection-current response in $\mathcal{T}$-symmetric systems \cite{Xu2021}.

The nonlinear photovoltaic response of topological materials has attracted considerable attention in recent years. In particular, topological Weyl semimetals exhibit pronounced enhancements of second-order charge photocurrents, including divergent responses near Weyl nodes and characteristic signatures tied to the underlying band topology \cite{PhysRevX.10.041041,PhysRevB.95.035134}. Another example is the circular injection current, which can be quantized in Weyl semimetals, with its magnitude determined by the topological charge of the Weyl node, providing a direct optical probe of Weyl-fermion chirality \cite{de_juan_quantized_2017}. Moreover, enhancement of the shift current near first-order topological phase transitions, driven by inversion-symmetry-breaking perturbations, has highlighted the intimate connection between nonlinear optical responses and band topology \cite{PhysRevLett.116.237402,PhysRevB.96.075421}. 
BPVE occurs in magnetic and nonmagnetic materials and has potential applications in photovoltaics, photodetection, and magnetic sensing technologies \cite{Zhang2019CrI3}. Recently, nonlinear photovoltaic effects have also been investigated in altermagnets, a newly identified class of unconventional magnetic materials that break time-reversal symmetry while exhibiting vanishing net magnetization \cite{PhysRevX.12.040501,PhysRevX.12.031042}. In these systems, unconventional transport \cite{LS1,LS2,SP} and optical responses emerge from the interplay between crystal symmetry and magnetic order \cite{PhysRevB.111.195210}. Furthermore, second-order charge photovoltaic conductivity has been predicted in AMs upon introduction of Rashba spin-orbit coupling, which breaks inversion symmetry and enables a finite nonlinear charge photocurrent response \cite{PhysRevB.111.L201405}. 

These developments motivate us to explore BPVE in first and second-order phase in topological materials. Although nonlinear optical responses have been shown to encode information about band topology in first-order topological phases, it remains unclear how they are affected by the emergence of higher-order topology. In particular, whether BPVE can provide a direct signature that distinguishes trivial and second-order topological phases is an important question that has received little attention. In this paper, we employ $d$-wave AM, which generates different topological phases and while maintaining $\mathcal{PT}$ symmetry in the system, allowing the generation of BPVEs \cite{PhysRevLett.133.106601, PhysRevB.109.L201109, Schindler2018}. The BPVE is computed in this $\mathcal{PT}$-symmetric degenerate system, which was previously discussed for charge conductivity \cite{PhysRevX.11.011001} and for spin conductivity \cite{PhysRevB.105.045201} in the non-degenerate system. In our work, we study two-dimensional heterostructure formed by $d$-wave AM and TI.  We show that in such a system the presence of $\mathcal{PT}$-enforced degenerate bands provides an ideal platform for realizing and controlling nonlinear spin photocurrents in the absence of conventional charge responses. Although our formalism mainly focuses on spin currents, the same approach can also be used to compute orbital currents, for which one needs to substitute the orbital current operator in place of the spin current operator, and the rest of the calculations are the same \cite{PhysRevB.105.L121407, tanaka2026nonlinearresponsetheoryorbital}. We note that for an orbital current, breaking the inversion symmetry is not a necessary condition.

The outline of the paper is presented in the following manner. In Sec. \ref{sec2}, we discuss the model, topological phases and symmetry properties of the Hamiltonian. In Secs. \ref{sec3} and \ref{sec4}, we discuss the BPVE and corresponding numerical results. Section \ref{Sec4b} is devoted to the discussion of the effect of N\'eel-vector rotation on the optical response. Finally, we provide the summary and conclusion of our  work in Sec. \ref{conclusion}. Detailed derivation of the optical conductivity calculation associated with our study is discussed in  Appendix \ref{sec5}, symmetry constraints on the matrix elements in Appendix \ref{AppB}, contour plots of response tensor in Appendix \ref{AppC} followed by emergence of FOTI when N\'eel vector is oriented out of plane in Appendix \ref{AppD}.

\begin{figure}
    \centering    \includegraphics[width=1\linewidth]{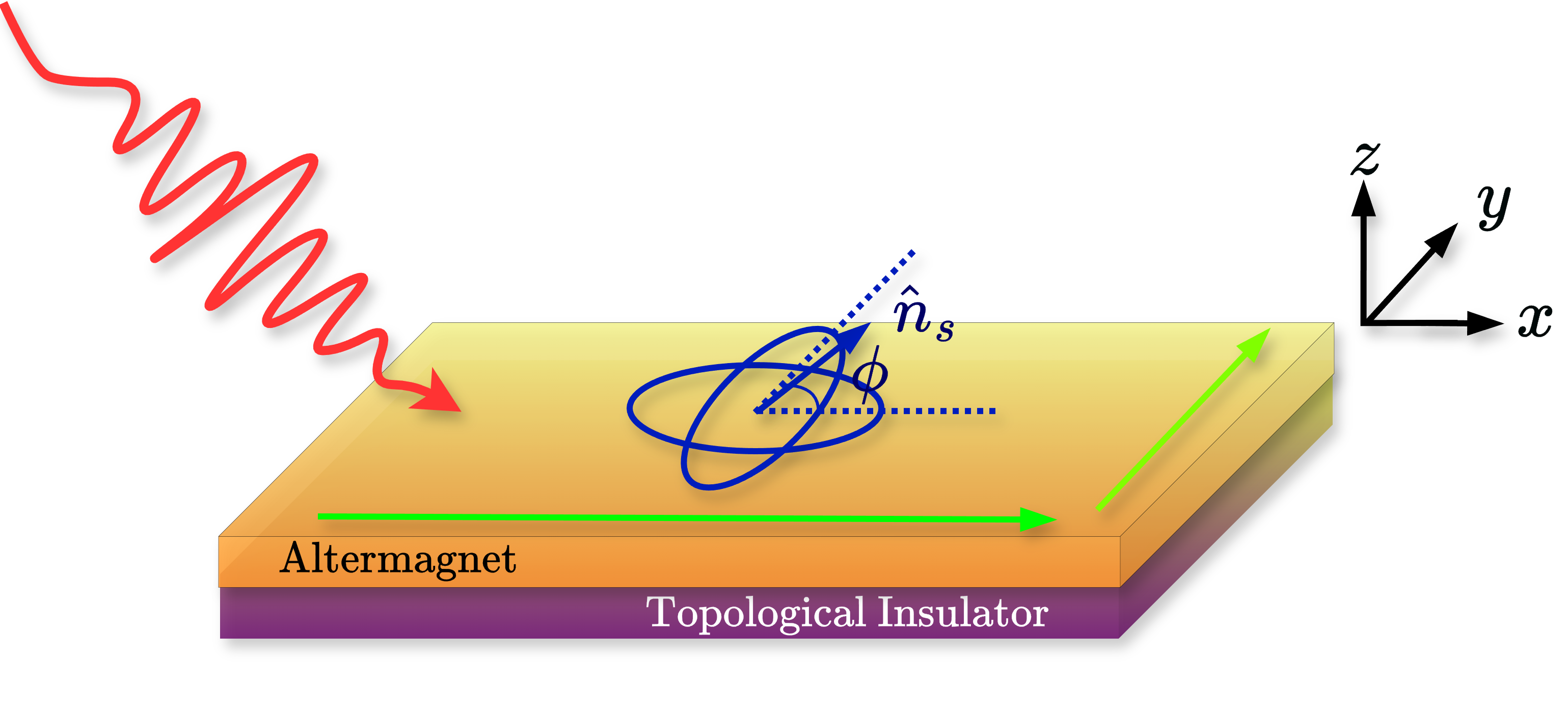}
    \caption{Schematic illustration of an AM-2D TI heterostructure and optically induced spin transport. The bilayer form consists of an AM layer on top of a 2D TI (bottom layer). When system is illuminated by an incident light wave (red wavy arrow), nonlinear spin shift currents are generated, which are indicated by the green arrow. The allowed spin-current components are determined by the orientation of the N\'eel vector, $\hat{n}_s$ (blue arrow), parametrized by the angle $\phi$.}
    \label{fig1}
\end{figure}
\section{Model}
\label{sec2}
We consider a bilayer heterostructure consisting of a two-dimensional topological insulator (2D TI) \cite{2DTI} coupled to a $d$-wave AM, as illustrated in Fig.~\ref{fig1}. The Hamiltonian of the system is given by
\begin{equation}
    \mathcal{H}(\mathbf{k}) = \mathcal{H}_{o}(\mathbf{k}) + J(\mathbf{k}),
    \label{eq:total_hamiltonian}
\end{equation}

where the individual terms has the following form
\begin{align}
    \mathcal{H}_{o}(\mathbf{k}) &= A(\mathbf{k}) + M(\mathbf{k}), \nonumber \\
    A(\mathbf{k}) &= A ( \sin k_x \kappa_z \sigma_x + \sin k_y \kappa_z \sigma_y), \nonumber \\
    M(\mathbf{k}) &= [B + t(4 - 2 (\cos k_x + \cos k_y))] \kappa_x, \nonumber \\
    J(\mathbf{k}) &= J_{a} (\cos k_x - \cos k_y) \kappa_z \sigma_z.
    \label{eq:hamiltonian}
\end{align}
The term $\mathcal{H}_{0}(\mathbf{k})$ describes a first-order topological insulator (FOTI) supporting gapless helical edge states. It consists of the spin-orbit coupling (SOC) term $A(\mathbf{k})$ and the mass term $M(\mathbf{k})$, the latter driving the band inversion responsible for the topological phase. The term $J(\mathbf{k})$ represents the $d$-wave AM exchange field with strength $J_a$, whose momentum dependence gives rise to the characteristic altermagnetic spin splitting \cite{PhysRevLett.133.106601, PhysRevB.109.L201109}. The parameters $A$ and $t$ determine the strengths of SOC and nearest-neighbor hopping, while $B$ controls the bare Dirac mass. In addition, $\kappa_i$ and $\sigma_i$ $(i=x,y,z)$ are Pauli matrices acting in the orbital and spin subspaces, respectively. 
 
The Hamiltonian in Eq. (\ref{eq:total_hamiltonian}) is written in the basis
$ \quad \frac{|E\uparrow\rangle - |H\downarrow\rangle}{\sqrt{2}},\; \frac{|E\downarrow\rangle + |H\uparrow\rangle}{\sqrt{2}},\; \frac{|E\uparrow\rangle + |H\downarrow\rangle}{\sqrt{2}},\; \frac{|E\downarrow\rangle - |H\uparrow\rangle}{\sqrt{2}}$ and can be transformed into the conventional BHZ \cite{BHZ} basis through the following unitary transformation
\begin{equation}
    U = \exp \left[ -\frac{i\pi}{4} (1 - \kappa_z) \otimes \sigma_y \right] \cdot \exp \left[ \frac{i\pi}{4} \kappa_y \otimes \sigma_0 \right].
\end{equation}
After applying the transformation, components of $\mathcal{H}(\mathbf{k})$ in the conventional BHZ basis $ |E\uparrow\rangle,|E\downarrow\rangle, |H\uparrow\rangle,|H\downarrow\rangle$, take the form
\begin{align}
    \mathcal{H}_{0}(\mathbf{k}) &= m(\mathbf{k}) \kappa_z \sigma_0 + A \sin k_x \kappa_x \sigma_z + A \sin k_y \kappa_y \sigma_0, \nonumber\\
    J(\mathbf{k}) &= J_a(\mathbf{k}) \kappa_x \sigma_x.
    \label{BHZ2}
\end{align}
where $m(\mathbf{k}) = B + t[4 - 2(\cos k_x + \cos k_y)]$ and $J_a(\mathbf{k}) = -J_{a} (\cos k_x - \cos k_y) $.
 If we take a general direction ($\hat{n}_s$) of a N\'eel vector  {\cite{PhysRevB.109.245306}, Eq. {\eqref{BHZ2}} reduces to
\begin{equation}
    J(\mathbf{k}) = -J_{a} (\cos k_x - \cos k_y) \left( \kappa_x \otimes (\hat{n}_s \cdot \vec{\sigma}) \right).
\end{equation}
In this section, we consider the N\'eel vector of the AM to be oriented along the $x$ axis. The corresponding doubly degenerate energy spectrum  is obtained by diagonalizing the Hamiltonian and is given by

\begin{figure}[t]
\centering
\includegraphics[width=1\linewidth]{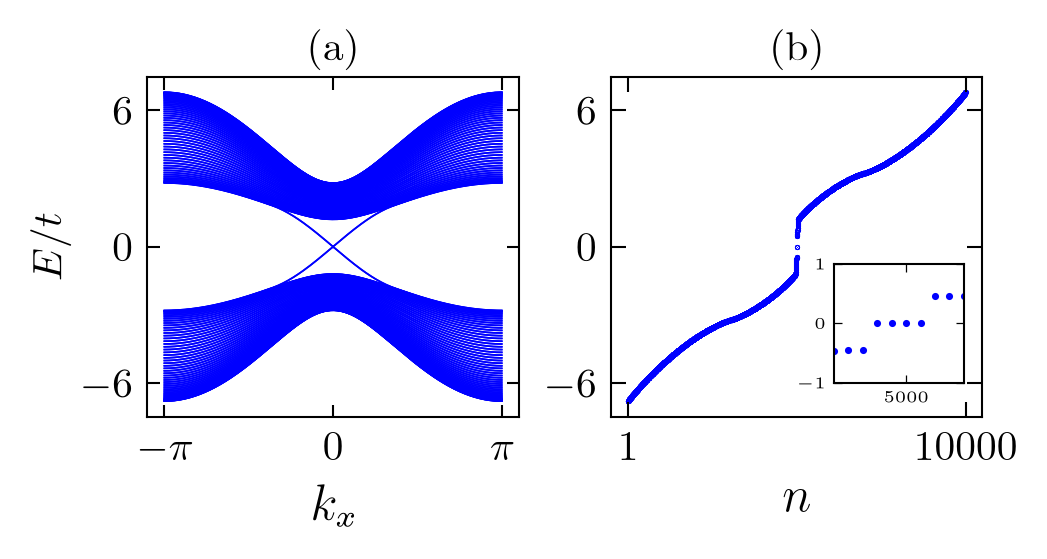}
\caption{Energy spectra of the system corresponding to the two topological phases. (a) First-order topological phase ($J_a$ = 0) with gapless helical edge mode traverses the bulk gap computed under open
boundary conditions taken along the $y$-direction. (b) Second-order
topological phase with  $J_a=0.8t$, considering open boundary conditions along both directions. Here, the edge modes are gapped by the AM, leaving four zero-energy corner states. Parameters are fixed at $B=-1.2t$, $A=2t$, $t=1$.}
\label{fig2}
\end{figure}
\begin{equation}
\varepsilon(\mathbf{k}) = \pm \sqrt{A^2(\sin^2 k_x + \sin^2 k_y) + m^2(\mathbf{k}) + J^2_a(\mathbf{k})}.
\label{eq:energy_eigenvalue}
\end{equation}
We now examine the bulk spectrum and the corresponding topological phase diagram as a function of the model parameters. The parameter $B$ controls the topological phase transitions. For $B\in[-8t,0]$, the system is in a topologically nontrivial regime. In the absence of altermagnetism $(J_a=0)$, the system realizes a first-order topological insulator (FOTI). For $J_a=0$, the bulk gap closes at three critical values of  $B$, corresponding to distinct high-symmetry points in the Brillouin zone: $B = 0$ at $\Gamma(0,0)$, $B = -4t$ at $X(\pi,0)$ and $Y(0,\pi)$, and $B = -8t$ at $M(\pi,\pi)$. Each gap closing is accompanied by a band inversion that changes the bulk topological invariant. The bulk gap closing at $B=-4t$ marks a shift in the momentum-space location of the helical edge modes in the spectrum under open boundary conditions along one direction (say $y$-direction). Specifically, the helical edge states appear around $k_x=0$ for $-4t<B<0$, whereas they are centered around $k_x=\pi$ for $-8t<B<-4t$. Upon introducing a finite altermagnetic exchange $(J_a\neq0)$, the system enters a second-order topological insulator (SOTI) phase. In this phase, the altermagnetic exchange gaps the helical edge states by inducing mass terms of opposite signs on edges oriented along the $x$ and $y$ directions. The resulting mass-domain wall structure, corresponding to the Jackiw–Rebbi mechanism \cite{Jackiw-Rebbi}, localizes four zero-energy corner states at the intersections of adjacent edges, as illustrated in Fig.~\ref{fig2}(b). 

The emergence of the SOTI phase can be understood from the symmetry properties of the Hamiltonian. Although the altermagnetic exchange preserves the bulk gap, it modifies the symmetry of the system in a manner that is essential for the formation of corner states. In particular, the altermagnetic term breaks time-reversal symmetry,
$\mathcal{T}=i\kappa_0\sigma_y\mathcal{K}$ where $\mathcal{K}$ denotes  complex conjugation operator. However, it preserves the $\mathcal{C}_{4z}\mathcal{T}$, where $\mathcal{C}_{4z} = e^{-i\frac{\pi}{4}\kappa_z\sigma_z}$. The Hamiltonian preserves the mirror symmetries $\mathsf{M}_x=i\kappa_o\sigma_x$ and $\mathsf{M}_y=i\kappa_z\sigma_y$, and therefore also the twofold rotational symmetry $\mathcal{C}_{2z}=\mathsf{M}_x \mathsf{M}_y$. Although the altermagnetic term breaks the BHZ inversion symmetry, $\mathcal{P}=\kappa_z\sigma_0$, the combined $\mathcal{PT}$ symmetry remains preserved. As a consequence, all Bloch bands remain doubly degenerate across the Brillouin zone. These symmetry operations satisfy the following relations:
\begin{align}
    \mathcal{P} \mathcal{H}(k_x, k_y) \mathcal{P}^{-1} &\neq \mathcal{H}(-k_x, -k_y), \label{eq:inversion_breaking} \\
    \mathsf{M}_x \mathcal{H}(k_x, k_y) \mathsf{M}_x^{-1} &= \mathcal{H}(-k_x, k_y), \label{eq:mirror_x} \\
    \mathsf{M}_y \mathcal{H}(k_x, k_y) \mathsf{M}_y^{-1} &= \mathcal{H}(k_x, -k_y).
    \label{eq:mirror_y}\\
    ({\mathcal{C}_{4z}\mathcal{T}}) \mathcal{H}(k_x, k_y)   ({\mathcal{C}_{4z}\mathcal{T}})^{-1} &= \mathcal{H}(k_y,- k_x)
    \label{eq:C4zT_Ham}
\end{align}

The symmetry properties of individual Hamiltonian terms are summarized in Table~\ref{tab:symmetry}. In the absence of altermagnetic exchange $(J_a=0)$, the FOTI phase is protected by time-reversal symmetry, giving rise to gapless helical edge states. A finite altermagnetic exchange breaks $\mathcal{T}$, thus gapping the edge spectrum, but preserves the combined $\mathcal{C}_{4z}\mathcal{T}$ symmetry. This symmetry protects the resulting SOTI phase and its zero-energy corner states.

\begin{table}[h]
\centering
\caption{Symmetry properties of the Hamiltonian terms. $+$ ($-$) indicates 
even (odd) under the corresponding symmetry operation.}
\label{tab:symmetry}
\begin{ruledtabular}
\begin{tabular}{l cccccccc}
Term & $\mathcal{P}$ & $\mathcal{T}$ & $\mathcal{PT}$ & $\mathcal{C}_{4z}$ & $\mathcal{C}_{4z}\mathcal{T}$ & $\mathsf{M}_x$ & $\mathsf{M}_y$ & $\mathcal{C}_{2z}$ \\
\hline
$\mathcal{H}_0(\mathbf{k})$ & $+$ & $+$ & $+$ & $+$ & $+$ & $+$ & $+$ & $+$ \\
$J(\mathbf{k})$             & $-$ & $-$ & $+$ & $-$ & $+$ & $+$ & $+$ & $+$ \\
\end{tabular}
\end{ruledtabular}
\end{table}

\section{ Second order Bulk Photovoltaic conductivity}
\label{sec3}

Having established the topological phases of the model, we now investigate its nonlinear optical response. We focus on the generation of a dc photocurrent through BPVE. The BPVE is a second-order nonlinear optical phenomenon in which a monochromatic electric field induces a dc photocurrent in the absence of an applied bias voltage.

The second-order charge BPVE is described by the third-rank conductivity tensor $\sigma^{\mu;\nu\lambda}$, where the electric field components are applied along the $\nu$ and $\lambda$ directions to the resulting current flowing along the $\mu$ direction. Our two-dimensional system preserves the orthogonal mirror symmetries $\mathsf{M}_x$ and $\mathsf{M}_y$ (Eqs.~(\ref{eq:mirror_x})-(\ref{eq:mirror_y})). Their product generates a twofold rotation symmetry about the out-of-plane axis, $\mathcal{C}_{2z}= \mathsf{M}_x\mathsf{M}_y$. The $\mathcal{C}_{2z}$ operation flips both in-plane spatial coordinates, meaning $(x, y)\to(-x,-y)$. Since our system is strictly 2D, the relevant spatial indices $\mu,\nu,\lambda$ can only be $x$ or $y$. Consequently, the charge conductivity tensor transforms as follows
\begin{equation} \sigma^{\mu;\nu\lambda}\xrightarrow{\mathcal{C}_{2z}}-\sigma^{\mu;\nu\lambda}.
\end{equation}
Therefore, second-order charge photocurrent responses are forbidden by symmetry and invariance of $\mathcal{C}_{2z}$
requires $\sigma^{\mu;\nu\lambda}=0$.  The situation is fundamentally different for a pure spin photocurrent. A dc spin current is characterized by the fourth-rank spin conductivity tensor $\sigma^{s,\mu;\nu\lambda}$ which relates the electric field components applied along the $\nu$ and $\lambda$ directions to the resulting current flowing along the $\mu$ direction with spin polarization $s$, where $s \in \{x,y,z\}$. Because the spin index transforms differently from the spatial indices under the crystal symmetries, the constraints on the spin conductivity tensor are distinct from those governing the charge response. Therefore, we proceed to analyze the symmetry-allowed components of $\sigma^{s,\mu;\nu\lambda}$.

Moreover, in topological systems, nonlinear responses can provide valuable information about the underlying band geometry and symmetry properties, while also offering a route to optically probe and manipulate spin transport. A distinctive feature of the present system is its $\mathcal{PT}$ symmetry, which guarantees a double degeneracy of every Bloch state throughout the Brillouin zone. As a result, the conventional band-resolved spin BPVE formalism \cite{PhysRevB.105.045201}, which is formulated for isolated non-degenerate bands, requires a non-Abelian generalization to describe the $\mathcal{PT}$-enforced degenerate states considered here. In the standard treatment, individual Bloch bands are assumed to be non-degenerate, and the response coefficients contain energy denominators of the form $1/\varepsilon_{ab}$, where $\varepsilon_{ab}=\varepsilon_a-\varepsilon_b$ denotes the energy difference between states $a$ and $b$. However, within a degenerate manifold $\varepsilon_a=\varepsilon_b$, these expressions become ill-defined.

To properly account for $\mathcal{PT}$-enforced degeneracy, we treat each degenerate set of bands as a $N$-fold degenerate manifold. Consequently, the matrix elements of any operator $\hat{O}$ are represented by $N\times N$ matrices $O_{ab}$ acting within the manifolds, and the ordinary momentum derivatives are replaced by the corresponding non-Abelian covariant derivatives $U(N)$. The complete derivation of the resulting nonlinear response theory is presented in Appendix~\ref{sec5}. In the limit of non-degenerate bands ($N=1$), our formalism reduces to the conventional expressions derived in Ref.~\cite{PhysRevB.105.045201}. For the $\mathcal{PT}$-symmetric system considered here, each Bloch state forms a twofold degenerate manifold $N=2$. The second-order spin photocurrent $J^{s,\mu}$ with frequency $\omega$ (in the dc limit $\omega \rightarrow 0$) induced by a monochromatic electric field of frequency $\Omega$ is related to the applied field through

\begin{equation}
J^{s,\mu}_{(2)}(\omega \rightarrow 0)=\sum_{\nu\lambda}{\sigma}^{s,\mu;\nu\lambda}(\omega \rightarrow 0;\Omega, - \Omega)\,
    E^\nu(\Omega)\,[E^\lambda(\Omega)]^*.
\end{equation}
Here, $\sigma^{s,\mu;\nu\lambda}$ is the symmetrized spin conductivity tensor. For dc spin conductivity $(\omega\rightarrow0)$, the interband contribution  can be decomposed into two distinct components, namely the spin shift conductivity and the spin injection conductivity (the detailed process is shown in Appendix \ref{sec5}),
\begin{equation}
    \sigma^{s,\mu;\nu\lambda}(0;\Omega,-\Omega) =\sigma^{s,\mu;\nu\lambda}_\text{shift}+\sigma^{s,\mu;\nu\lambda}_\text{inj}.
\end{equation}
The spin shift conductivity is a geometric contribution that is independent of the relaxation time and originates from the real-space displacement of an electronic wave packet during an optical interband transition. In contrast, the spin injection conductivity arises from the asymmetric population of photoexcited carriers and is proportional to the relaxation time $\tau$, diverging in the clean limit as the broadening parameter $\eta \rightarrow 0$ \cite{PhysRevX.11.011001}.

\subsection{Spin Shift Conductivity}
The shift conductivity tensor for the degenerate system is derived in Appendix \ref{sec5} and has the following form:
\begin{equation}
\sigma^{s,\mu;\nu\lambda}_\text{shift} = -
\frac{i\pi q^3}{2}
\int\frac{d^dk}{(2\pi)^d}
\sum_{a\neq b}
\frac{f_{ab}\,}{\varepsilon^2_{ab}}
\;\mathcal{M}^{s,\mu;\nu\lambda}_{ab}\delta(\varepsilon_{ab}-\Omega).
\label{eq:response}
\end{equation}
Here 
\begin{equation}
\mathcal{M}^{s,\mu;\nu\lambda}_{ab} = \text{Tr}\left[W^{s,\mu;\nu}_{ab} v^\lambda_{ba} - W^{s,\mu;\lambda}_{ba} v^\nu_{ab}\right],
\label{eq:shift2}
\end{equation}
and 
\begin{equation}
W^{s,\mu;\nu}_{ab} = 
S^{s,\mu;\nu}_{ab} 
- \frac{S^{s,\mu}_{ab}\,\Delta^\nu_{ab}}{\varepsilon_{ab}},
\qquad
\Delta^\nu_{ab} = v^\nu_{aa} - v^\nu_{bb}.
\label{eq:shift3}
\end{equation}
Throughout this work, the indices $a, b$ label distinct degenerate manifolds, each containing $N$ degenerate states. Here, $q$ denotes the electronic charge and the occupation difference is defined as $f_{ab}=f_a-f_b$, where $f_{a(b)}$ denotes the Fermi--Dirac occupation of manifold $a (b)$. The velocity matrix elements are given by $v_{ab}^\mu(\mathbf{k})=\langle a,\mathbf{k} |\hat{v}^{\mu}| b,\mathbf{k} \rangle$ where $\hat{v}^{\mu}=\frac{\partial \hat{H}(\mathbf{k})}{\partial k_{\mu}}$. We employ the conventional spin-current operator,  $ \hat{J}^{s,\mu} = \frac{1}{2} \{ \hat{S}^s, \hat{v}^\mu \},    \text{ whose matrix elements are} \quad S_{ab}^{s,\mu} = \langle a,\mathbf{k} | \frac{1}{2} \{ \hat{S}^s, \hat{v}^\mu \} | b,\mathbf{k} \rangle$ and  $S_{ab}^{s,\mu;\nu} = \langle a,\mathbf{k} | \frac{1}{2} \{ \hat{S}^s, \hat{v}^{\mu\nu}\} | b,\mathbf{k} \rangle$ with $\hat{v}^{\mu\nu}=\frac{\partial \hat{^2H}(\mathbf{k})}{\partial k_{\mu}\partial k_\nu}$. In the presence of spin–orbit coupling, spin is generally not a conserved quantity, and therefore the definition of spin current is not unique. A conserved spin-current operator has been proposed in Ref. \cite{PhysRevLett.96.076604}; however, its conservation is valid only in the absence of bulk spin generation. This assumption is violated in nonlinear optical processes as follows. A second-order response requires broken inversion symmetry, and in the presence of spin–orbit coupling this same symmetry breaking allows the driving field to generate spin polarization directly \cite{Xu2021}. As a result, the conserved spin-current operator is no longer strictly conserved under these conditions. We therefore use the conventional spin-current operator, which is standard in the spin-photocurrent literature \cite{PhysRevB.105.045201}. This choice is justified on two grounds; first, its non-conservation corresponds to a physically measurable effect rather than a formal inconsistency \cite{PhysRevB.77.035327}, and second, the conventional and conserved definitions have been shown to agree quantitatively to within approximately 10-20\% \cite{Xu2021}. Notably, in the limit $s=0$, the resulting tensor reduces to the conventional charge shift conductivity \cite{PhysRevX.11.011001}.

We next analyze the symmetry constraints on the spin-conductivity tensor. For this purpose, Eq.~(\ref{eq:response}) is decomposed into its real and imaginary components as
\begin{equation}
   \sigma^{s,\mu;\nu\lambda}_{\text{shift}}=\  \sigma^{s,\mu;\nu\lambda}_{L}+i\sigma^{s,\mu;\nu\lambda}_{C}.
\end{equation}
Here, the real part, $\sigma^{s,\mu;\nu\lambda}_{L}$, describes the response to LPL, while the imaginary part, $\sigma^{s,\mu;\nu\lambda}_{C}$, describes the response to CPL. In the FOTI phase, the presence of inversion symmetry forces all second-order bulk responses to vanish. By contrast, in the SOTI phase, inversion symmetry is broken, allowing finite second-order responses such as the bulk photovoltaic effect.

In our $\mathcal{PT}$-symmetric degenerate system, the symmetry constraints on spin photocurrents are complementary to those governing charge photocurrents, resulting in an interchange of the responses to LPL and CPL. Under $\mathcal{PT}$, the velocity and spin-current matrix inside the trace transform as

\begin{equation}
v^\mu_{ab} \xrightarrow{\mathcal{PT}} (v^\mu_{ab})^* \quad \text{and} \quad S^{s,\mu}_{ab} \xrightarrow{\mathcal{PT}} -(S^{s,\mu}_{ab})^*,
\end{equation}

\begin{equation}
v^{\mu\nu}_{ab} \xrightarrow{\mathcal{PT}} (v^{\mu\nu}_{ab})^* \quad \text{and} \quad S^{s,\mu;\nu}_{ab} \xrightarrow{\mathcal{PT}} -(S^{s,\mu;\nu}_{ab})^*.
\end{equation}

Consequently, the tensor $\mathcal{M}^{s,\mu;\nu\lambda} $  transforms as   
\begin{equation}
\mathcal{M}^{s,\mu;\nu\lambda}_{ab}(\mathbf{k}) 
 \xrightarrow{\mathcal{PT}} -\left(\mathcal{M}^{s,\mu;\nu\lambda}_{ab}(\mathbf{k})\right)^*,
\label{eq:PT_M}
\end{equation}
which implies that $\mathcal{M}^{s,\mu;\nu\lambda}_{ab}$ is purely imaginary. Since Eq.~(\ref{eq:response}) contains an overall factor of $-i$, the spin shift conductivity is purely real,

\begin{equation}
\sigma^{s,\mu;\nu\lambda}_\text{shift} = \left(\sigma^{s,\mu;\nu\lambda}_\text{shift}\right)^*,
\label{eq:PT_shift}
\end{equation}
or equivalently,
\begin{equation}
\sigma^{s,\mu;\nu\lambda}_\text{shift} \in \mathbb{R} \quad \forall\; s,\mu,\nu,\lambda.
\label{eq:shift_is_real}
\end{equation}

We now consider the constraints imposed by the mirror symmetries $\mathsf{M}_x$ and $\mathsf{M}_y$. Under these operations, the velocity and spin-current matrix elements  transform as

\begin{equation}
v^\mu_{ab}(\mathbf{k}) \xrightarrow{\;\mathsf{M}_\alpha\;} (-1)^{\delta_{\mu\alpha}} \, v^\mu_{ab}(\mathsf{M}_\alpha \mathbf{k}),
\end{equation}

\begin{equation}
S^{s,\mu}_{ab}(\mathbf{k}) \xrightarrow{\;\mathsf{M}_\alpha\;} (-1)^{1+\delta_{s\alpha}+\delta_{\mu\alpha}} \, S^{s,\mu}_{ab}(\mathsf{M}_\alpha \mathbf{k}),
\end{equation}
\begin{equation}
v^{\mu \nu}_{ab}(\mathbf{k}) \xrightarrow{\;\mathsf{M}_\alpha\;} (-1)^{\delta_{\mu\alpha}+\delta_{\nu\alpha}} \, v^{\mu \nu}_{ab}(\mathsf{M}_\alpha \mathbf{k}),
\end{equation}
\begin{equation}
S^{s,\mu;\nu}_{ab}(\mathbf{k}) \xrightarrow{\;\mathsf{M}_\alpha\;} (-1)^{1+\delta_{s\alpha}+\delta_{\mu\alpha}+\delta_{\nu\alpha}} \, S^{s,\mu;\nu}_{ab}(\mathsf{M}_\alpha \mathbf{k}),
\end{equation}

where $\alpha \in \{x, y\}$. Combining these transformation properties, the conductivity tensor written in Eq. \eqref{eq:response} transforms under mirror symmetries as
\begin{equation}
\mathcal{M}^{s,\mu;\nu\lambda}_{ab}(\mathbf{k}) 
\xrightarrow{\;\mathsf{M}_x\;} (-1)^{1+n_x}\,
\mathcal{M}^{s,\mu;\nu\lambda}_{ab}(-k_x,k_y),
\end{equation}
\begin{equation}
\mathcal{M}^{s,\mu;\nu\lambda}_{ab}(\mathbf{k}) 
\xrightarrow{\;\mathsf{M}_y\;} (-1)^{1+n_y}\,
\mathcal{M}^{s,\mu;\nu\lambda}_{ab}(k_x,-k_y).
\end{equation}
This motivates us to write the general form of the mirror symmetry operation corresponding to the $\mathcal{M}^{s,\mu;\nu\lambda}_{ab}(\mathbf{k})$ matrix as
\begin{equation}
\mathcal{M}^{s,\mu;\nu\lambda}_{ab}(\mathbf{k}) 
\xrightarrow{\;\mathsf{M}_\alpha\;} (-1)^{1+n_\alpha}\,
\mathcal{M}^{s,\mu;\nu\lambda}_{ab}(\mathsf{M_\alpha}\mathbf{k}).
\label{eq:mirror_on_M}
\end{equation}
Here, we have defined $n_\alpha= \delta_{s\alpha}+\delta_{\mu \alpha}+\delta_{\nu \alpha}+\delta_{\lambda \alpha}$. From mirror symmetry constraint, the conductivity vanishes whenever $n_\alpha$ is even. Since our system preserves both $\mathsf{M}_x$ and $\mathsf{M}_y$, a non-zero response requires $n_x$ and $n_y$ to be odd simultaneously.  Consequently, for $s=x$, the allowed conductivity components are $\sigma^{x,y;yy}$, $\sigma^{x,y;xx}$, $\sigma^{x,x;yx}$, $\sigma^{x,x;xy}$, while for $s=y$, non vanishing components are
$\sigma^{y,x;xx}$, $\sigma^{y,x;yy}$, 
$\sigma^{y,y;xy}$, $\sigma^{y,y;yx}$. 
For $s=z$, the spin index does not contribute to either $n_x$ or $n_y$, since $\delta_{zx}=\delta_{zy}=0$. In a two-dimensional $xy$ system, the remaining indices $\mu,\nu$, and $\lambda$ can take only the values $x$ or $y$. Therefore, $n_x+n_y=3$, implying that $n_x$ and $n_y$ cannot be odd simultaneously. As a result, all $z$-polarized spin-conductivity components vanish identically throughout the Brillouin zone.

The system also possesses chiral symmetry $\mathbb{S}_x=\kappa_x \sigma_y$. Under  which the spin shift conductivity tensor transforms as (see Appendix~\ref{AppB} for details):
\begin{equation}
\sigma^{s,\mu;\nu\lambda} \xrightarrow{\mathbb{S}_x} (-1)^{1 + \delta_{sx}+\delta_{sz}} \sigma^{s,\mu;\nu\lambda}.
\end{equation}
Consequently, chiral and mirror symmetries protect the existence of the pure spin photovoltaic response along $x$ direction,  while suppressing the components along the $y$ and $z$ directions, respectively.

\subsection{Spin Injection Conductivity}
In addition to the shift current, the second-order spin photovoltaic response contains an injection-current contribution as well. For a $\mathcal{PT}$-symmetric degenerate manifold, the injection conductivity is proportional to the scattering lifetime and given by (see Appendix~\ref{sec5} for more details)

\begin{align}
\sigma^{s,\mu;\nu\lambda}_{\text{inj}} = \frac{-\pi q^{3}}{\hbar\gamma}\int\frac{d^{d}k}{(2\pi)^{d}}\sum_{a\neq b}\mathcal{I}^{s,\mu;\nu\lambda}_{ab}\,\delta(\hbar\Omega-\varepsilon_{ba})\,f_{ab}, 
\label{eq:Injection main}
\end{align}

where 
\begin{align}
\mathcal{I}^{s,\mu;\nu\lambda}_{ab}=\text{Tr}\Bigl[\bigl(S^{s,\mu}_{aa}A^{\lambda}_{ab}-A^{\lambda}_{ab}S^{s,\mu}_{bb}\bigr)A^{\nu}_{ba}\Bigr].
\label{eq:Injeq2}
\end{align}

Here, $\tau=1/\gamma$ denotes the relaxation time and
$A^\lambda_{ab}=v_{ab}/i\varepsilon_{ab}$ is the interband Berry connection between the degenerate manifolds $a$ and $b$. The matrices $S^{s,\mu}_{aa}$ and $S^{s,\mu}_{bb}$ are the intraband matrix elements of the spin-current operator within the respective manifolds and play a role analogous to the diagonal band velocities appearing in the conventional charge injection current.\cite{PhysRevX.11.011001} . The quantity $(S^{s,\mu}_{aa}A^{\lambda}_{ab}-A^{\lambda}_{ab}S^{s,\mu}_{bb})$ therefore measures the asymmetry of the spin-current carried by the photoexcited states across the optical transition. It reduces to the familiar difference of band velocities that governs the charge injection current~\cite{PhysRevX.11.011001}. For degenerate manifolds, however, these quantities become matrix-valued and generally do not commute, requiring the gauge-covariant expression. Eq.~(\ref{eq:Injection main}) therefore provides a natural generalization of the spin injection-current formalism of Ref.~\cite{PhysRevB.105.045201} to the systems with $\mathcal{PT}$-enforced band degeneracies.  

Under $\mathcal{PT}$ symmetry, the tensor $\mathcal{I}^{s,\mu;\nu\lambda}_{ab}$ is purely imaginary, implying that the injection conductivity in Eq.~(\ref{eq:Injection main}) is also purely imaginary. Consequently, only the antisymmetric part of the conductivity tensor under the interchange of the optical field indices, $\nu \leftrightarrow \lambda$ survives. Since antisymmetric tensor components couple to CPL, the spin injection response is intrinsically a CPL-driven effect.
The mirror and chiral symmetries impose the same selection rules as those derived for the spin shift conductivity, eliminating all $y$ and $z$ polarized spin responses. As a result, only two independent nonvanishing components remain: $\sigma^{x,x;yx}_{\mathrm{inj}}$ and $\sigma^{x,x;xy}_{\mathrm{inj}}$
which satisfy $\sigma^{x,x;yx}_{\mathrm{inj}}=-\sigma^{x,x;xy}_{\mathrm{inj}}$.
Therefore, in the present $\mathcal{PT}$-symmetric system, the spin injection current is generated exclusively by CPL and exhibits the characteristic linear dependence on the relaxation time $\tau$, reflecting its ballistic nonequilibrium origin.

\section{Numerical computation and Results}
\label{sec4}
\begin{figure}
\centering
\begin{tabular}{c c}
\includegraphics[width=0.49\linewidth]{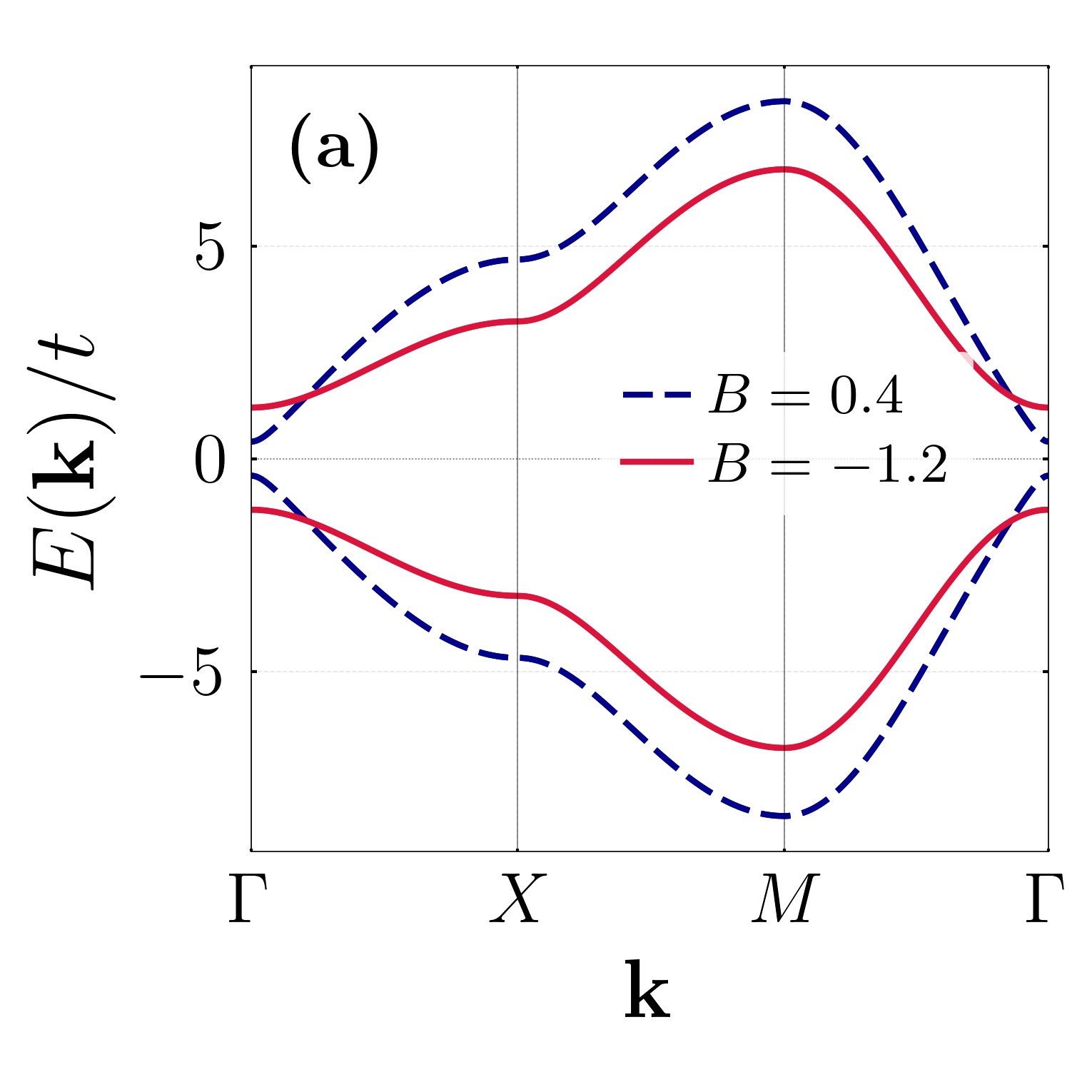}
& \includegraphics[width=0.49\linewidth]{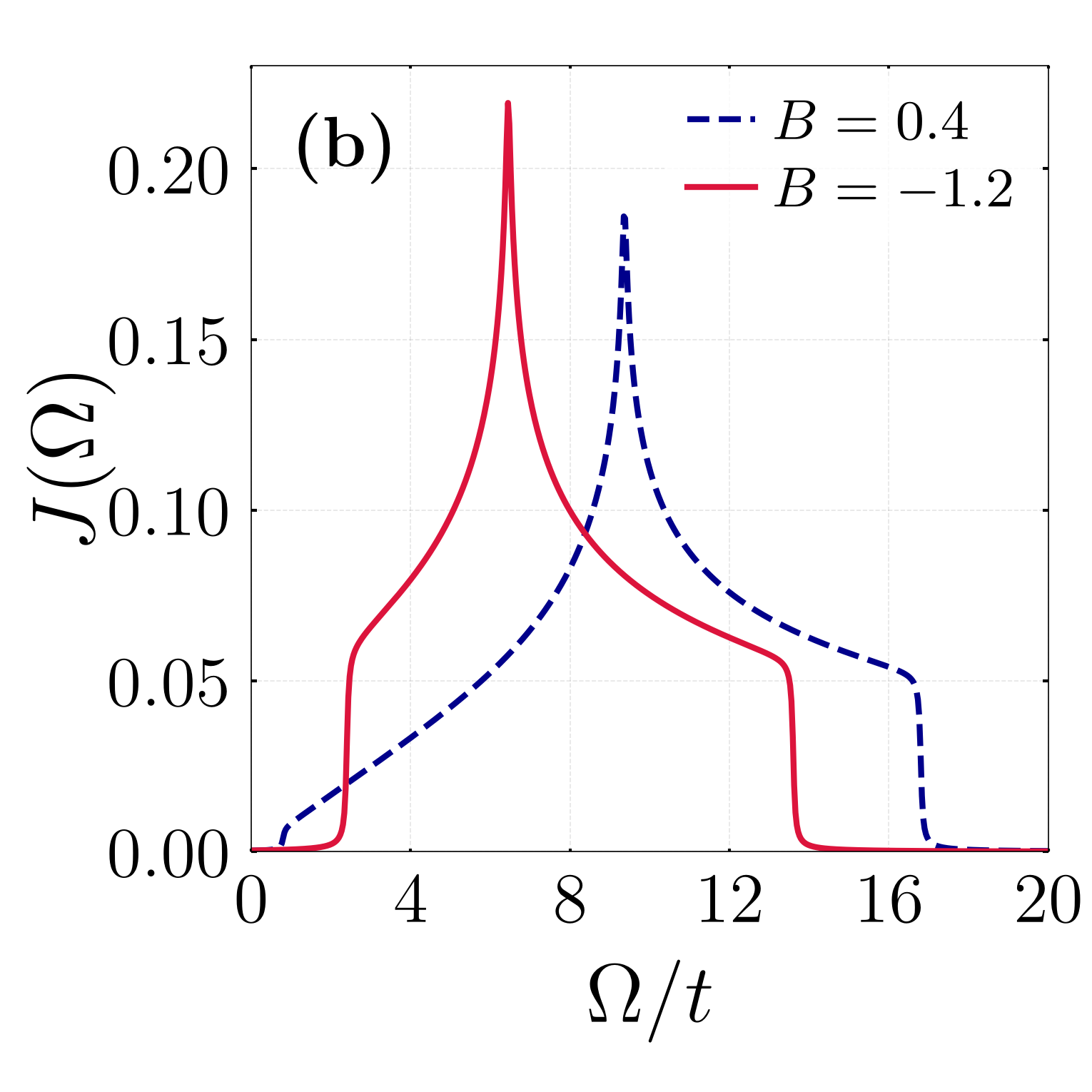}
    \end{tabular}
    \caption{Bulk band structure and joint density of states (JDOS) across two different topological phases. (a) Normalized bulk energy dispersion, $E(\mathbf{k})/t$, plotted along the high-symmetry path $\Gamma(0,0)\!-\!X(\pi,0)\!-\!M(\pi,\pi)\!-\!\Gamma(0,0)$ of the square-lattice Brillouin zone. This dispersion is compared between the topological phase ($B=-1.2t$, red solid line) and the trivial ($B=0.4t$, blue dashed line). In both regions, the minimum band gap is at the $\Gamma$ point. (b) The corresponding JDOS spectrum, $J(\Omega)$, is mapped against the normalized photon energy $\Omega/t$. The prominent divergent peaks observed in the spectra indicate the logarithmic Van~Hove singularities from the saddle points at the symmetry-equivalent corners $X$ or $Y$, which are present in both regions. The remaining parameters are the same as fixed in Fig.~\ref{fig2}.}
    \label{fig3}
\end{figure} 

In this section, we compute the spin shift and injection conductivities numerically. The pronounced features in the optical conductivity often originate from critical points of the band structure, where the joint density of states is enhanced. Such critical points satisfy $\nabla_{\mathbf{k}}\varepsilon(\mathbf{k})=0$,
and can be classified as minima, maxima, or saddle points. In two dimensions, saddle points give rise to logarithmically divergent contributions to the density of states, known as Van Hove singularities \cite{Van_Hove}.
The JDOS ($J(\Omega)$) is calculated by using 
\begin{equation}
J(\Omega) = \sum_{a\neq b}\int_{\mathrm{BZ}} \frac{d^2k}{(2\pi)^2}\,
\delta\!\left(\hbar\Omega - \varepsilon_{ba}\right),
\label{eq:jdos}
\end{equation}
where the Dirac-delta function is regularized by the factor $\eta$ and given by
\begin{equation}
\delta(\hbar\Omega - \varepsilon_{ba}) = \frac{1}{\pi} \frac{\eta}{(\hbar\Omega - \varepsilon_{ba})^2 + \eta^2},
\end{equation}
where $\eta=\hbar/\tau$ and $\tau$ denotes the relaxation time. For the present model, the critical-point condition is obtained using Eq. (\ref{eq:energy_eigenvalue}) and written as
\begin{align}
    \sin k_x \Big[ 2A^2\cos k_x &- 2J_a^2\big(\cos k_x - \cos k_y\big) \nonumber \\
    &+ 4tm(\mathbf{k}) \Big] = 0, \nonumber\\
    \sin k_y \Big[ 2A^2\cos(k_y) &+ 2J_a^2\big(\cos k_x - \cos k_y\big) \nonumber \\
    &+ 4tm(\mathbf{k}) \Big]  = 0.
\end{align}
\begin{figure}[htbp]
\centering

  \includegraphics[width=0.5\linewidth]{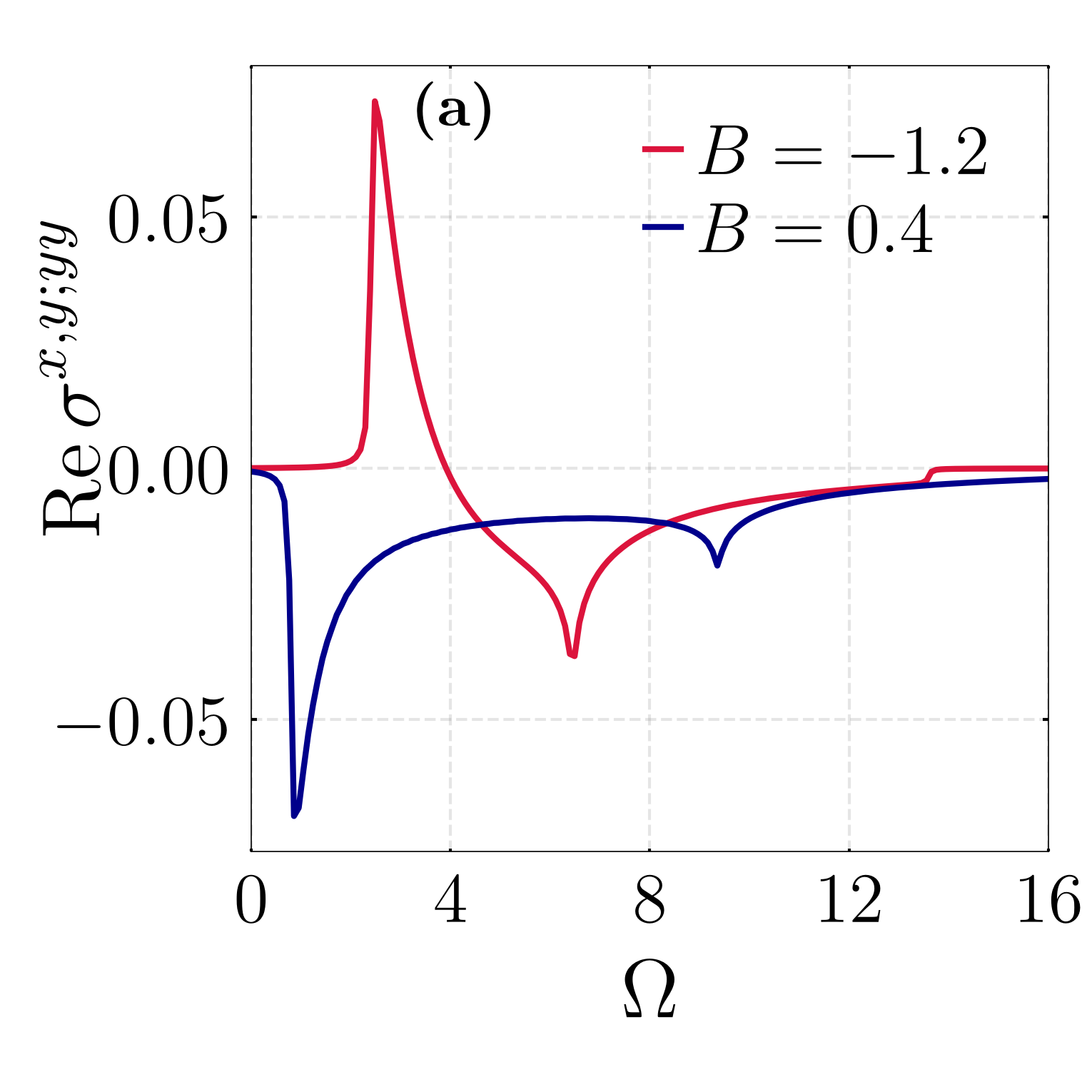}\hfill
  \includegraphics[width=0.5\linewidth]{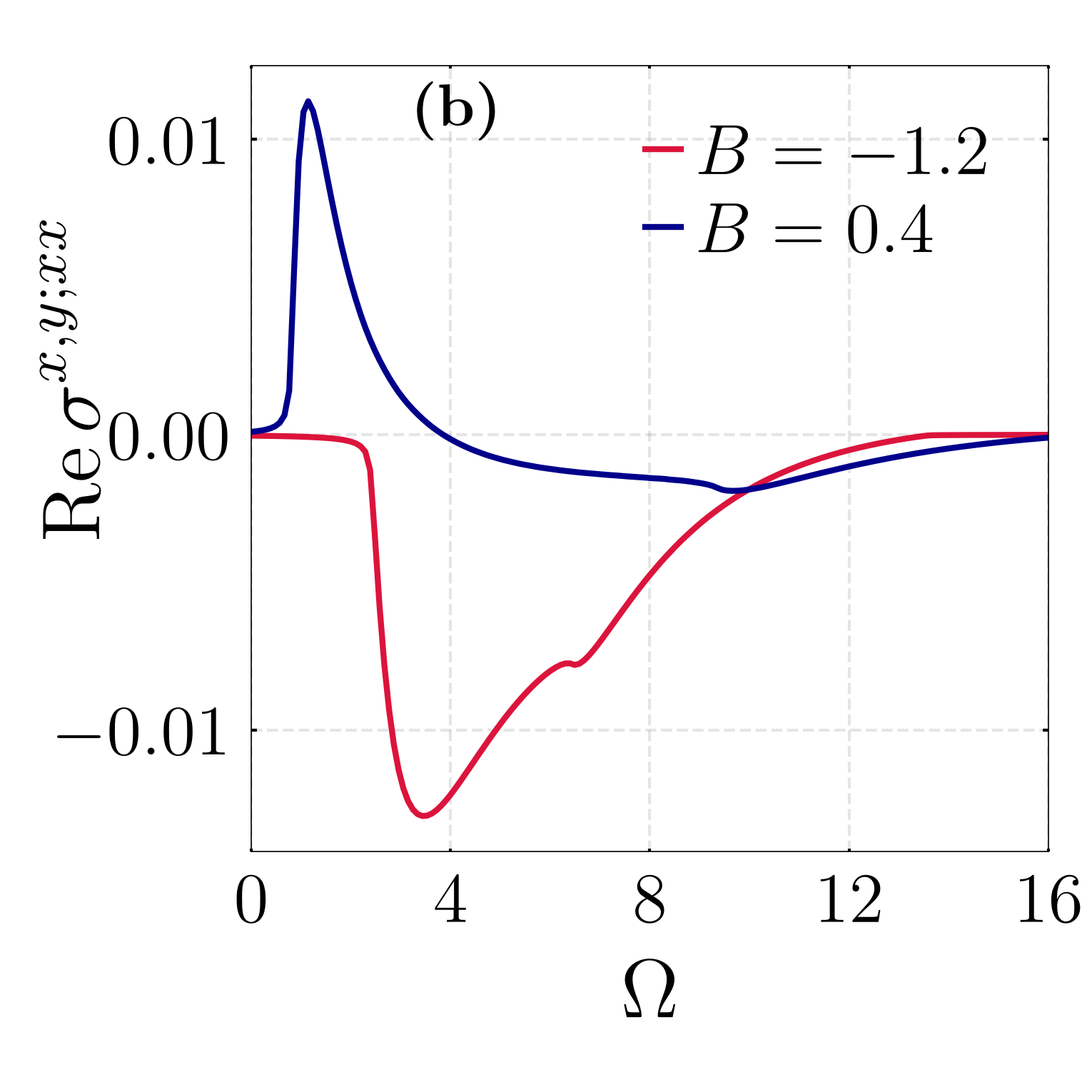}
  
  \vspace{0.15cm}

  \includegraphics[width=0.5\linewidth]{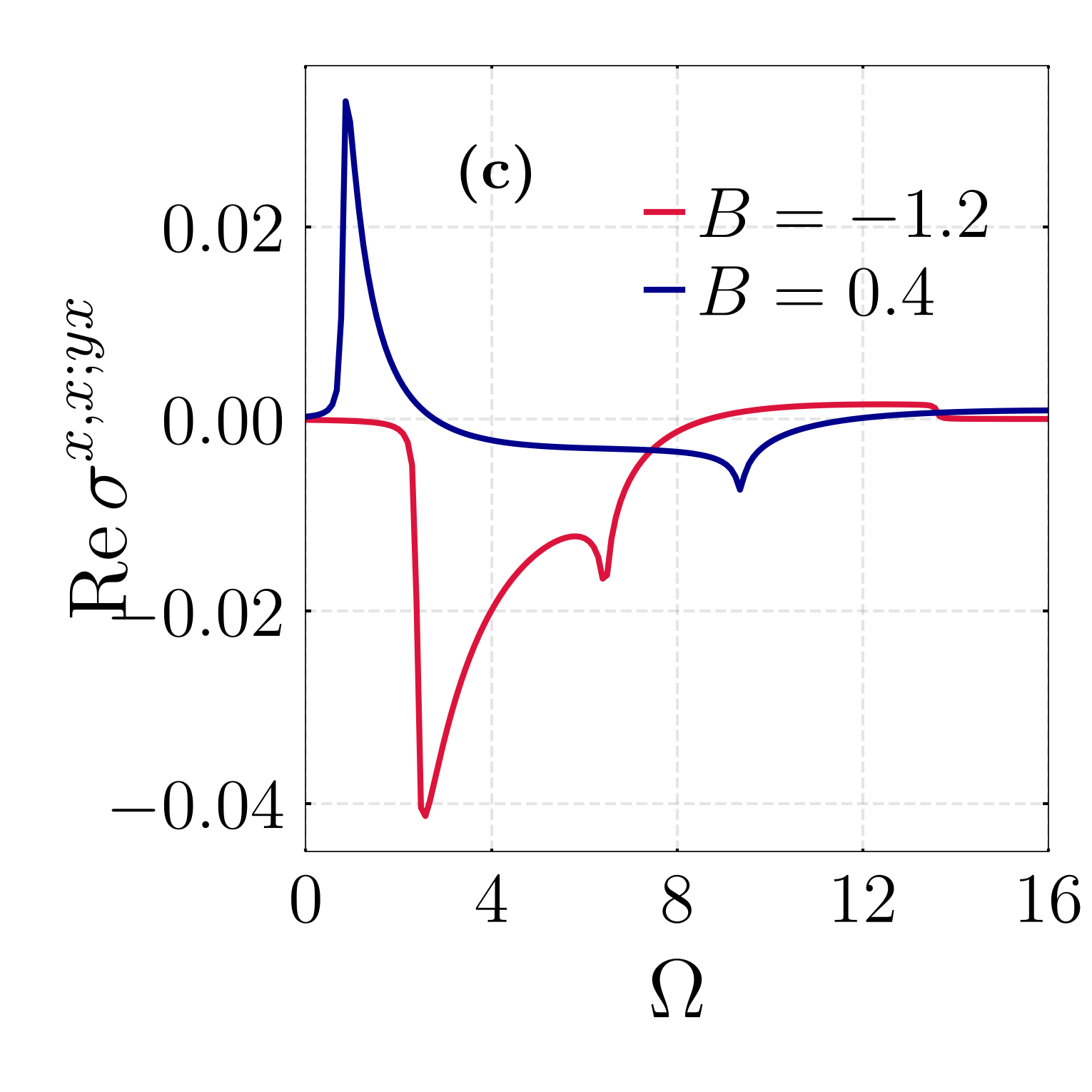}\hfill
  \includegraphics[width=0.5\linewidth]{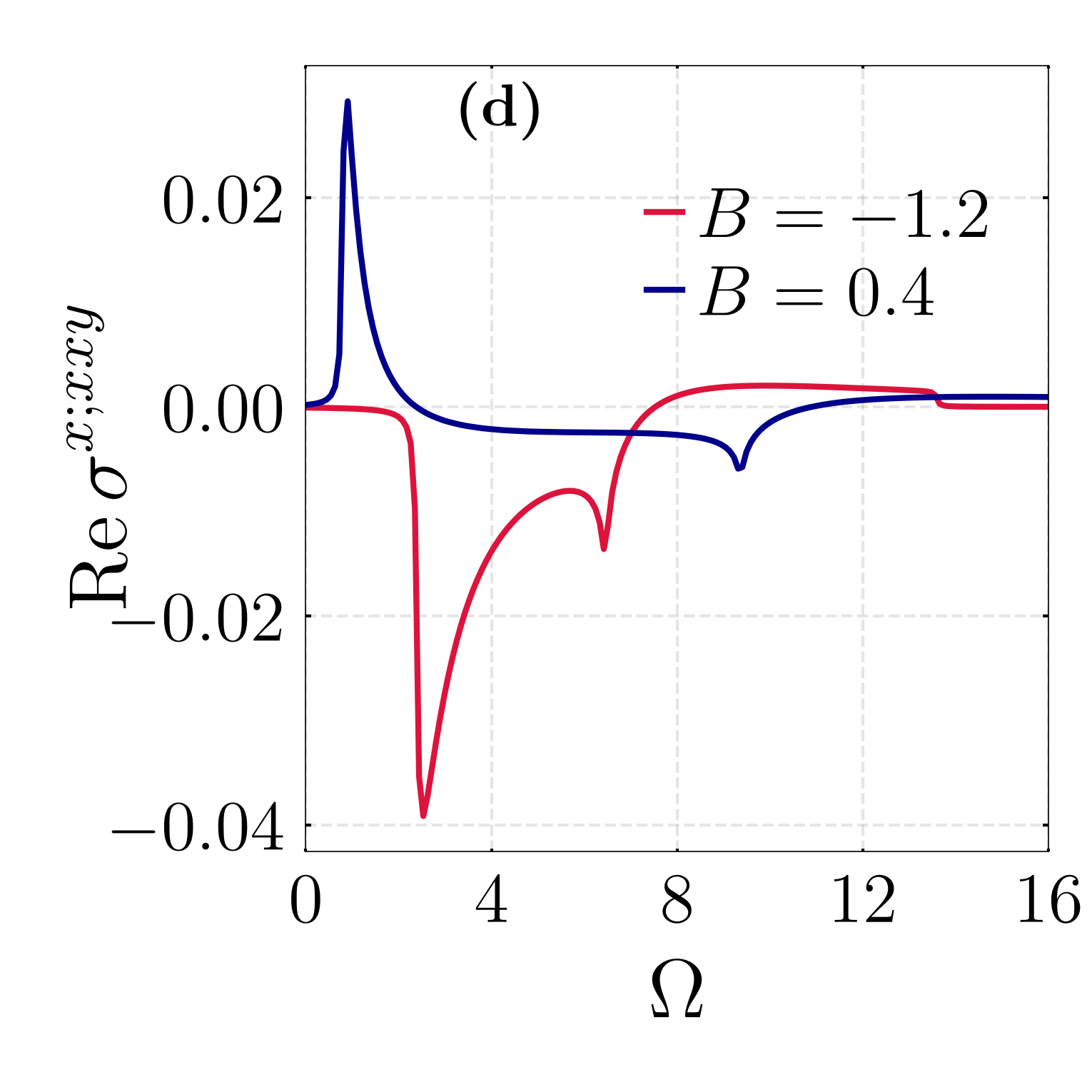}

  \includegraphics[width=0.5\linewidth]{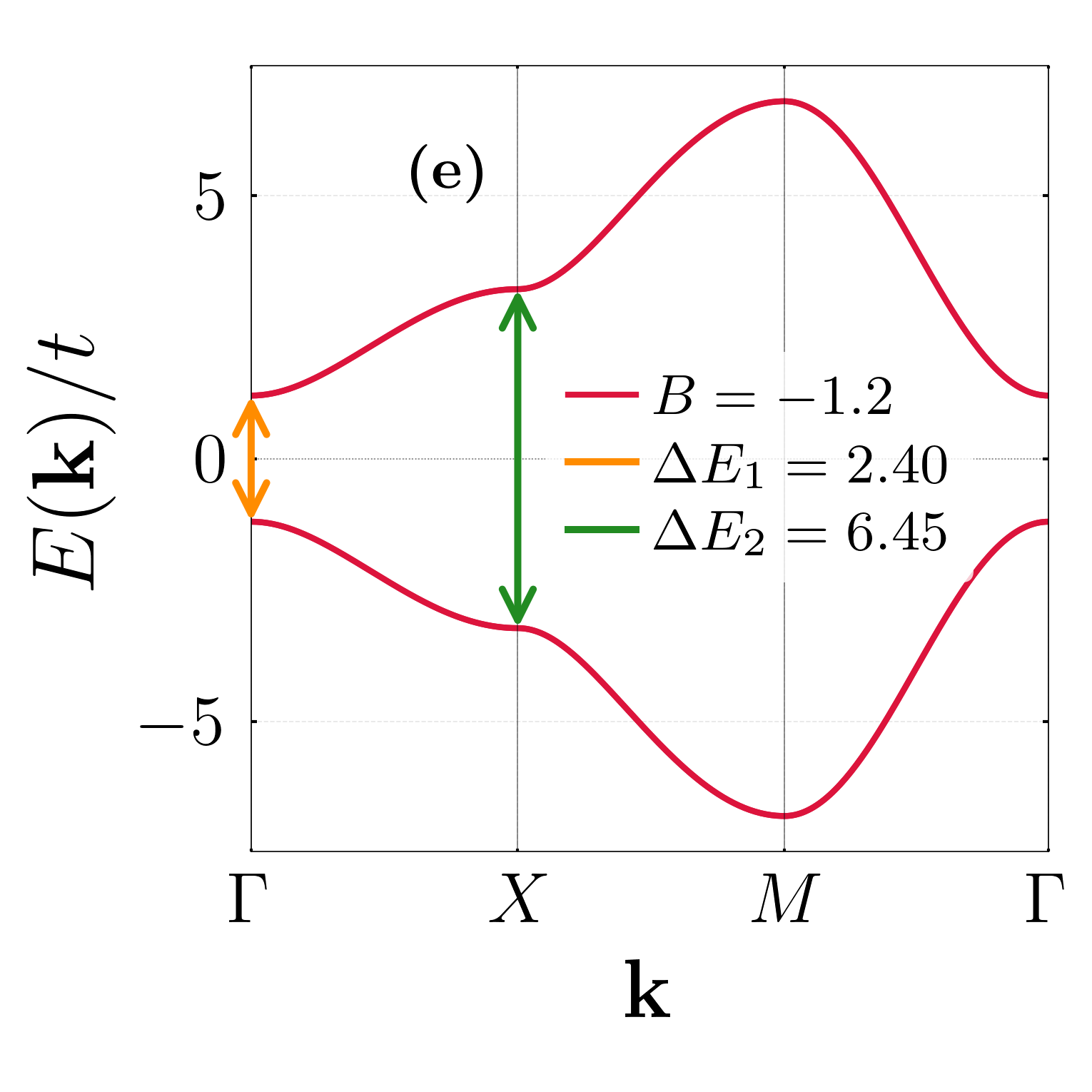}\hfill
  \includegraphics[width=0.5\linewidth]{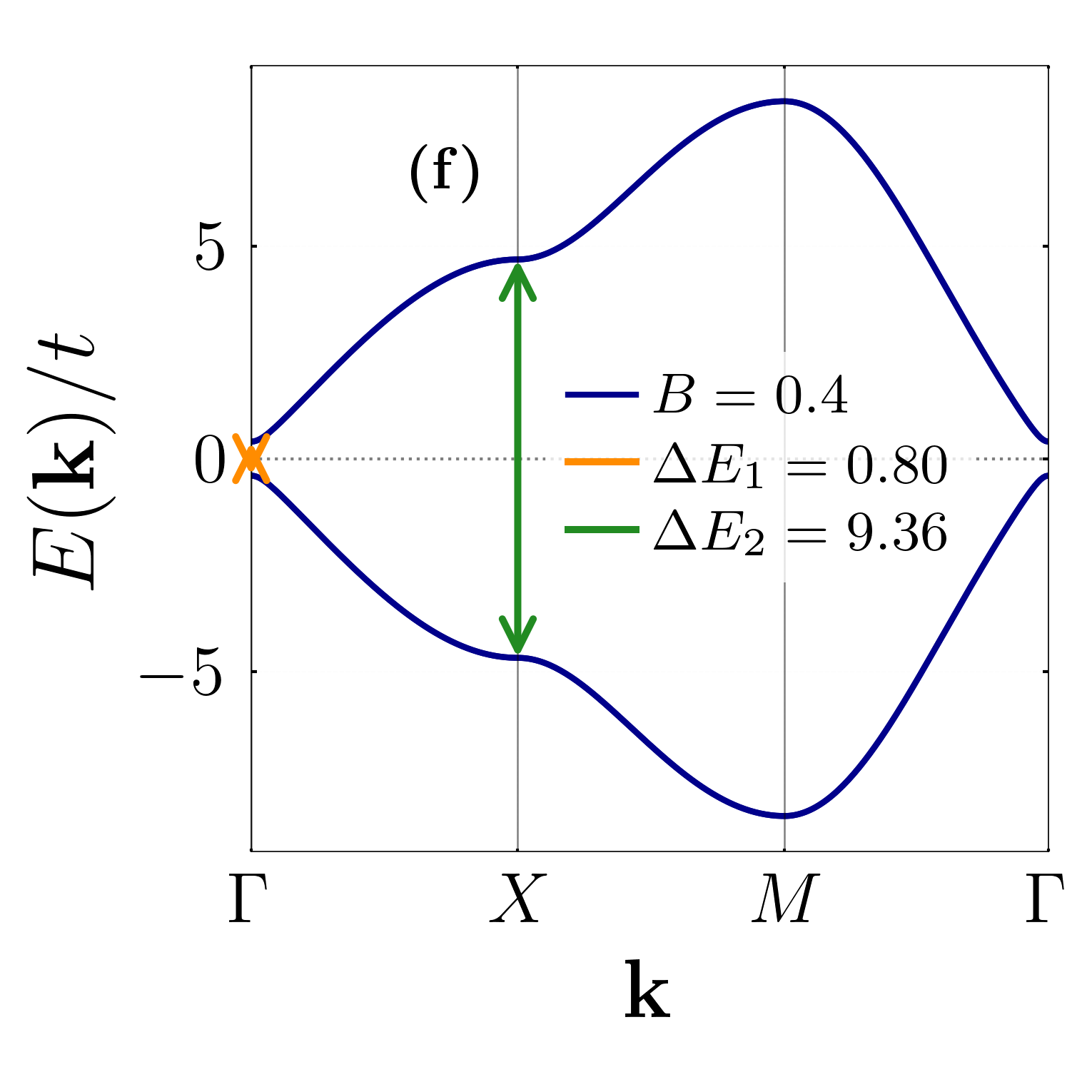}
\caption{Nonlinear pure spin shift conductivity across the mass parameter $B=0$: (a) $\mathrm{Re}\,\sigma^{x,y;yy}$, (b) $\mathrm{Re}\,\sigma^{x,y;xx}$,
(c) $\mathrm{Re}\,\sigma^{x,x;yx}$, and (d) $\mathrm{Re}\,\sigma^{x,x;xy}$ are plotted as a function of the driving frequency $\Omega$  during the topological phase transition near the
$\Gamma$ point, and (e) $\&$ (f) are the normalized energy band structure along the high-symmetry path $\Gamma(0,0)\!-\!X(\pi,0)\!-\!M(\pi,\pi)\!-\!\Gamma(0,0)$ of the Brillouin zone corresponding to the second-order topological (red color) and trivial phases (blue color), respectively. The energy difference between conduction and valence at the minima ($\Gamma$) and the saddle points $X$ or $Y$ in both phases gives the lowest- and highest-frequency peaks, respectively. The lowest-frequency peak exhibits a sign flip between the topological ($B=-1.2t$) and trivial ($B=0.4t$) regions, corresponding to the transition at the $\Gamma$ point. Remaining parameters are the same as Fig~\ref{fig2} with $J_a=0.8t$.}
\label{fig4}
\end{figure}
\begin{figure}
\centering
\begin{tabular}{c c}
\includegraphics[width=0.49\linewidth]{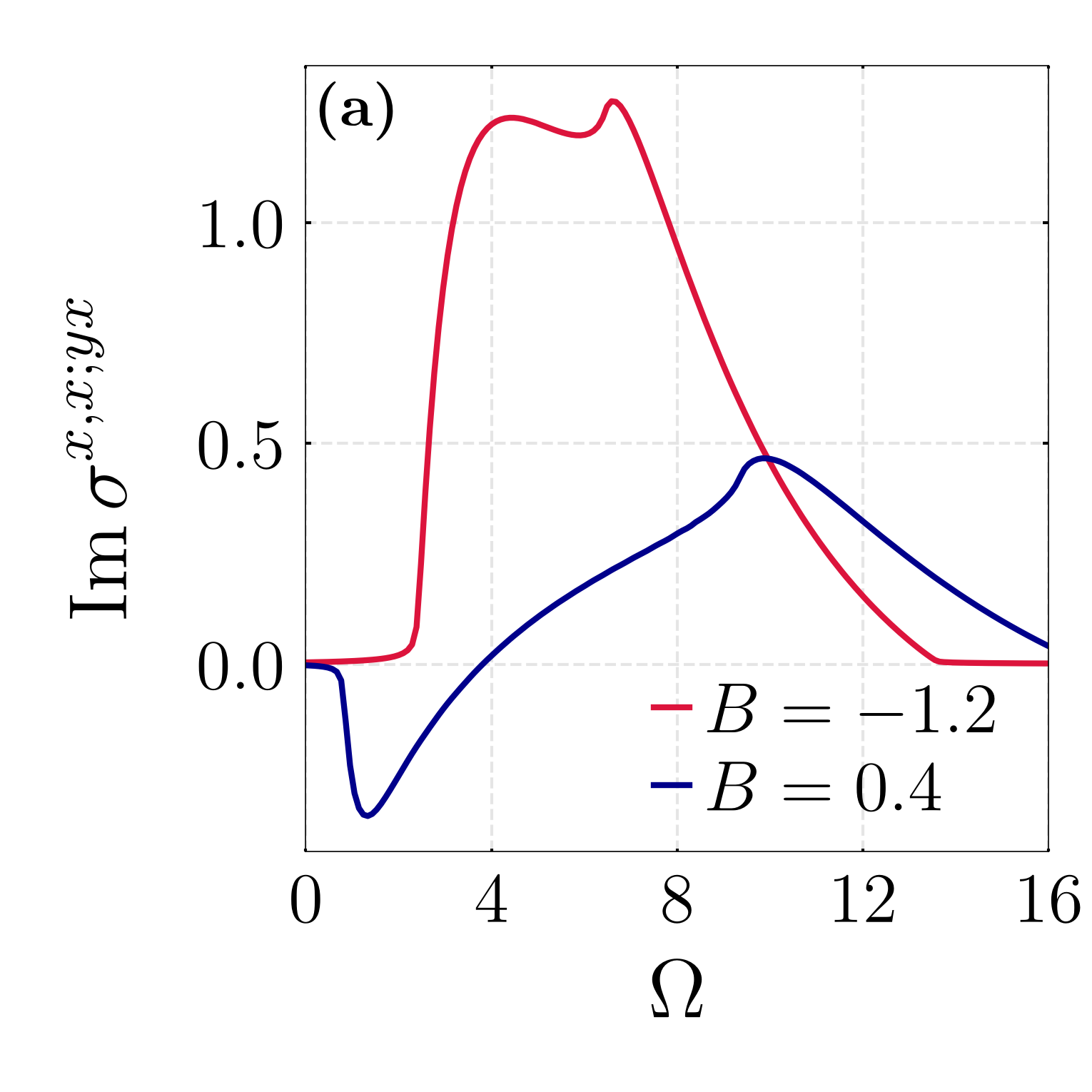}
& \includegraphics[width=0.49\linewidth]{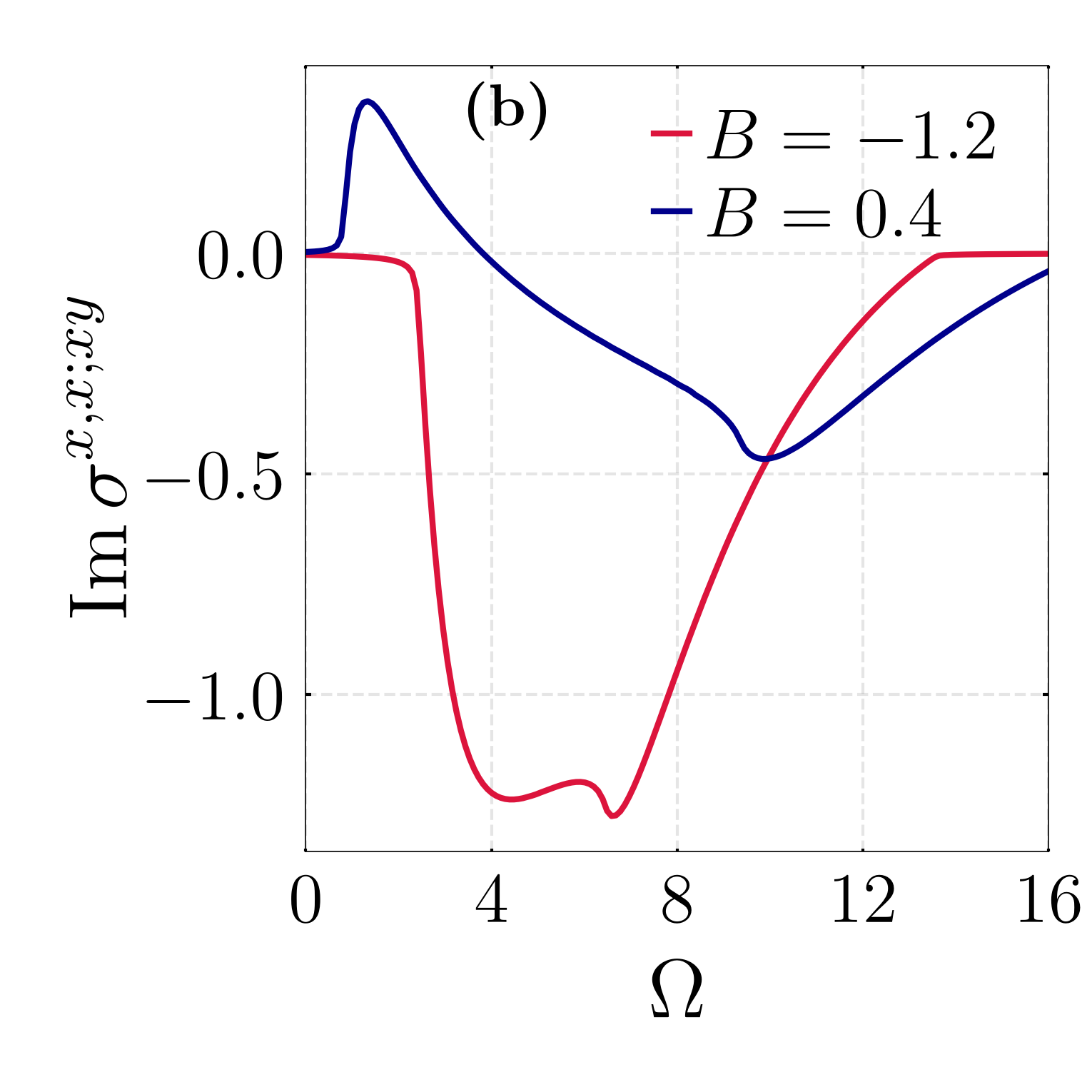}
\end{tabular}
\caption{Behavior of nonlinear pure spin injection conductivity across the topological transition around $B=0$ has been shown. Imaginary parts of the injection conductivity tensors (a) $\mathrm{Im}\,\sigma^{x,x;yx}$ and (b) $\mathrm{Im}\,\sigma^{x,x;xy}$ are plotted as a function of frequency. The results are compared for topological ($B=-1.2t$, red color) and trivial ($B=0.4t$, blue color) phases. The response is purely imaginary, and the lowest frequency peak undergoes a distinct sign change across the topological phase transition at the $\Gamma$ point. All other system parameters are fixed as in Fig.~\ref{fig2}.}
\label{fig5} 
\end{figure}

\begin{figure}[htbp]
\centering
\begin{tabular}{c c}
\includegraphics[width=0.5\linewidth]{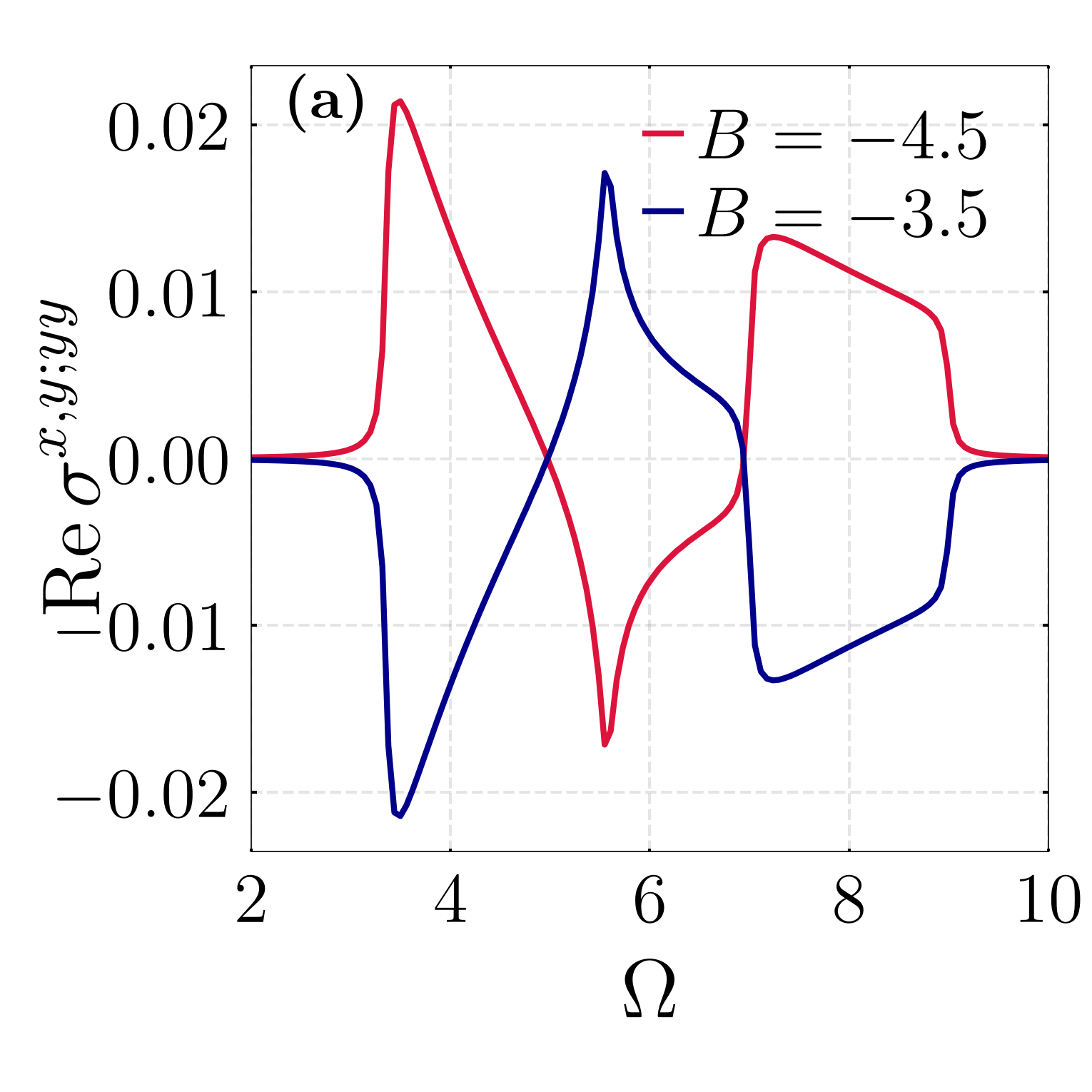} & \includegraphics[width=0.5\linewidth]{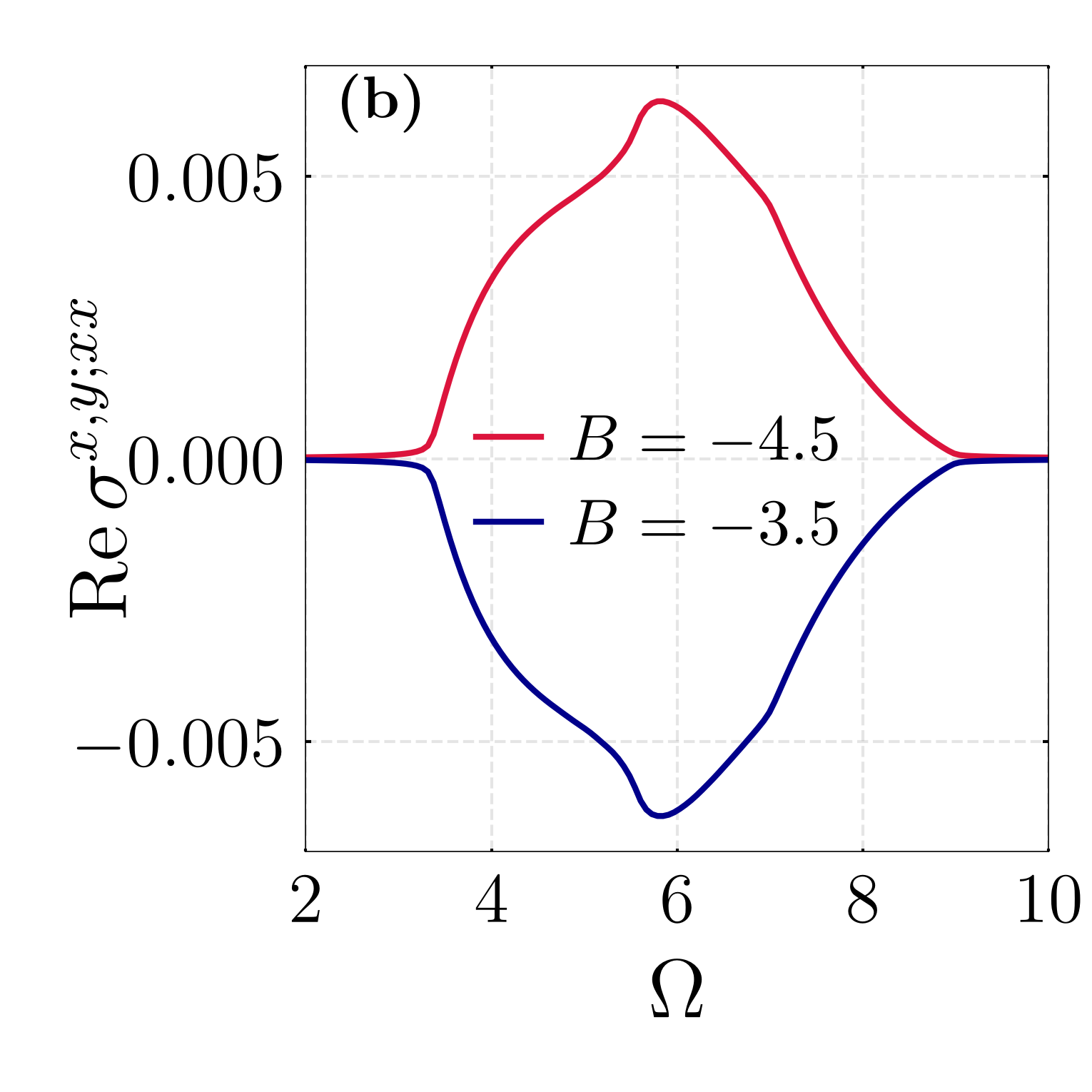} \\
\includegraphics[width=0.5\linewidth]{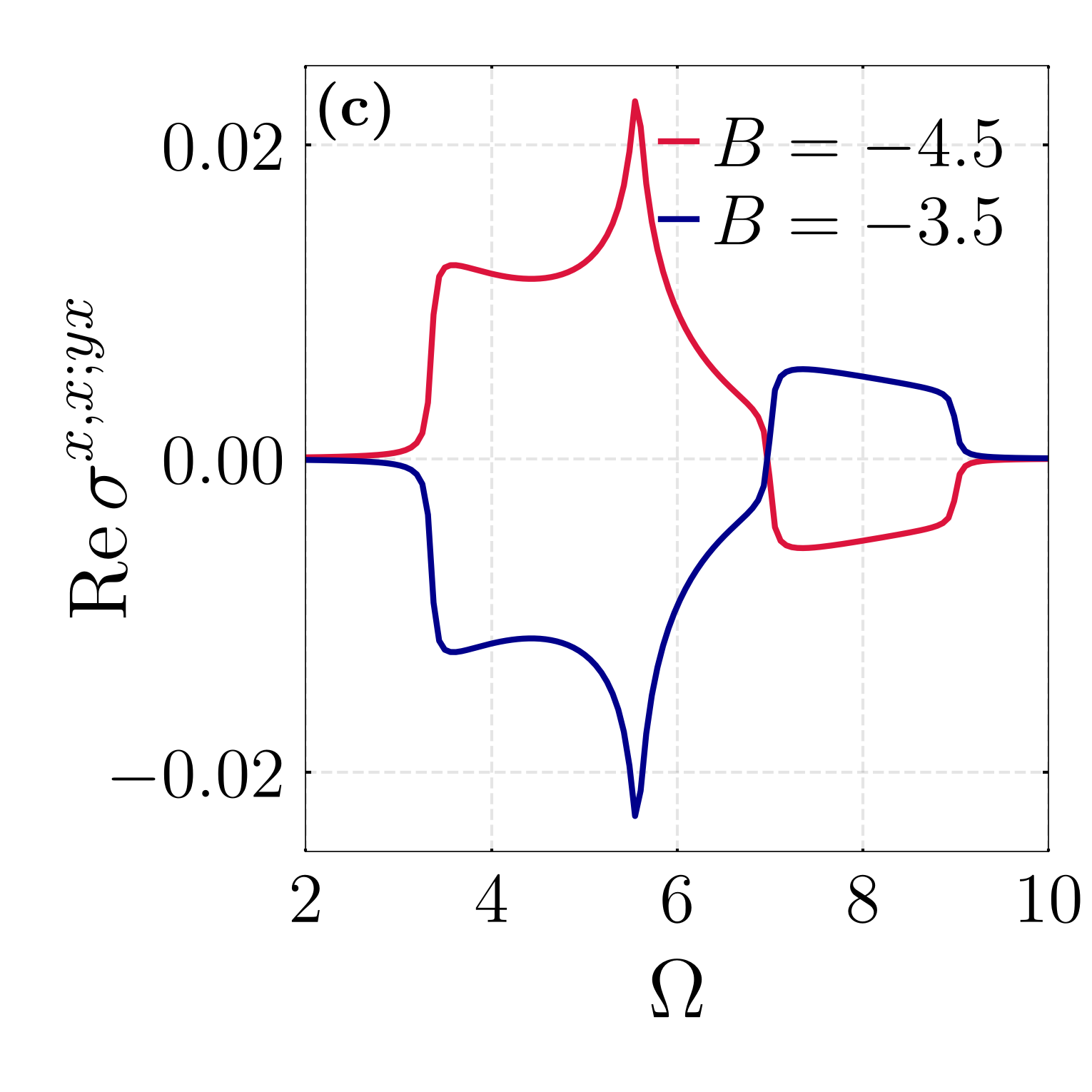} & \includegraphics[width=0.5\linewidth]{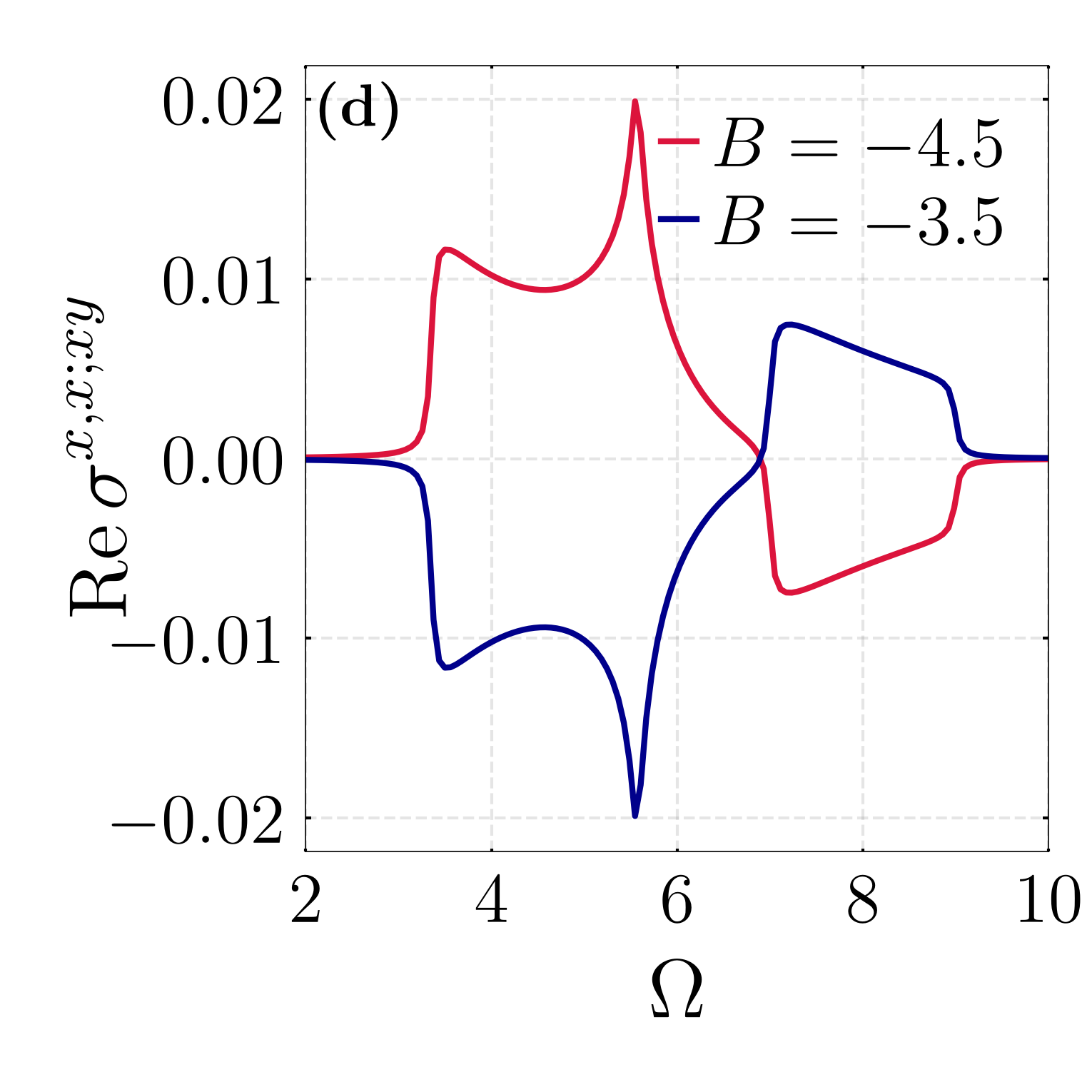}\\
\includegraphics[width=0.5\linewidth]{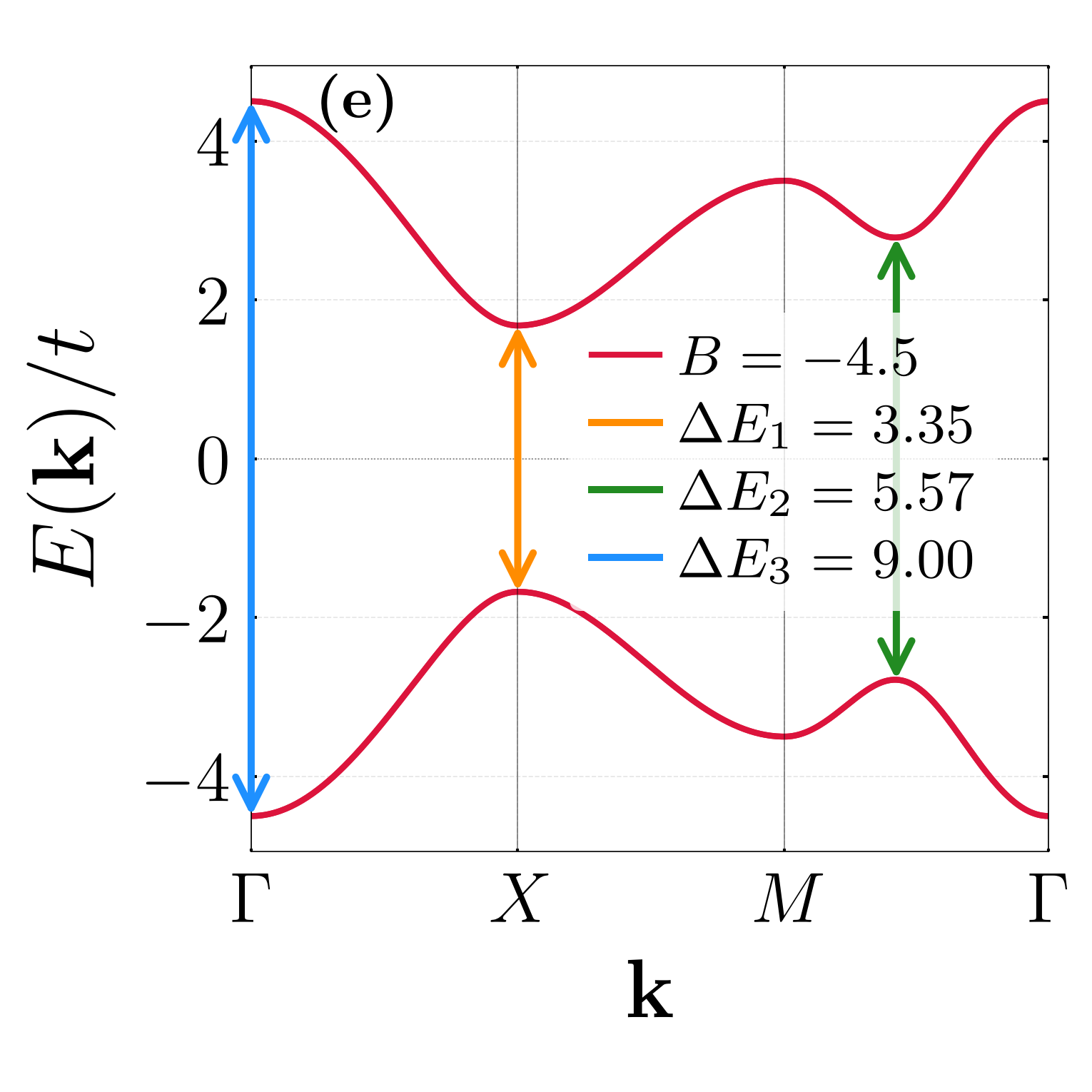} & \includegraphics[width=0.5\linewidth]{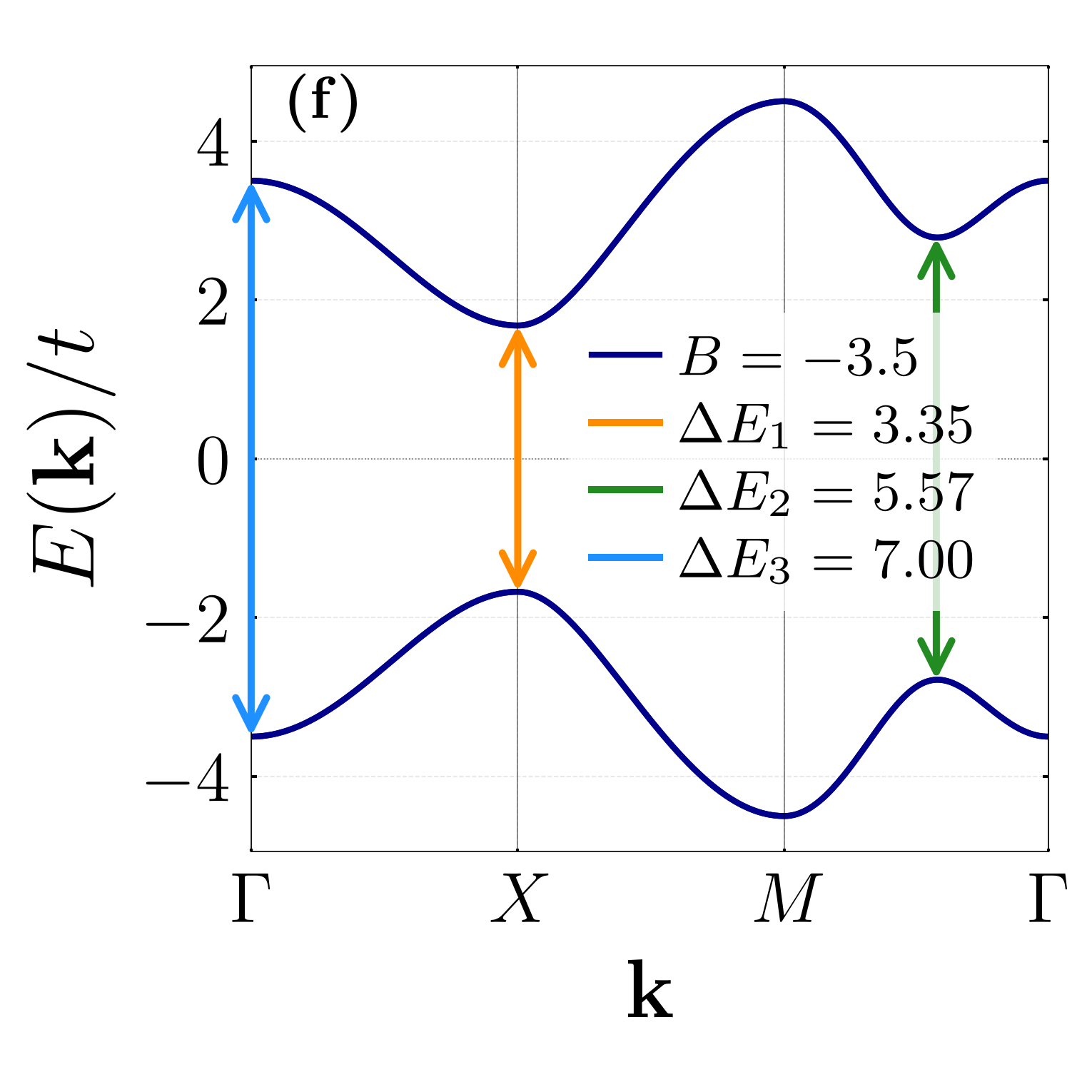}
\end{tabular}
\caption{Nonlinear pure spin shift conductivity across the mass parameter $B=-4t$: Panels
(a) $\mathrm{Re}\,\sigma^{x,y;yy}$, (b) $\mathrm{Re}\,\sigma^{x,y;xx}$,
(c) $\mathrm{Re}\,\sigma^{x,x;yx}$, and (d) $\mathrm{Re}\,\sigma^{x,x;xy}$ show the frequency dependence of the symmetry-allowed conductivity components. The red curve denotes the response for $B=-4.5t$  and the blue curve is for $B=-3.5t$. (e) and (f) show the normalized energy band structure along the high-symmetry path $\Gamma(0,0)\!-\!X(\pi,0)\!-\!M(\pi,\pi)\!-\!\Gamma(0,0)$ of the Brillouin zone for $B=-4.5t$ and $B=-3.5t$, respectively. The vertical colored arrows denote the key transition energy gaps. In contrast to the transition in
Fig. ~\ref{fig4}, where the dominant response originates at the $\Gamma$ point,
here the lowest-frequency contribution comes from the $X=(\pi,0)$ or $Y=(0,\pi)$ points. Also the lowest-frequency peak exhibits a sign flip as well.
The remaining parameters are the same as in Fig. ~\ref{fig2}.} 
\label{fig6}
\end{figure}

We now demonstrate analytically that both the spin shift and spin injection conductivities change sign across a band inversion. In the BHZ basis, the Hamiltonian in Eq.~(\ref{eq:total_hamiltonian}) can be written as a 
four-$\Gamma$ model, $\mathcal{H(}\mathbf{k})=\sum_{i=1}^{4}d_i(\mathbf{k})\Gamma_i$, with
$\Gamma_1=\kappa_z\sigma_0$, $\Gamma_2=\kappa_x\sigma_z$,
$\Gamma_3=\kappa_y\sigma_0$, $\Gamma_4=\kappa_x\sigma_x$, and 
$d_1=m(\mathbf{k})$ (band-inversion mass),  $d_2=A \sin k_x$, $d_3=A \sin k_y$, and  $d_4=J_a(\mathbf{k}) = -J_{a} (\cos k_x - \cos k_y) $. Notably, $\Gamma_i$'s matrices mutually anticommute. We define the operator $\mathcal{O}$ such that
\begin{equation}
\mathcal{O}=\Gamma_2\Gamma_3\Gamma_4=-i\,\kappa_y\sigma_y.
\label{Ope}
\end{equation}
It anticommutes with $\Gamma_1$ and commutes with $\Gamma_{2,3,4}$, consequently
\begin{equation}
\mathcal{O}\,\mathcal{H}(\mathbf{k})\,\mathcal{O}^{-1}=\mathcal{H}(\mathbf{k})\big|_{m\to-m},
\label{eq:massflip}
\end{equation}
which flips the sign of the mass term without altering the crystal momentum. Because the energy spectrum is invariant under $m\rightarrow -m$, the transition energies $\varepsilon_{ab}$, occupation differences $f_{ab}$, and the optical resonance factor $\delta(\varepsilon_{ba}-\Omega)$ are all even functions of $m$. Also, the velocity operators and Berry connections transform covariantly under the same mapping, $\mathcal{O} \ v^{\mu}(m)\ \mathcal{O}^{-1}=v^{\mu}(-m)$ and
$\mathcal{O}A^{\nu}(m)\mathcal{O}^{-1}=A^{\nu}(-m)$. The spin operator, however, changes sign. Using $\sigma_y\sigma_x\sigma_y=-\sigma_x$, one has
$\mathcal{O} S^{x}\mathcal{O}^{-1}=-S^{x}$, and hence every spin-current vertex
acquires a negative sign,
\begin{equation}
\mathcal{O} S^{x,\mu}(m) \mathcal{O}^{-1} = -S^{x,\mu}(-m),
\end{equation}
\begin{equation}
\mathcal{O}\,S^{x,\mu;\nu}(m)\,\mathcal{O}^{-1}=-S^{x,\mu;\nu}(-m).
\label{eq:spinflip}
\end{equation}
The shift
integrand $\mathcal{M}^{x,\mu;\nu\lambda}_{ab}$
[Eqs. (\ref{eq:shift2})-(\ref{eq:shift3})] and the injection integrand $I^{x,\mu;\nu\lambda}_{ab}$ [Eq. (\ref{eq:Injeq2})] 
each contain a single spin-current vertex, while all remaining factors are even under $m\rightarrow -m$. By inserting $\mathcal{O}^{-1}\mathcal{O}=\mathbb{I}$ into the trace expressions and using cyclicity of the trace, both kernels acquire an overall minus sign under the mass-flip transformation. As a result, the corresponding spin-photovoltaic conductivities are odd functions of the mass term inversion,

\begin{equation}
\sigma^{x,\mu;\nu\lambda}_{\rm shift/inj}(m)
=-\,\sigma^{x,\mu;\nu\lambda}_{\rm shift/inj}(-m).
\label{eq:oddsigma}
\end{equation}
Therefore, both the spin shift and spin injection conductivities reverse sign across a band inversion, providing a direct nonlinear optical signature of the underlying topological transition.

Fig. \ref{fig4} illustrates the four symmetry-allowed spin shift conductivity components with spin polarization along the $x$ direction. Due to $\mathcal{PT}$ symmetry, the spin shift conductivity tensor is purely real, indicating that the response is generated exclusively by LPL. For $B=0$, the bulk gap closes at the $\Gamma (0,0)$ point, marking the phase boundary between distinct topological phases. Consequently, the nonlinear optical response is dominated by transitions near the $\Gamma$ point (see Fig.~\ref{fig3}). As the system is tuned across the transition point $B=0$, the conductivity peak reverses sign, directly reflecting the band inversion occurring at the $\Gamma$ point. This sign change is clearly visible at the lowest frequency in spin shift (injection) conductivity, see Fig.~\ref{fig4} (a)-(d) (Fig.~\ref{fig5} (a)-(b)). Furthermore, because the shift conductivity tensor is purely real, it is symmetric under the interchange of the electric-field indices. As a result, the tensor components $\sigma^{x,x;yx}$ and $\sigma^{x,x;xy}$ are exactly equal.  

The bulk gap also closes at the $X(\pi,0)$ and $Y(0,\pi)$ points when $B =-4t$ in the absence of altermagnetism ($J_a=0$), since the local mass term $m(\mathbf{k})=B+4t$ vanishes at these momenta. In the SOTI phase, however, the altermagnetic term prevents a true gap closing and maintains a finite excitation gap of magnitude $2J_a$. Nevertheless, the effective mass $m(\mathbf{k})$ still changes sign across $B = -4t$, leading to a corresponding sign reversal of the spin dc conductivity as shown in Fig. \ref{fig6}. A similar sign change is observed at $B=-8t$, where the bulk gap closes at the $M(\pi,\pi)$ point and a band inversion drives a topological phase transition. As in the case of the spin shift conductivity, upon crossing $B=-4t$ and $B=-8t$, the injection conductivity changes sign at the lowest frequency.

\section{Effect of N\'eel Vector Rotation on the optical response}
\label{Sec4b}
In this section, we investigate how the optical response evolves under the rotation of the N\'eel vector. For a general in-plane orientation of the altermagnetic order parameter, the altermagnetic term in Eq.~(\ref{BHZ2}) can be written as

\begin{align}
J(\mathbf{k}) &=- J_{a} (\cos k_x - \cos k_y)\, (\kappa_x \otimes (\hat{n}_s\cdot\vec{\sigma}))   . 
\end{align}
where the N\'eel vector is parameterized as $\hat{n}_s=(\cos\phi,\sin\phi,0)$. As long as the N\'eel vector remains within the $xy$ plane, the combined symmetry $\mathcal{C}_{4z}\mathcal{T}$ is preserved. Consequently, the system remains in the SOTI phase and supports zero-energy corner states for arbitrary in-plane orientations of the N\'eel vector. The mirror symmetries for a general direction of the N\'eel vector can be given as 
\begin{equation}
\mathsf{M}_x=\kappa_0\otimes(\hat{n}_s\cdot\vec\sigma) , \quad \mathsf{M}_y= \kappa_z\otimes(\hat{n}^\perp_s\cdot \vec \sigma ), 
\end{equation}
where 
\begin{equation}
\hat{n}^\perp_s=(-\sin\phi,\cos\phi,0).
\end{equation}
The product of the two mirror symmetries, $\mathcal{C}_{2z}=\mathsf{M}_{x} \mathsf{M}_{y}$,
remains a symmetry of the system for any in-plane orientation of the N\'eel vector. Since $\mathcal{C}_{2z}$ reverses both in-plane coordinates, it forbids all second-order charge-photocurrent responses, while still allowing pure spin-photocurrent responses \cite{PhysRevB.109.245306}.

\begin{figure}
\centering
\begin{tabular}{c c}
\includegraphics[width=0.49\linewidth]{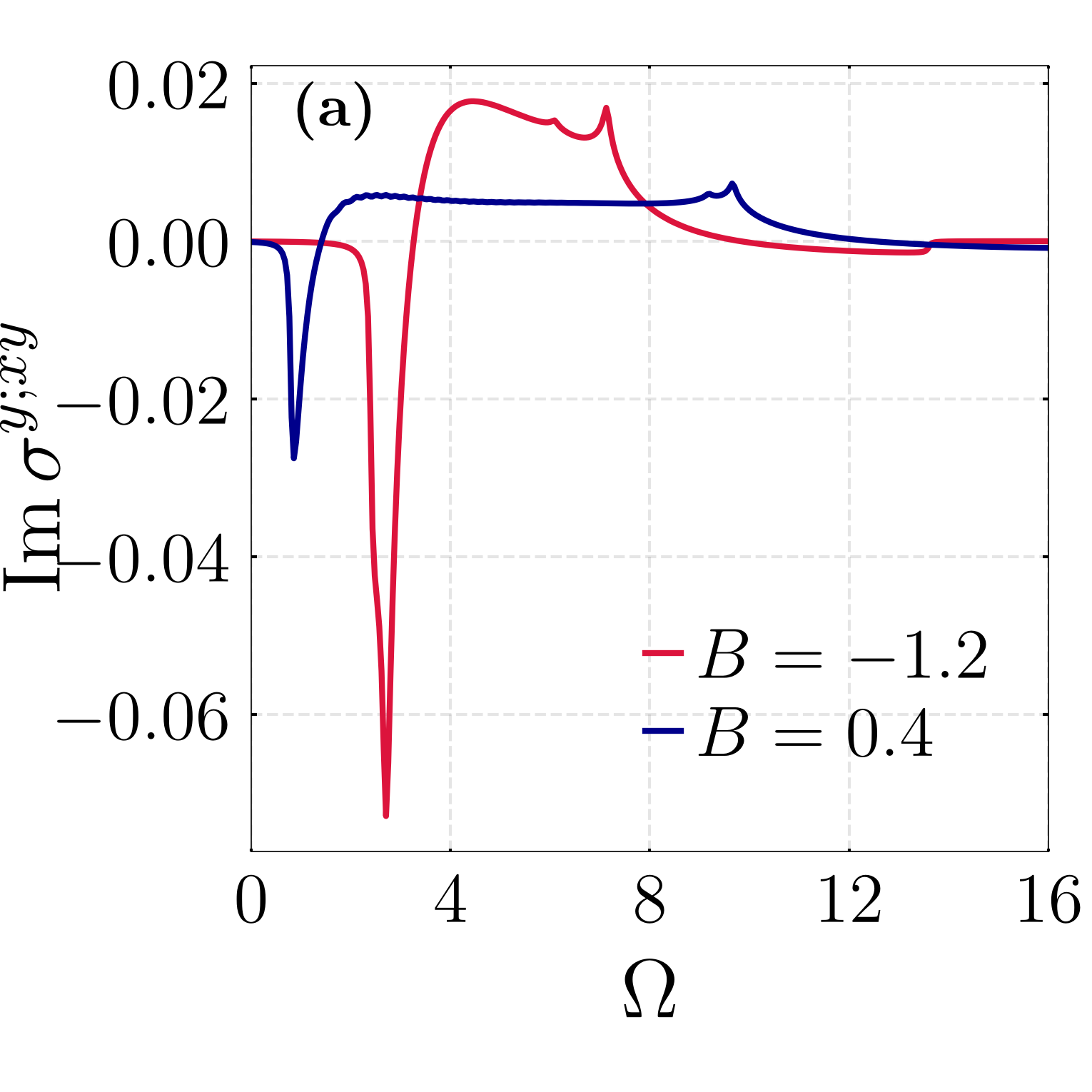}
& \includegraphics[width=0.49\linewidth]{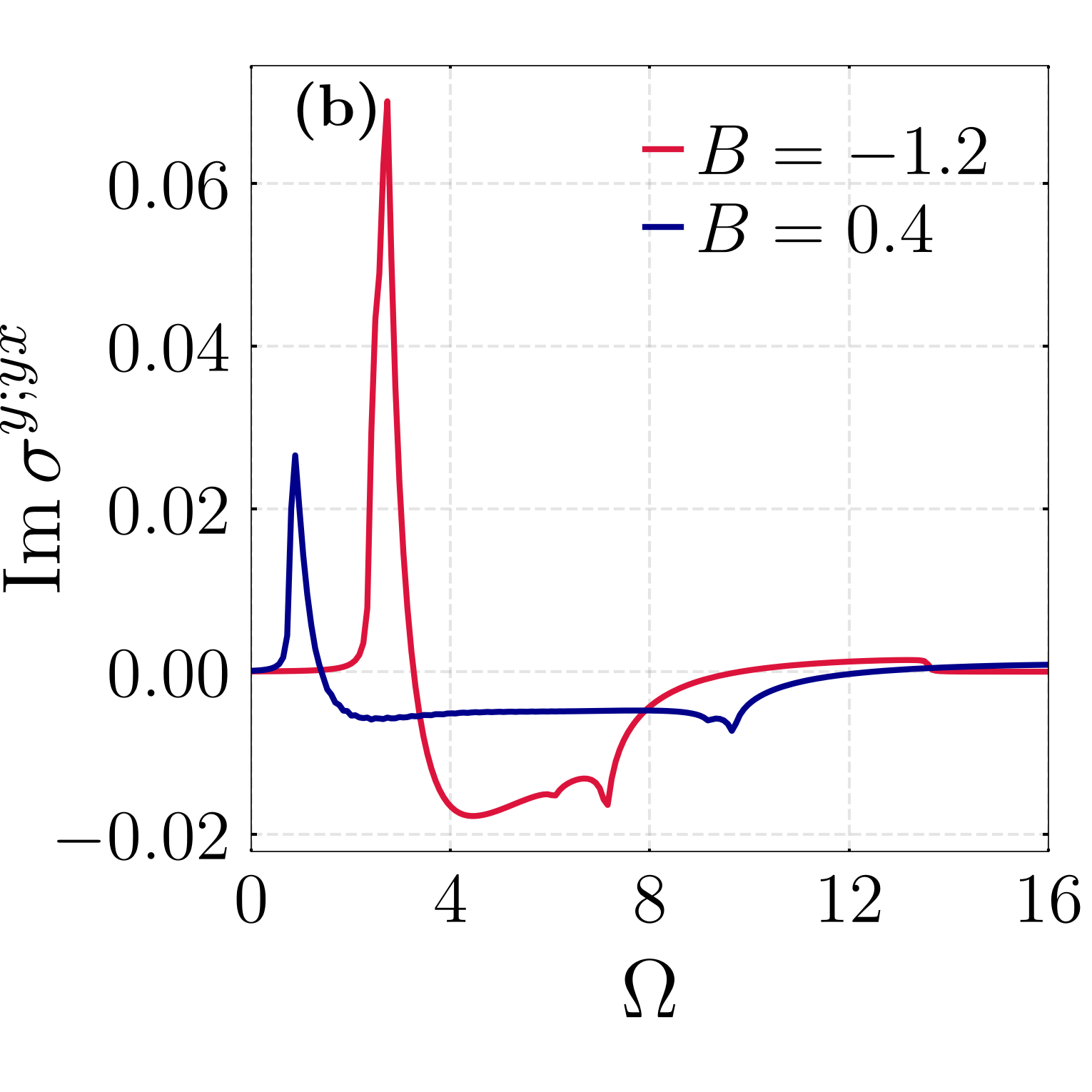}
    \end{tabular}
     \caption{Frequency spectrum of nonlinear charge shift (gyration) conductance under an out-of-plane
(z-aligned) Néel configuration. The imaginary components of the nonlinear conductivity tensors, (a) $\mathrm{Im}\,\sigma^{yxy}$,
(b) $\mathrm{Im}\,\sigma^{yyx}$, are evaluated as a function of the driving frequency $\Omega$ for both the topological phase ($B=-1.2t$, red curve) and the trivial phase ($B=0.4t$, blue curve). The out-of-plane alignment of the Néel vector unlocks a charge response that is symmetrically forbidden in the in-plane configuration. These imaginary components correspond to the gyration current generated by CPL in the $\mathcal{PT}$- symmetric system. Notably, unlike the spin shift response, the gyration current does not undergo a sign reversal across the topological transition. The remaining model parameters are the same as Fig.~\ref{fig2}.}
\label{fig7} 
\end{figure}

The chiral symmetry also rotates together with the N\'eel vector and is given by $\mathbb{S}(\phi)=\kappa_x\otimes(\hat{n}^\perp_s\cdot \vec{\sigma})$.
Consequently, the symmetry constraints on the spin-conductivity tensor rotate with the magnetic order parameter. The only symmetry-allowed spin-photocurrent polarization is parallel to the N\'eel vector $\hat{n}_s$
, whereas the component polarized along the orthogonal direction $\hat{n}^\perp_s$ is forbidden. For example, when the altermagnetic order is oriented along the y-axis $\phi=\pi/2$, the nonvanishing spin-conductivity components are polarized exclusively along $y$ (see Table~\ref{tab:transposed_symmetry}), while the $x$-polarized components vanish by symmetry.

We now consider the case where the N\'eel vector is oriented towards the $z$ axis. In this case, $\mathcal{C}_{2z}$ is broken, allowing a nonzero second-order charge photocurrent response. As the system remains $\mathcal{PT}$ symmetric, which implies that the charge shift conductivity is purely imaginary and therefore couples to CPL, while the charge injection conductivity is purely real and couples to LPL.  We can recover the charge shift-current response from Eq.~(\ref{eq:response}) by replacing the spin-current vertices with the corresponding charge-current vertices. In this limit, the symmetrized operators reduce to the bare velocity operators, $S^{s,\mu}\to v^{\mu}$ and
$S^{s,\mu\nu}\to v^{\mu\nu}$, and response kernel become
\begin{equation}
\mathcal{M}^{s,\mu;\nu\lambda}_{ab}\to \mathcal{M}^{\mu;\nu\lambda}_{ab}
=\mathrm{Tr}\!\big[W^{\mu;\nu}_{ab}v^{\lambda}_{ba}-W^{\mu;\lambda}_{ba}v^{\nu}_{ab}\big],
\end{equation}
where
\begin{equation}
W^{\mu;\nu}_{ab}=v^{\mu\nu}_{ab}-v^{\mu}_{ab}\frac{\Delta^{\nu}_{ab}}{\varepsilon_{ab}},
\end{equation}
and the conductivity therefore reduces to the charge shift-current response,
\begin{equation}
\sigma^{\mu;\nu\lambda}_{\text{shift}}=-\frac{i\pi q^{3}}{2}
\int\frac{d^dk}{(2\pi)^d}\sum_{a\neq b}\frac{f_{ab}}{\varepsilon_{ab}^{2}}\,
\mathcal{M}^{\mu;\nu\lambda}_{ab}\,\delta(\varepsilon_{ab}-\Omega),
\label{eq:charge_shift}
\end{equation}

In the nondegenerate limit, each manifold contains only a single band. Consequently,  all quantities reduce to scalar band-resolved matrix elements. In this limit, the present formalism recovers the conventional two-band expression for the charge shift conductivity reported in Ref.~\cite{Cook2017}. Further, the mirror symmetry $\mathsf{M}_y$($=i\kappa_z\sigma_y$) remains intact, under which the charge conductivity   Eq.~(\ref{eq:charge_shift}) and spin shift conductance   Eq.~(\ref{eq:response}) transform as
\begin{equation}
\mathcal{M}^{\mu;\nu\lambda}_{ab}\xrightarrow{\mathsf{M}_y}(-1)^{\delta_{\mu y}+\delta_{\nu y}+\delta_{\lambda y}} \, \mathcal{M}^{\mu;\nu\lambda}_{ab},
\end{equation}
\begin{equation}  \mathcal{M}^{s,\mu;\nu\lambda}_{ab}\xrightarrow{\mathsf{M}_y}(-1)^{1+\delta_{sy}+ \delta_{\mu y}+\delta_{\nu y}+\delta_{\lambda y }}\, \mathcal{M}^{s,\mu;\nu\lambda}_{ab}.
\end{equation}
We note that a nonzero charge conductivity requires the quantity $\delta_{\mu y}+\delta_{\nu y}+\delta_{\lambda y} $ to be even. Consequently, only those tensor components containing an even number of $y$-indices are symmetry allowed. The remaining charge-conductivity components are therefore $\sigma^{y:yx},\sigma^{y:xy},\sigma^{x:yy}$, and $\sigma^{x:xx}$. Furthermore, $\mathcal{PT}$ symmetry constrains the charge conductivity tensor to be purely imaginary, implying that the response couples exclusively to CPL. As a result, the antisymmetric components $\sigma^{y;yx},\sigma^{y;xy}$ remain nonvanishing, whereas the symmetric components do not contribute.

However,  non-zero spin shift conductivity components allowed by $\mathsf{M}_y$ are:
\begin{align}
 s=x:\quad & \sigma^{x;xxy},\ \sigma^{x;xyx},\ \sigma^{x;yxx},\ \sigma^{x;yyy},\nonumber\\
 s=y:\quad & \sigma^{y;xxx},\ \sigma^{y;xyy},\ \sigma^{y;yxy},\ \sigma^{y;yyx},\nonumber\\
s=z:\quad & \sigma^{z;xxy},\ \sigma^{z;xyx},\ \sigma^{z;yxx},\ \sigma^{z;yyy}.
\end{align}

The system also possesses chiral symmetries ($\mathbb{S}_x=\kappa_x\sigma_y \ \text{and} \ \mathbb{S}_y =-\kappa_x\sigma_x $) such that 
\begin{equation}
\sigma^{s,\mu;\nu\lambda} \xrightarrow{\mathbb{S}_x} (-1)^{1 + \delta_{sx}+\delta_{sz}} \sigma^{s,\mu;\nu\lambda},
\end{equation}
\begin{equation}
\sigma^{s,\mu;\nu\lambda} \xrightarrow{\mathbb{S}_y} (-1)^{1 + \delta_{sy}+\delta_{sz}} \sigma^{s,\mu;\nu\lambda}.
\end{equation}
The chiral symmetry $\mathbb{S}_x$ suppresses all spin-conductivity components with spin polarization along the $y$ direction, while $\mathbb{S}_y$ eliminates all components polarized along the $x$ direction. Consequently, the only remaining spin response is polarized along the $z$ axis. Together with the mirror-symmetry constraints, this leaves four nonvanishing spin conductivities exclusively along the $z$ direction, as summarized in Table~\ref{tab:transposed_symmetry}.

As shown in Figs.~\ref{fig7} and \ref{fig8}, the spin shift conductivity continues to exhibit a sign reversal across the topological transition, whereas the charge conductivity does not. This contrasting behavior can be understood analytically as follows. For a N\'eel vector aligned along the $z$-direction, the Hamiltonian in Eq.~(\ref{eq:total_hamiltonian}) simplifies to a three-$\Gamma$ model, $\mathcal{H(}\mathbf{k})=\sum_{i=1}^{3}d_i(\mathbf{k})\Gamma_i$, with
$\Gamma_1=\kappa_z\sigma_0$, $\Gamma_2=\kappa_x\sigma_z$,
and $\Gamma_3=\kappa_y\sigma_0$; and 
$d_1=m(\mathbf{k})$ (band-inversion mass),  $d_2=A \sin k_x - J_{a} (\cos k_x - \cos k_y)$, $d_3=A \sin k_y$. Now, we can define an analogous operator defined in Eq. (\ref{Ope}), $\mathcal{O}=\kappa_y\sigma_x$ or $\mathcal{O}=\kappa_y\sigma_y$, such that 
\begin{equation}
\mathcal{O}\,\mathcal{H}(\mathbf{k})\,\mathcal{O}^{-1}=\mathcal{H}(\mathbf{k})\big|_{m\to-m},
\label{eq:massflipz}
\end{equation}
\begin{equation}
    \mathcal{O} \ v^{\mu}(m)\ \mathcal{O}^{-1}=v^{\mu}(-m).
\end{equation}
Also as  
$\sigma_x\sigma_z\sigma_x=-\sigma_z$, one has
$\mathcal{O} S^{z}\mathcal{O}^{-1}=-S^{z}$, and hence every spin-current vertex
acquires a negative sign,
\begin{equation}
\mathcal{O} S^{z,\mu}(m) \mathcal{O}^{-1} = -S^{z,\mu}(-m),
\end{equation}
\begin{equation}
\mathcal{O}\,S^{z,\mu;\nu}(m)\,\mathcal{O}^{-1}=-S^{z,\mu;\nu}(-m).
\label{eq:spinflipz}
\end{equation}
As the charge conductivity depends only on the velocity vertex, $\sigma^{\mu;\nu \lambda}(m)=\sigma^{\mu;\nu \lambda}(-m)$, thus there is no sign change for the mass term inversion.  However, spin conductivity contains a spin-current vertex, and thus, similar to the previous section, shows a sign flip for mass term inversion. In a similar way, injection conductivity written in Eq.~(\ref{eq:Injeq2}), for a charge case, reduces to
\begin{align}
\mathcal{I}^{\mu;\nu\lambda}_{ab}=\text{Tr}\Bigl[\bigl(v^{\mu}_{aa}A^{\lambda}_{ab}-A^{\lambda}_{ab}v^{\mu}_{bb}\bigr)A^{\nu}_{ba}\Bigr].
\end{align}
For a non-degenerate limit, all quantities reduce to scalar matrix elements which recover the conventional expression for charge injection conductivity \cite{PhysRevX.11.011001}.

\begin{figure}
\centering
\begin{tabular}{c c}
\includegraphics[width=0.49\linewidth]{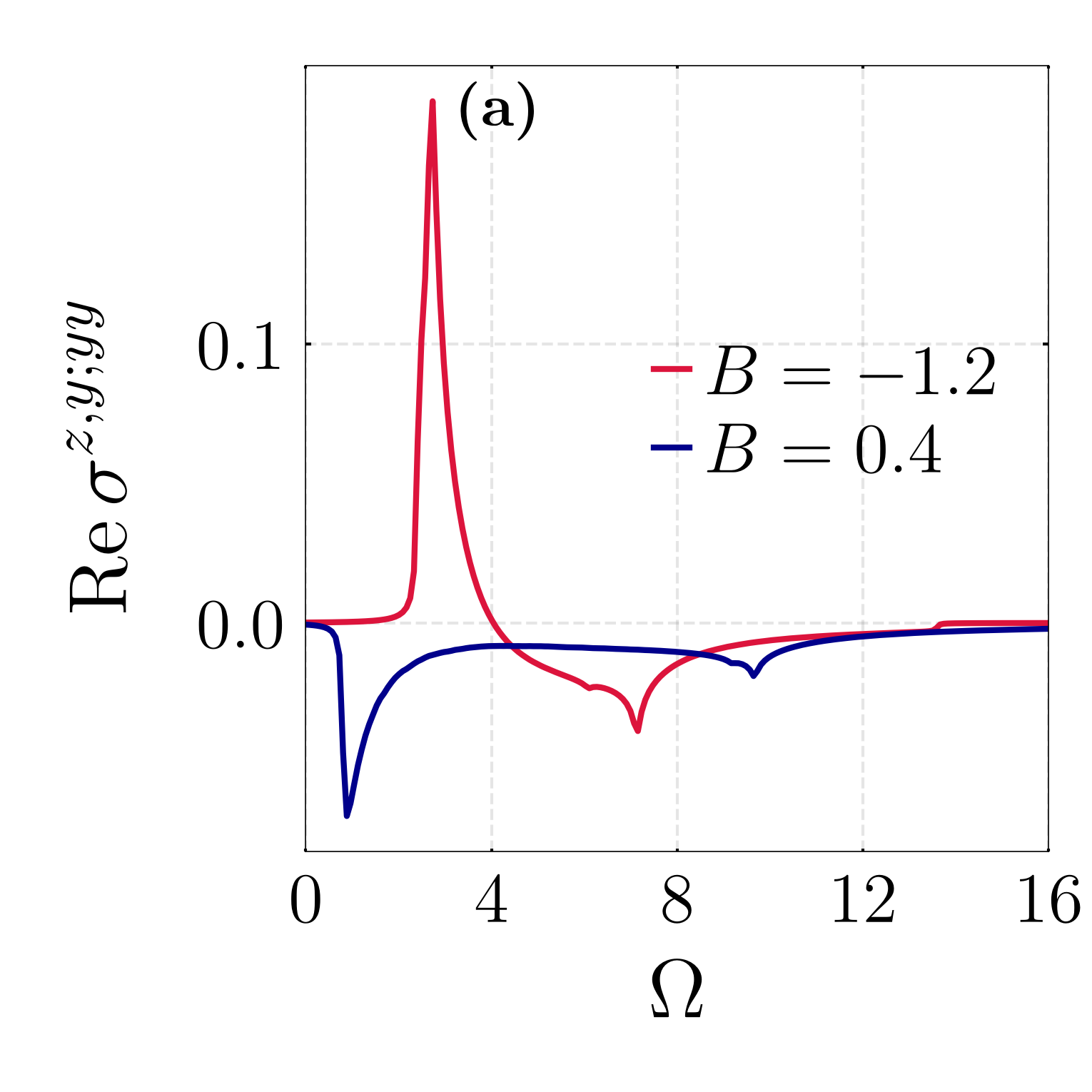}
& \includegraphics[width=0.49\linewidth]{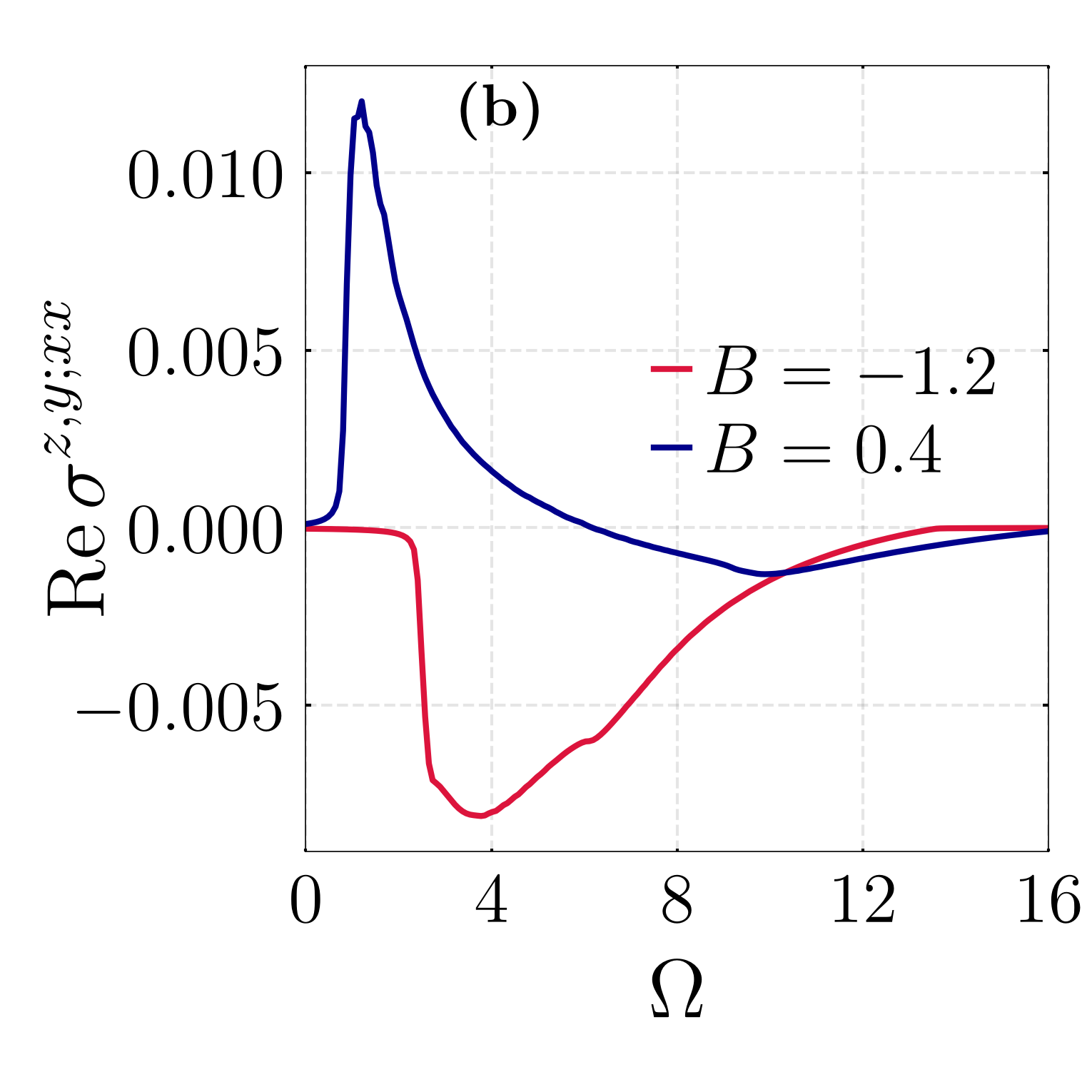}
    \end{tabular}
  \caption{Behavior of nonlinear spin shift conductivity components (a) the real part $\mathrm{Re}\, \sigma^{z,y;yy}$ and (b) $\mathrm{Re} \,\sigma^{z,y;xx}$ as a function of the driving frequency $\Omega$ across the topological phase transition. The optical response in the topological phase ($B=-1.2t$, red line) is contrasted against the trivial phase ($B=0.4t$, blue line). The model parameters are fixed as in Fig.~\ref{fig2}.}
\label{fig8} 
\end{figure}

\begin{table}[t] 
\caption{\label{tab:transposed_symmetry}
Symmetry-allowed second-order tensor components ($\sigma^{\mu\nu\lambda}$) decomposed by current channel and altermagnetic Néel vector configuration.}
\begin{ruledtabular}
\footnotesize
\begin{tabular}{lp{2.4cm}p{2.4cm}}
Current Channel & Néel Orientation ($\phi$) & Allowed Tensors ($\mu\nu\lambda$) \\ 
\midrule
\textbf{Spin} $\sigma_{s_x}$ & Along $x$ ($\phi=0$) & $yyy, xxy, xyx, yxx$ \\
                             & Along $y$ ($\phi=\pi/2$) & $0$ \\
                             
                             & Out-of-Plane ($z$) & $0$ \\ 
\midrule
\textbf{Spin} $\sigma_{s_y}$ & Along $x$ ($\phi=0$) & $0$ \\
                             & Along $y$ ($\phi=\pi/2$) & $yyy, xxy, xyx, yxx$ \\
                          
                             & Out-of-Plane ($z$) & $0$ \\ 
\midrule
\textbf{Spin} $\sigma_{s_z}$ & Along $x$ ($\phi=0$) & $0$ \\
                             & Along $y$ ($\phi=\pi/2$) & $0$ \\
                         
                             & Out-of-Plane ($z$) & $yyy, xxy, xyx, yxx$ \\
\midrule
\textbf{Charge} $\sigma_c$ & Along $x$ ($\phi=0$) & $0$ \\
                           & Along $y$ ($\phi=\pi/2$) & $0$ \\
                       
                           & Out-of-Plane ($z$) & $xxx, xyy, yxy, yyx$ \\ 
\end{tabular}
\end{ruledtabular}
\end{table}

\begin{table}[h!]
\caption{\label{tab:topology_symmetry_compact}Symmetry groups and topological phase classification across different altermagnetic N\'eel vector orientations ($\phi$).}
\begin{ruledtabular}
\begin{tabular}{lcc}
N\'eel Axis & Surviving Symmetries & Topological Phase \\ 
\midrule
$x$-axis ($\phi = 0$) & $\mathcal{C}_{2z}$, $\mathbb{S}_x$,$\mathsf{M}_x$, $\mathsf{M}_y$, $\mathcal{C}_{4z}\mathcal{T}$ & HOTI \\ 
\addlinespace
$y$-axis ($\phi = \pi/2$) & $\mathcal{C}_{2z}$, $\mathbb{S}_y$, $\mathsf{M}_x$,$\mathsf{M}_y$ $\mathcal{C}_{4z}\mathcal{T}$ & HOTI \\ 
\addlinespace
General In-Plane & $\mathcal{C}_{2z}$, Adaptive $\mathbb{S}(\phi)$, $\mathcal{C}_{4z}\mathcal{T}$ & HOTI \\ 
\addlinespace
Out-of-Plane ($z$) & $\mathsf{M}_y$, $\mathbb{S}_x$, $\mathbb{S}_y$ & FOTI \\ 
\end{tabular}
\end{ruledtabular}
\end{table}
\newpage
\section{Summary and Conclusion}
\label{conclusion}
In this work, we have investigated the BPVE in a $\mathcal{PT}$-symmetric two-dimensional heterostructure composed of a 2D TI coupled to a $d$-wave altermagnet. The $\mathcal{PT}$-enforced band degeneracy provides a unique setting for realizing nonlinear optical responses with a pronounced spin character. To describe such degenerate manifolds, we developed a non-Abelian generalization of the conventional spin BPVE formalism, extending the theory from isolated nondegenerate bands to systems with symmetry-enforced band degeneracies.

For a N\'eel vector confined to the $xy$-plane, the heterostructure realizes a SOTI phase protected by $\mathcal{C}_{4z}\mathcal{T}$ symmetry. In this regime, the chiral symmetry operator co-rotates with the N\'eel vector and remains preserved for arbitrary in-plane orientations, thereby pinning corner states to zero energy. The system also possesses $\mathcal{C}_{2z}$, which completely suppresses all second-order charge photocurrents while allowing finite spin photocurrents. Consequently, the BPVE in this phase is intrinsically a pure spin photovoltaic effect, generating a dc spin current in the absence of any accompanying charge current. We find that LPL drives a spin shift current, whereas CPL generates a spin injection current, with the spin polarization locked to the direction of the N\'eel vector. Our work also demonstrates that the spin shift and spin injection conductivities reverse sign whenever the local Dirac mass changes sign. We establish this behavior analytically and also confirm it numerically.

We further examined the effect of rotating the N\'eel vector out of the plane. When the N\'eel vector is oriented along the $z$-direction, $\mathcal{C}_{4z}\mathcal{T}$ symmetry is broken and the system no longer supports the SOTI phase, instead exhibiting conventional TI edge states. At the same time, breaking of $\mathcal{C}_{2z}$ removes the symmetry constraint that forbids charge photocurrents, allowing nonlinear charge and spin responses to coexist. This demonstrates that the nature of the photovoltaic response can be continuously tuned through magnetic reorientation, which simultaneously drives the transition between the FOTI and SOTI phases (Table~\ref{tab:topology_symmetry_compact}). In summary, our results establish $\mathcal{PT}$-symmetric altermagnetic–TI heterostructures as a promising platform for the generation and manipulation of pure spin photocurrents. More broadly, we show that nonlinear spin transport can serve as a sensitive probe of band inversions, higher-order topology, and magnetic symmetry,  therefore opening new avenues for opto-spintronic applications based on spin-current generation in altermagnetic systems.

\section{ACKNOWLEDGMENTS}
For financial support, S. U. and A. B. thank UGC, India, and M.T. gratefully acknowledges Indian Institute of Technology Hyderabad, India.
\onecolumngrid
\appendix
\section{Bulk Photovoltaic Effect in $\mathcal{PT}$ Degenerate System}

\label{sec5}
\renewcommand{\theequation}{A\arabic{equation}}
\setcounter{equation}{0}
In this section, we discuss the formalism for the spin bulk photovoltaic effect for a degenerate system. We consider a system whose unperturbed degenerate Hamiltonian is $\mathcal{H}_{o}$ and the electromagnetic coupling is given by $\mathcal{H}_{p}=-q \textbf{r.E($t$)}$. Therefore the total Hamiltonian of the system is $\mathcal{H}(t)=\mathcal{H}_{o}+\mathcal{H}_{p}$. We have derived all optical responses using the reduced density matrix $\rho_{ab}$. The equation of motion for a single  particle density operator can be written as
\begin{equation}
i\hbar\frac{d\rho_{ab}(t)}{dt}=\hbar\omega_{ab}\rho_{ab}(t)-qE^{\mu}(t)[r^{\mu},\rho(t)]_{ab},
\end{equation}

where $r^\mu$ and $E^\mu$ are the components of the position vector and the Electric field vector along the $\mu$ direction.  
Using the Fourier transformation, the above equation can be written in the frequency domain as  
\begin{equation}
    (\hbar\omega - \varepsilon_{ab})\rho_{ab}(\omega) = -q \int \frac{d\Omega}{2\pi} E^\mu(\Omega) [r^\mu, \rho(\omega-\Omega)]_{ab}.
\end{equation}

The single-particle Bloch states at each crystal momentum $\mathbf{k}$, including internal degrees of freedom such as spin, are grouped into degenerate subspaces (folds). Throughout this appendix, the indices a,b label these folds. We write $a\sim b$ when $\varepsilon_a = \varepsilon_b$ (indicating the same subspace) and $a \neq b$ when $\varepsilon_a \neq \varepsilon_b$ (indicating distinct subspaces). Let us introduce a propagator as follows
\begin{equation}
  d_{ab}^\omega = \frac{1}{\hbar\omega+i0^+-\varepsilon_{ab}}.
\end{equation}

Now the position operator $r^\mu$ can be split into the intraband ($r_i$) and the interband ($r_e$) parts as shown below.
\begin{equation}
    (r_{\text{i}})^\mu_{ab} = i\partial_{\mu }\mathbb{I}_a  \delta_{ab} + \alpha^\mu_{ab} \delta_{ab}, \qquad (r_{\text{e}})^\mu_{ab} = A^\mu_{ab}(1-\delta_{ab}) .
\end{equation}
For the states within the same degenerate subspace, the $\alpha_{ab}=i\langle u_a | \partial_\mu | u_b \rangle$ term represents the intraband Berry connection matrix and  $A^\mu_{ab}$ represents the interband Berry connection between the two folds, where $A^\mu_{ab} = i\langle u_a | \partial_\mu | u_b \rangle$ where $\ket{u_a} \equiv \left( \ket{u_{a,1}},\ \ket{u_{a,2}},\ \dots,\ \ket{u_{a,N}} \right)$. 
The density operator can be expanded in terms of the applied electric field, so $\rho=\sum \rho^{(n)}$. So the $n^{th}$ order of the density matrix is given by a recursive relation as follows
\begin{equation}
  \rho_{ab}^{(n+1)}(\omega)
  = -q\int\frac{d\Omega}{2\pi}\,d_{ab}^\omega\,E^\mu(\Omega)\,[r^\mu,\rho^{(n)}(\omega-\Omega)]_{ab}.
  \label{dme}
\end{equation}
 Now starting from the equilibrium state $\rho^{(0)}_{ab}=2\pi\delta(\omega)f_a \mathbb{I}_N \delta_{ab}$ (where $\mathbb{I}$ is an N-fold identity matrix), the first order response for intraband and interband contribution follows from Eq.~(\ref{dme}). Since a single photon is involved, the input frequency is equal to output frequency $(\omega=\Omega)$.
\begin{equation}
\rho_{aa}^{(1,i)}(\omega) = [r^{\mu}_{i}, \rho^{(0)}]_{aa} = -q E^{\mu}(\omega) \, d_{aa}^{\omega} \, (i\partial_{\mu} f_a) \, \mathbb{I}_N \qquad \text{and} \qquad \rho_{ab}^{(1,e)}(\omega) = [r^{\mu}_{e},\rho^{(0)}]_{ab} = qE^{\mu}(\omega)\,d_{ab}^\omega\,f_{ab}A^\mu_{ab}.
\end{equation}
So, from the first-order response, we now have four second-order responses from commutation relations of the position vector and the first-order density matrix as follows
\begin{equation}
\rho^{(2)}=\rho^{(ii)}+\rho^{(ei)}+\rho^{(ie)}+\rho^{(ee)}.
\end{equation}
Now for the interband transition, we will focus on the last two terms which involve the first-order density matrix $\rho^{(1,e)}$, namely $[r_{i}^\nu,\rho^{(1,e)}]$ and $[r_{e}^\nu,\rho^{(1,e)}]$. From the intraband operator part, we have
\begin{align}
[r_i^\nu, \rho^{(1,e)}]_{ab} &= [i\partial_\nu, \rho^{(1,e)}]_{ab} + [\alpha^\nu, \rho^{(1,e)}]_{ab}.
\end{align}
Again, using the shift property of the delta function, we can write the first-order response using Eq.~(\ref{dme}) with the electric field $ E^\lambda(\Omega')$  having input frequency $\Omega'$ as 
\begin{equation}
\rho^{(1,e)}_{ab}(\omega-\Omega) = q\int\frac{d\Omega'}{2\pi}d^{\omega-\Omega}_{ab}E^\lambda(\Omega')A^\lambda_{ab}f_{ab}\cdot 2\pi\delta(\omega-\Omega-\Omega').
\end{equation}
Now, using the recursive relation, we can write the second-order response by substituting the first-order response from Eq.~(\ref{dme}) and using another electric field  $E^\nu(\Omega)$ as 
\begin{equation}
\rho^{(2,ie)}_{ab}(\omega) = -q\int\frac{d\Omega}{2\pi}d^\omega_{ab}E^\nu(\Omega)[r^\nu_i,\rho^{(1,e)}(\omega-\Omega)]_{ab},
\end{equation}
\begin{equation}
[i\partial_\nu, \rho^{(1,e)}(\omega-\Omega)]_{ab} = iq \int \frac{d\Omega'}{2\pi} E^\lambda(\Omega')  \partial_\nu \left( d^{\omega-\Omega}_{ab} f_{ab} A^{\lambda}_{ab} \right) \cdot 2\pi\delta(\omega-\Omega-\Omega'),
\end{equation}
\begin{equation}
[\alpha^\nu,  \rho^{(1,e)}(\omega-\Omega)]_{ab} = q \int \frac{d\Omega'}{2\pi} E^\lambda(\Omega') 
   \Big( \alpha^\nu_{aa} d^{\omega-\Omega}_{ab} f_{ab} A^\lambda_{ab} 
 - d^{\omega-\Omega}_{ab} f_{ab} A^\lambda_{ab} \alpha^\nu_{bb} \Big) 
  2\pi\delta(\omega-\Omega-\Omega').
\end{equation}
The final expression from the intraband part becomes 
\begin{equation}
\rho^{(2,ie)}_{ab}(\omega) = -q^2 \int \frac{d\Omega \, d\Omega'}{(2\pi)^2} E^\nu(\Omega) E^\lambda(\Omega') d^\omega_{ab} \Big[ i\partial_\nu \left( d^{\omega-\Omega}_{ab} f_{ab} A^\lambda_{ab} \right) +  \Big( \alpha^\nu_{aa} d^{\omega-\Omega}_{ab} f_{ab} A^\lambda_{ab} - d^{\omega-\Omega}_{ab} f_{ab} A^\lambda_{ab} \alpha^\nu_{bb} \Big) \Big] 2\pi\delta(\omega-\Omega-\Omega').
\end{equation}
From the interband position operator, we have 
\begin{equation}
[r_e^\nu, \rho^{(1,e)}(\omega-\Omega)]_{ab} = \sum_{c\neq a,b} \left(A^\nu_{ac}\rho^{(1,e)}_{cb}(\omega-\Omega) - \rho^{(1,e)}_{ac}(\omega-\Omega)A^\nu_{cb}\right),
\end{equation}

\begin{align}
[r_e^\nu, \rho^{(1,e)}(\omega-\Omega)]_{ab} &= q\int\frac{d\Omega'}{2\pi}E^\lambda(\Omega')\sum_{c\neq a,b}\left(A^\nu_{ac}d^{\omega-\Omega}_{cb}f_{cb}A^\lambda_{cb} - d^{\omega-\Omega}_{ac}f_{ac}A^\lambda_{ac}A^\nu_{cb}\right)\cdot 2\pi\delta(\omega-\Omega-\Omega'),
\end{align}

\begin{align}
\rho^{(2,ee)}_{ab}(\omega) &= -q\int\frac{d\Omega}{2\pi}d^\omega_{ab}E^\nu(\Omega)[r^\nu_e,\rho^{(1,e)}(\omega-\Omega)]_{ab},
\end{align}

\begin{align}
\rho^{(2,ee)}_{ab}(\omega) = -q^2\int\frac{d\Omega\,d\Omega'}{(2\pi)^2}E^\nu(\Omega)E^\lambda(\Omega')\,d^\omega_{ab}
\sum_{c\neq a,b}\left(A^\nu_{ac}d^{\omega-\Omega}_{cb}f_{cb}A^\lambda_{cb} 
- d^{\omega-\Omega}_{ac}f_{ac}A^\lambda_{ac}A^\nu_{cb}\right)
2\pi\delta(\omega-\Omega-\Omega'),
\end{align}

\begin{align}
\rho^{(2,ee)}_{ba}(\omega) = -q^2\int\frac{d\Omega\,d\Omega'}{(2\pi)^2}E^\nu(\Omega)E^\lambda(\Omega')\,d^\omega_{ba}
\sum_{c\neq a,b}\left(A^\nu_{bc}d^{\omega-\Omega}_{ca}f_{ca}A^\lambda_{ca} 
- d^{\omega-\Omega}_{bc}f_{bc}A^\lambda_{bc}A^\nu_{ca}\right)
2\pi\delta(\omega-\Omega-\Omega').
\end{align}
Therefore, the current can be written as 
\begin{equation}
J^{s,\mu}_{(2)}(\omega) = q\int\frac{d^dk}{(2\pi)^d}\sum_{ab}\mathrm{Tr}\left[ S^{s,\mu}_{ab}\,\rho^{(2)}_{ba}(\omega) \right],
\end{equation}

\begin{multline}
J^{s,\mu}_{(2)}(\omega) = q^3\int\frac{d^dk}{(2\pi)^d}\int\frac{d\Omega\,d\Omega'}{(2\pi)^2} E^\nu(\Omega)E^\lambda(\Omega')\,2\pi\delta(\omega-\Omega-\Omega') \times \sum_{ab}\mathrm{Tr}\Bigg\{ S^{s,\mu}_{ab}\,d^\omega_{ba}\bigg[ i\partial_\nu\!\left(d^{\omega-\Omega}_{ba}f_{ab}A^\lambda_{ba}\right) \\
+ \left(\alpha^\nu_{bb}\,d^{\omega-\Omega}_{ba}f_{ab}A^\lambda_{ba} - d^{\omega-\Omega}_{ba}f_{ab}A^\lambda_{ba}\,\alpha^\nu_{aa}\right) + \sum_{c\neq a,b}\left(A^\nu_{bc}\,d^{\omega-\Omega}_{ca}f_{ac}A^\lambda_{ca} - d^{\omega-\Omega}_{bc}f_{cb}A^\lambda_{bc}A^\nu_{ca}\right) \bigg]\Bigg\}.
\label{current_full}
\end{multline}

The second-order conductivity tensor $\tilde{\sigma}^{s,\mu;\nu\lambda}$ is
defined as the response coefficient relating this current to the two
driving fields,
\begin{equation}
J^{s,\mu}_{(2)}(\omega)=\int\frac{d\Omega\,d\Omega'}{(2\pi)^{2}}\,
\tilde{\sigma}^{s,\mu;\nu\lambda}(\omega;\Omega,\Omega')\,
E^{\nu}(\Omega)\,E^{\lambda}(\Omega').
\label{eq:sigma_def}
\end{equation}
Comparing with Eq. ~\eqref{current_full}, the tensor
$\tilde{\sigma}$ carries an overall energy-conserving delta function.
Following the convention of Refs.~\cite{PhysRevX.11.011001,PhysRevB.105.045201} we factor
it out,
\begin{equation}
\tilde{\sigma}^{s,\mu;\nu\lambda}(\omega;\Omega,\Omega')
=2\pi\,\delta(\omega-\Omega-\Omega')\,
\sigma^{s,\mu;\nu\lambda}(\omega;\Omega,\Omega').
\label{eq:sigma_convention}
\end{equation}

For a monochromatic field of frequency $\Omega$,
$\mathbf{E}(t)=\mathbf{E}(\Omega)e^{-i\Omega t}+\mathrm{c.c.}$, the
frequency integrals  collapses and the delta function is factored out so that the
symmetrized form of the conductivity, reduces to dc conductivity with $\Omega'=-\Omega$ with $\omega=0$
\begin{equation}
J^{s,\mu}_{(2)}(\omega)=\sum_{\nu\lambda}
\sigma^{s,\mu;\nu\lambda}(\omega;\Omega,-\Omega)\,
E^{\nu}(\Omega)\,[E^{\lambda}(\Omega)]^{*}.
\label{eq:...}
\end{equation}

The nonlinear spin conductivity tensor $\sigma^{s,\mu;\nu\lambda}(\omega;\Omega,\Omega')$ can be decomposed into the intraband-interband ($\sigma_{ie}$) and the purely interband ($\sigma_{ee}$) contributions as follows
\begin{equation}
    \sigma^{s,\mu;\nu\lambda}(\omega;\Omega,\Omega') = \sigma^{s,\mu;\nu\lambda}_{ie}(\omega;\Omega,\Omega') + \sigma^{s,\mu;\nu\lambda}_{ee}(\omega;\Omega,\Omega').
\end{equation}
The purely interband contribution is 
\begin{equation}
\begin{split}
\sigma^{s,\mu;\nu\lambda}_{ee} (\omega;\Omega,\Omega') &= \frac{q^3}{2}\int\frac{d^dk}{(2\pi)^d}\sum_{ab} \mathrm{Tr}\Bigg\{ S^{s,\mu}_{ab}\,d^\omega_{ba} 
 \times \bigg[ \sum_{c\neq a,b}\left(A^\nu_{bc}d^{\omega-\Omega}_{ca}f_{ac}A^\lambda_{ca} - d^{\omega-\Omega}_{bc}f_{cb}A^\lambda_{bc}A^\nu_{ca}\right) \bigg] \Bigg\} 
 + \Big[ (\nu,\Omega)\leftrightarrow(\lambda,\Omega') \Big].
\end{split}
\end{equation}

The intraband-interband contribution, which contains the momentum-space derivative and the non-Abelian Berry connection terms, is given by
\begin{multline}
  \sigma^{s,\mu;\nu\lambda}_{ie} (\omega;\Omega,\Omega') = \frac{q^3}{2}\int\frac{d^dk}{(2\pi)^d}\sum_{ab} \mathrm{Tr}\Bigg\{ S^{s,\mu}_{ab}\,d^\omega_{ba} 
 \bigg[ i\partial_\nu\left(d^{\omega-\Omega}_{ba}f_{ab}A^\lambda_{ba}\right) 
 + \left(\alpha^\nu_{bb}d^{\omega-\Omega}_{ba}f_{ab}A^\lambda_{ba} - d^{\omega-\Omega}_{ba}f_{ab}A^\lambda_{ba}\alpha^\nu_{aa}\right) \bigg] \Bigg\} \\
 + \Big[ (\nu,\Omega)\leftrightarrow(\lambda,\Omega') \Big].
 \label{eq:64}
\end{multline}
Through partial integration of the first term and taking the dc limit ($\omega \to 0$) into consideration, Eq. (\ref{eq:64}) boils down as follows, taking $d^{\omega}_{ba} = \dfrac{1}{\varepsilon_{ab}}$

\begin{align}
\int\frac{d^dk}{(2\pi)^d}\sum_{ab}\frac{iS^{s,\mu}_{ab}}{\varepsilon_{ab}}\partial_\nu\left(d^{-\Omega}_{ba}f_{ab}A^\lambda_{ba}\right) &= -\int\frac{d^dk}{(2\pi)^d}\sum_{ab}\partial_\nu\left(\frac{iS^{s,\mu}_{ab}}{\varepsilon_{ab}}\right)d^{-\Omega}_{ba}f_{ab}A^\lambda_{ba},
\end{align}

\begin{align}
\partial_\nu \left( \frac{i S^{s,\mu}_{ab}}{\varepsilon_{ab}} \right) &= \frac{i \partial_\nu S^{s,\mu}_{ab}}{\varepsilon_{ab}} - \frac{i S^{s,\mu}_{ab} \Delta^\nu_{ab}}{\varepsilon^2_{ab}}.
\end{align}
The second term of Eq.~(\ref{eq:64}) can also be simplified as follows. If we write $X_{ba}=d^{-\Omega}_{ba}f_{ab}A^\lambda_{ba}$,
the connection commutator inside the trace is given by 
\begin{align}
\sum_{ab}\mathrm{Tr}\!\left[\frac{S^{s,\mu}_{ab}}{\varepsilon_{ab}}
\left(\alpha^\nu_{bb}X_{ba} - X_{ba}\alpha^\nu_{aa}\right)\right]
= -\sum_{ab}\mathrm{Tr}\!\left[\frac{1}{\varepsilon_{ab}}
\left(\alpha^\nu_{aa}S^{s,\mu}_{ab} - S^{s,\mu}_{ab}\alpha^\nu_{bb}\right)X_{ba}\right].
\end{align}
Here, we used the cyclic property of the trace to bring the intraband connection
$\alpha^\nu$ next to $S^{s,\mu}_{ab}$. Combined with the partial-integration result
for the first term, the derivative and connection contributions assemble into the
$U(2)$-covariant derivative as follows,
\begin{align}
& [\mathcal{D}_\nu S^{s,\mu}]_{ab} = \partial_\nu S^{s,\mu}_{ab} - i  \alpha^\nu_{aa} S^{s,\mu}_{ab} + i  S^{s,\mu}_{ab} \alpha^\nu_{bb}, 
\label{eq:70}
\end{align}

\begin{align}
\sigma^{s,\mu;\nu\lambda}_{ie}  &= \frac{q^3}{2} \int \frac{d^dk}{(2\pi)^d} \sum_{ab} \text{Tr}\left[\left( \frac{-i [\mathcal{D}_\nu S^{s,\mu}]_{ab}}{\varepsilon_{ab}} + \frac{i S^{s,\mu}_{ab} \Delta^\nu_{ab}}{\varepsilon^2_{ab}} \right) d^{-\Omega}_{ba} f_{ab} A^\lambda_{ba}\right].
\label{eq:sigma_ie}
\end{align}
If we take the non diagonal part of $\sigma^{s,\mu;\nu\lambda}_{ee}$ we have 

\begin{equation}
\sigma^{s,\mu;\nu\lambda}_{ee, \text{od}} (0;\Omega,-\Omega) = \frac{q^3}{2}\int\frac{d^dk}{(2\pi)^d}\sum_{c\neq a,b} \mathrm{Tr}\Bigg\{ S^{s,\mu}_{ab} d^\omega_{ba} 
\bigg[ \left( A^\nu_{bc} d^{-\Omega}_{ca} f_{ac} A^\lambda_{ca} - d^{-\Omega}_{bc} f_{cb} A^\lambda_{bc} A^\nu_{ca} \right) \bigg] \Bigg\} + \Big[ (\nu,\Omega) \leftrightarrow (\lambda,-\Omega) \Big].
\label{eq:od}
\end{equation}
We can rearrange the terms of the RHS since $a,b,c$ are dummy indices. Therefore, we can write the above expression as

\begin{align}
\sum_{ab}\frac{S^{s,\mu}_{ab}}{\varepsilon_{ab}}\sum_{c\neq a,b}\left(A^\nu_{bc}d^{-\Omega}_{ca}f_{ac}A^\lambda_{ca} - d^{-\Omega}_{bc}f_{cb}A^\lambda_{bc}A^\nu_{ca}\right) = \sum_{c\neq a,b}
\left(\frac{S^{s,\mu}_{ac}A^\nu_{cb}}{\varepsilon_{ac}} -\frac{A^\nu_{ac}S^{s,\mu}_{cb}}{\varepsilon_{cb}}\right)d^{-\Omega}_{ba}f_{ab}A^\lambda_{ba}.
\end{align}

Adding $\sigma^{s,\mu;\nu\lambda}_{ie}$ and $\sigma^{s,\mu;\nu\lambda}_{ee,\text{od}}$ we can get

\begin{multline}
\sigma^{s,\mu;\nu\lambda}_{ie+ ee,\text{od}} = \frac{q^3}{2}\int\frac{d^dk}{(2\pi)^d} \sum_{a\neq b}\text{Tr}\Bigg[\left(\frac{-i[\mathcal{D}_\nu S^{s,\mu}]_{ab}}{\varepsilon_{ab}}+\frac{iS^{s,\mu}_{ab}\Delta^\nu_{ab}}{\varepsilon_{ab}^2}\right) d^{-\Omega}_{ba}\, f_{ab}\, A^\lambda_{ba} +\sum_{c\neq a,b} \left( \frac{S^{s,\mu}_{ab}A^\nu_{cb}}{\varepsilon_{ac}} -\frac{A^\nu_{ac}S^{s,\mu}_{cb}}{\varepsilon_{cb}} \right) d^{-\Omega}_{ba}\, f_{ab}\, A^\lambda_{ba} \Bigg] \\+\Big[(\nu,\Omega)\leftrightarrow(\lambda,-\Omega)\Big].
\label{eq:74}
\end{multline}
The shift component of the spin conductivity tensor $\sigma^{s,\mu;\nu\lambda}(\omega;\Omega,\Omega')$ is constructed by the combination of these two distinct second-order optical responses, which are the intraband-interband cross term $(\sigma_{ie})$ (Eq. \ref{eq:sigma_ie}), and the off-diagonal interband term $({\sigma_{ee,\text{od}}})$ (Eq.~(\ref{eq:od})) as expressed below
\begin{equation}
    \sigma^{s,\mu;\nu\lambda}_{\text{shift}} = \sigma^{s,\mu;\nu\lambda}_{ie+ ee,\text{od}}.
\end{equation}

We can expand the U(2) covariant derivative of the above Eq. (\ref{eq:74}) as follows
\begin{align}
& [\mathcal{D}_\nu S^{s,\mu}]_{ab} = \partial_\nu S^{s,\mu}_{ab} - i\alpha^\nu_{aa} S^{s,\mu}_{ab} + iS^{s,\mu}_{ab}\alpha^\nu_{bb} .
\end{align}
The derivative of the matrix $S^{s,\mu}_{ab} $ can be written as
\begin{equation}
\partial_\nu S^{s,\mu}_{ab} = S^{s,\mu;\nu}_{ab}
+ i\big(\alpha^\nu_{aa}S^{s,\mu}_{ab} - S^{s,\mu}_{ab}\alpha^\nu_{bb}\big)
+ i\big(A^\nu_{ab}S^{s,\mu}_{bb} - S^{s,\mu}_{aa}A^\nu_{ab}\big)
+ i\sum_{c\neq a,b}\big(A^\nu_{ac}S^{s,\mu}_{cb} - S^{s,\mu}_{ac}A^\nu_{cb}\big),
\end{equation}

\begin{equation}
[\mathcal{D}_\nu S^{s,\mu}]_{ab} = S^{s,\mu;\nu}_{ab}
+ i\big(A^\nu_{ab}S^{s,\mu}_{bb} - S^{s,\mu}_{aa}A^\nu_{ab}\big)
+ i\sum_{c\neq a,b}\big(A^\nu_{ac}S^{s,\mu}_{cb} - S^{s,\mu}_{ac}A^\nu_{cb}\big),
\end{equation}

\begin{align}
i A^\nu_{ab} = \frac{v^\nu_{ab}}{E_a - E_b} = \frac{v^\nu_{ab}}{\varepsilon_{ab}},
\end{align}

\begin{equation}
\frac{-i[\mathcal{D}_\nu S^{s,\mu}]_{ab}}{\varepsilon_{ab}}
= \frac{-iS^{s,\mu;\nu}_{ab}}{\varepsilon_{ab}}
+ \frac{i\big(S^{s,\mu}_{aa}v^\nu_{ab} - v^\nu_{ab}S^{s,\mu}_{bb}\big)}{\varepsilon^2_{ab}}
+ \frac{1}{\varepsilon_{ab}}\sum_{c\neq a,b}\big(A^\nu_{ac}S^{s,\mu}_{cb} - S^{s,\mu}_{ac}A^\nu_{cb}\big).
\end{equation}

The substitution in Eq. (\ref{eq:74}) is simplified by grouping the coefficients of the three-band terms. Specifically, we utilize the identity
\begin{equation}
    \frac{1}{\varepsilon_{ab}} \left( A^\nu_{ac}S^{s,\mu}_{cb} - S^{s,\mu}_{ac}A^\nu_{cb} \right)+\frac{S^{s,\mu}_{ac}A^\nu_{cb}}{\varepsilon_{ac}} - \frac{A^\nu_{ac}S^{s,\mu}_{cb}}{\varepsilon_{cb}}\ = 
    -i \left( \frac{S^{s,\mu}_{ac}v^\nu_{cb}}{\varepsilon_{ab}\varepsilon_{ac}} - \frac{v^\nu_{ac}S^{s,\mu}_{cb}}{\varepsilon_{ab}\varepsilon_{cb}} \right).
\end{equation}
This leads to the consolidated expression for the conductivity, where we have used $A^\lambda_{ba} = v^\lambda_{ba}/i\varepsilon_{ba}$ and $\varepsilon_{ba}=-\varepsilon_{ab}$
\begin{multline}
\sigma^{s,\mu;\nu\lambda}_{\text{shift}} = \frac{q^3}{2}\int\frac{d^dk}{(2\pi)^d}\sum_{a\neq b} \text{Tr}\Biggl[ \Biggl\{
    \frac{-iS^{s,\mu;\nu}_{ab}}{\varepsilon_{ab}} 
    + \frac{i(S^{s,\mu}_{aa}v^\nu_{ab} - v^\nu_{ab}S^{s,\mu}_{bb})}{\varepsilon^2_{ab}} 
    + \frac{iS^{s,\mu}_{ab}\Delta^\nu_{ab}}{\varepsilon^2_{ab}} 
    -\frac{i}{\varepsilon_{ab}}\sum_{c\neq a,b} \left(\frac{S^{s,\mu}_{ac}v^\nu_{cb}}{\varepsilon_{ac}} - \frac{v^\nu_{ac}S^{s,\mu}_{cb}}{\varepsilon_{cb}}\right) \Biggr\}d^{-\Omega}_{ba}f_{ab}A^\lambda_{ba}\Biggr] \\
    + [(\nu,\Omega)\leftrightarrow(\lambda,-\Omega)],
\end{multline}
\begin{multline}
\sigma^{s,\mu;\nu\lambda}_{\text{shift}} = \frac{q^3}{2}\int\frac{d^dk}{(2\pi)^d}\sum_{a\neq b} \text{Tr}
\left[\left\{\frac{S^{s,\mu;\nu}_{ab}}{\varepsilon^2_{ab}} 
- \frac{S^{s,\mu}_{aa}v^\nu_{ab} - v^\nu_{ab}S^{s,\mu}_{bb}}{\varepsilon^3_{ab}} 
- \frac{S^{s,\mu}_{ab}\Delta^\nu_{ab}}{\varepsilon^3_{ab}}\right. 
\left.+\frac{1}{\varepsilon^2_{ab}}\sum_{c\neq a,b}
\left(\frac{S^{s,\mu}_{ac}v^\nu_{cb}}{\varepsilon_{ac}} 
+ \frac{v^\nu_{ac}S^{s,\mu}_{cb}}{\varepsilon_{bc}}\right)\right\}
f_{ab}\,v^\lambda_{ba}\,d^{-\Omega}_{ba} \right]\\
+ \left[(\nu,\Omega)\leftrightarrow(\lambda,-\Omega)\right].
\label{eq:41}
\end{multline}
The term $-S^{s,\mu}_{ab}\Delta^\nu_{ab}/\varepsilon^3_{ab}$ provides the missing $c=b$ and $c=a$ 
boundary terms to extend the restricted sums into $\mathcal{D}^{s,\mu;\nu}_{ab}$. 
The cross-manifold spin-current difference is carried out separately
\begin{equation}
\mathcal{D}^{s,\mu;\nu}_{ab}
= S^{s,\mu;\nu}_{ab}
+ \sum_{c\neq a}\frac{S^{s,\mu}_{ac}\,v^\nu_{cb}}{\varepsilon_{ac}}
+ \sum_{c\neq b}\frac{v^\nu_{ac}\,S^{s,\mu}_{cb}}{\varepsilon_{bc}}.
\label{eq:DSmunu}
\end{equation}
Using this we can define the spin-velocity generalized derivative in the degenerate case as $\mathcal{D}^{s,\mu;\nu}_{ab}$ (in Eq.  (\ref{eq:DSmunu})), which contains a matrix $S_{ab}^{s,\mu;\nu} = \langle a,\mathbf{k} | \frac{1}{2} \{ \hat{S}^s, \hat{v}^{\mu\nu}\} | b,\mathbf{k} \rangle$ with the intermediate-state sums over $c$   which encodes virtual interband processes connecting distinct degenerate manifolds.
We can substitute the propagator $d^{-\Omega}_{ba}$ using the Sokhotski-Plemelj theorem~\cite{Kanwal2004} to seperate delta function (resonance part) and principal value part, equation as follows
\begin{equation}
d^{-\Omega}_{ba} = \frac{1}{\varepsilon_{ab} - \hbar\Omega + i\eta} = \mathcal{P}\frac{1}{\varepsilon_{ab} - \hbar\Omega} - i\pi\delta(\varepsilon_{ab} - \hbar\Omega).
\end{equation}

For a degenerate system without an intermediate state c, this expression can further be simplified by taking out only the delta part of the Eq. (\ref{eq:41}) as follows

\begin{equation}
\sigma^{s,\mu;\nu\lambda}_{\text{shift}}
= -\frac{i\pi q^3}{2}\int\frac{d^dk}{(2\pi)^d}\sum_{a\neq b}\text{Tr}
\left[\left\{
\frac{S^{s,\mu;\nu}_{ab}}{\varepsilon^2_{ab}}
- \frac{S^{s,\mu}_{aa}v^\nu_{ab} - v^\nu_{ab}S^{s,\mu}_{bb}}{\varepsilon^3_{ab}}
- \frac{S^{s,\mu}_{ab}\Delta^\nu_{ab}}{\varepsilon^3_{ab}}
\right\}
f_{ab}\,v^\lambda_{ba}\,\delta(\varepsilon_{ab}-\Omega)\right]+ \left[(\nu,\Omega)\leftrightarrow(\lambda,-\Omega)\right].
\label{eq:A46}
\end{equation}
Again, we can consider 
\begin{equation}
\label{eq:W}
W^{s,\mu\nu}_{ab} =
S^{s,\mu;\nu}_{ab} 
- \frac{S^{s,\mu}_{ab}\,\Delta^\nu_{ab}}{\varepsilon_{ab}}
\qquad \text{and} \qquad
\Delta^\nu_{ab} = v^\nu_{aa} - v^\nu_{bb}.
\end{equation}
Therefore, Eq. (\ref{eq:A46}) can be expressed in compact form as follows
\begin{equation}
\sigma^{s,\mu;\nu\lambda}_\text{shift} = - 
\frac{i\pi q^3}{2}
\int\frac{d^dk}{(2\pi)^d}
\sum_{a\neq b}
\frac{f_{ab}\,\delta(\varepsilon_{ab}-\Omega)}{\varepsilon^2_{ab}}
\;\mathcal{M}^{\mu;\nu\lambda}_{ab},
\end{equation}
where 
\begin{equation}
\mathcal{M}^{s,\mu;\nu\lambda}_{ab} =\text{Tr}\left[W^{s,\mu;\nu}_{ab} v^\lambda_{ba} - W^{s,\mu;\lambda}_{ba} v^\nu_{ab}
- \frac{S^{s,\mu}_{aa} v^\nu_{ab} v^\lambda_{ba} - v^\nu_{ab} S^{s,\mu}_{bb} v^\lambda_{ba}}{\varepsilon_{ab}} 
- \frac{S^{s,\mu}_{bb} v^\lambda_{ba} v^\nu_{ab} - v^\lambda_{ba} S^{s,\mu}_{aa} v^\nu_{ab}}{\varepsilon_{ab}}\right]
\label{eq:matrix_element}.
\end{equation}
Using the trace property of the matrices, the last two terms in Eq. (\ref{eq:matrix_element}) cancel out, so we are left with
\begin{equation}
\mathcal{M}^{s,\mu;\nu\lambda}_{ab} =\text{Tr}\left[  W^{s,\mu;\nu}_{ab} v^\lambda_{ba} - W^{s,\mu;\lambda}_{ba} v^\nu_{ab}\right].
\end{equation}
Now, we can express the injection conductivity tensor ($\sigma^{s,\mu;\nu\lambda}_{\text{inj}}$) by considering the diagonal part of the $\sigma^{s,\mu;\nu\lambda}_{ee}$ tensor as follows
\begin{equation}
\sigma^{s,\mu;\nu\lambda}_{ee,\text{d}} = \frac{q^3}{2}\int\frac{d^dk}{(2\pi)^d}\sum_{a,b}\text{Tr}\left[S^{s,\mu}_{ab}\,d^\omega_{ba}\left\{\sum_c d^{\omega-\Omega}_{ca}A^\nu_{bc}A^\lambda_{ca}f_{ac} - \sum_c d^{\omega-\Omega}_{bc}A^\lambda_{bc}A^\nu_{ca}f_{cb}\right\}\right] 
 + [(\nu,\Omega)\leftrightarrow(\lambda,-\Omega)].
 \label{eq:A51}
\end{equation}
When we take $a=b$, we have divergence at $\omega \to 0$. This can be avoided by introducing $\Gamma$ in the denominator, which is the reciprocal of phenomenological relaxation time $\tau$ in the above expression, so the propagator can be written as $d_{ba}^\omega=(\hbar\omega+i\hbar\gamma)^{-1}$. Therefore, Eq. (\ref{eq:A51}) takes the form as follows
\begin{equation}
\sigma^{s,\mu;\nu\lambda}_{\text{inj}} = \frac{q^3}{2(\hbar\omega+i\hbar\gamma)}\int\frac{d^dk}{(2\pi)^d}\sum_{a,c}\text{Tr}\left[S^{s,\mu}_{aa}\left\{A^\nu_{ac}\,d^{\omega-\Omega}_{ca}\,f_{ac}\,A^\lambda_{ca} - d^{\omega-\Omega}_{ac}\,f_{ca}\,A^\lambda_{ac}\,A^\nu_{ca}\right\}\right] 
+ [(\nu,\Omega)\leftrightarrow(\lambda,-\Omega)].
\label{eq:A53}
\end{equation}
Taking symmetrization under consideration from the above equation, we can write the four terms as follows.
\begin{align} 
\sigma^{s,\mu;\nu\lambda}_{\text{inj}} &= \frac{q^3}{2(\hbar\omega+i\hbar\gamma)}\int\frac{d^dk}{(2 \pi)^d}\sum_{a,c} \text{Tr}\Biggl(S_{aa}^{s,\mu}A_{ac}^\nu d_{ca}^{\omega - \Omega}f_{ac}A_{ca}^\lambda - S_{aa}^{s,\mu}A_{ac}^\lambda d_{ac}^{\omega - \Omega}f_{ca}A_{ca}^\nu \nonumber\\
&\qquad + S_{aa}^{s,\mu}A_{ac}^\lambda d_{ca}^{\omega +\Omega}f_{ac}A_{ca}^\nu - S_{aa}^{s,\mu}A_{ac}^\nu d_{ac}^{\omega + \Omega}f_{ca}A_{ca}^\lambda\Biggr).\nonumber
\end{align}
After relabeling a and c by interchanging them in the second and fourth term, and also if we utilize the cyclic property of the trace, Eq. (\ref{eq:A53}) can be boiled down as follows.

\begin{equation}
\sigma^{s,\mu;\nu\lambda}_{\text{inj}} = \frac{q^{3}}{2i\hbar\gamma}\int\frac{d^dk}{(2\pi)^d}\sum_{a\neq c}\text{Tr}\Bigl[\bigl(S^{s,\mu}_{aa}A^{\nu}_{ac}-A^{\nu}_{ac}S^{s,\mu}_{cc}\bigr)A^{\lambda}_{ca}\,d^{\omega-\Omega}_{ca} 
+\bigl(S^{s,\mu}_{aa}A^{\lambda}_{ac}-A^{\lambda}_{ac}S^{s,\mu}_{cc}\bigr)A^{\nu}_{ca}\,d^{\omega +\Omega}_{ca}\Bigr]f_{ac}.
\end{equation}
We can do the
Taylor series expansion of both propagators around $\omega=0$ as shown below
\begin{align}
d^{\omega-\Omega}_{ca} &= {d^{-\Omega}_{ca}}+{\omega\,\partial_{\omega}d^{\omega-\Omega}_{ca}\big|_{\omega=0}}+O(\omega^{2}), \\
d^{\omega+\Omega}_{ca} &= {d^{\Omega}_{ca}}+{\omega\,\partial_{\omega}d^{\omega+\Omega}_{ca}\big|_{\omega=0}}+O(\omega^{2}).
\end{align}
After taking only the diverging term of the propagators, we can write,
\begin{equation}
\sigma^{s,\mu;\nu\lambda}_{\text{inj}} = \frac{q^{3}}{2i\hbar\gamma}\int\frac{d^dk}{(2\pi)^d}\sum_{a\neq c}\text{Tr}\Bigl[\bigl(S^{s,\mu}_{aa}A^{\nu}_{ac}-A^{\nu}_{ac}S^{s,\mu}_{cc}\bigr)A^{\lambda}_{ca}\,(-i\pi\delta(\hbar\Omega + \varepsilon_{ca}))
+\bigl(S^{s,\mu}_{aa}A^{\lambda}_{ac}-A^{\lambda}_{ac}S^{s,\mu}_{cc}\bigr)A^{\nu}_{ca}\,(-i\pi\delta(\hbar\Omega - \varepsilon_{ca}))\Bigr]f_{ac}.
\label{eq:Inj57}
\end{equation}
Again, we can rewrite the first term by relabeling $a$ and $c$ and from Eq. (\ref{eq:Inj57}) as follows
\begin{align}
    \bigl(S^{s,\mu}_{aa}A^{\nu}_{ac}-A^{\nu}_{ac}S^{s,\mu}_{cc}\bigr)A^{\lambda}_{ca}\,\bigl(-i\pi\delta(\hbar\Omega + \varepsilon_{ca})\bigr)f_{ac} &= \bigl(S^{s,\mu}_{cc}A^{\nu}_{ca}-A^{\nu}_{ca}S^{s,\mu}_{aa}\bigr)A^{\lambda}_{ac}\,\bigl(-i\pi\delta(\hbar\Omega + \varepsilon_{ac})\bigr)f_{ca}, \nonumber\\
    &= \bigl(A^{\nu}_{ca}S^{s,\mu}_{aa}A^{\lambda}_{ac}-S^{s,\mu}_{cc}A^{\nu}_{ca}A^{\lambda}_{ac}\bigr)\,\bigl(i\pi\delta(\hbar\Omega + \varepsilon_{ac})\bigr)f_{ca},\nonumber\\
    &= -\bigl(A^{\nu}_{ca}S^{s,\mu}_{aa}A^{\lambda}_{ac}-S^{s,\mu}_{cc}A^{\nu}_{ca}A^{\lambda}_{ac}\bigr)\,\bigl(i\pi\delta(\hbar\Omega -\varepsilon_{ca})\bigr)f_{ac}.
\end{align}
Now, using the trace property, the term can be written as follows
\begin{equation}
    -\bigl(A^{\nu}_{ca}S^{s,\mu}_{aa}A^{\lambda}_{ac}-S^{s,\mu}_{cc}A^{\nu}_{ca}A^{\lambda}_{ac}\bigr)\,\bigl(i\pi\delta(\hbar\Omega -\varepsilon_{ca})\bigr)f_{ac} = -\bigl(S^{s,\mu}_{aa}A^{\lambda}_{ac}-A^{\lambda}_{ac}S^{s,\mu}_{cc}\bigr)\,\bigl(i\pi\delta(\hbar\Omega -\varepsilon_{ca})\bigr)A^{\nu}_{ca}f_{ac}\nonumber .
\end{equation}
Therefore, using this simplified form of the first term, we can rewrite Eq. (\ref{eq:Inj57}) as
\begin{equation}
  \sigma^{s,\mu;\nu\lambda}_{\text{inj}} = \frac{-\pi q^{3}}{2\hbar\gamma}\int\frac{d^{d}k}{(2\pi)^{d}}\sum_{a\neq c}\text{Tr}\Bigl[\Bigl\{\bigl(S^{s,\mu}_{aa}A^{\lambda}_{ac}-A^{\lambda}_{ac}S^{s,\mu}_{cc}\bigr)A^{\nu}_{ca}+\bigl(S^{s,\mu}_{aa}A^{\lambda}_{ac}-A^{\lambda}_{ac}S^{s,\mu}_{cc}\bigr)A^{\nu}_{ca}\Bigr\}\,\delta(\hbar\Omega-\varepsilon_{ca})\,f_{ac}\Bigr].
\end{equation}
Now, the injection current takes the final form as follows:
\begin{align}
\sigma^{s,\mu;\nu\lambda}_{\text{inj}} = \frac{-\pi q^{3}}{\hbar\gamma}\int\frac{d^{d}k}{(2\pi)^{d}}\sum_{a\neq c}\text{Tr}\Bigl[\biggl\{\bigl(S^{s,\mu}_{aa}A^{\lambda}_{ac}-A^{\lambda}_{ac}S^{s,\mu}_{cc}\bigr)A^{\nu}_{ca}\Bigr\}\,\delta(\hbar\Omega-\varepsilon_{ca})\Bigr]\,f_{ac}.
\end{align}
Since $c$ is a dummy index, we relabel $c\rightarrow b$ for consistency, and the final expression of injection conductivity takes the form
\begin{align}
\sigma^{s,\mu;\nu\lambda}_{\text{inj}} = \frac{-\pi q^{3}}{\hbar\gamma}\int\frac{d^{d}k}{(2\pi)^{d}}\sum_{a\neq b}\text{Tr}\Bigl[\Bigl\{\bigl(S^{s,\mu}_{aa}A^{\lambda}_{ab}-A^{\lambda}_{ab}S^{s,\mu}_{bb}\bigr)A^{\nu}_{ba}\Bigr\}\,\delta(\hbar\Omega-\varepsilon_{ba})\Bigr]\,f_{ab}.
\end{align}

\section{Symmetry Constraints on the Matrix Elements}
\twocolumngrid
\label{AppB}
\renewcommand{\theequation}{B\arabic{equation}}
\setcounter{equation}{0}
\subsection{$\mathcal{PT}$ Symmetry Constraint}
In this section, we show exactly how the $\mathcal{PT}$ symmetry forces the spin shift conductivity to be purely real as shown in Eq.~(\ref{eq:PT_shift}). We here bridge the gap between the gauge-dependent matrix transformation laws and the simplified element-wise trace relations used in the main text.

The space-inversion ($\mathcal{PT}$) symmetry is an antiunitary operator, which satisfies $(\mathcal{PT})^2 = -1$. It also commutes with the Hamiltonian ([$\mathcal{PT}$, $\mathcal{H}(\mathbf{k})$] = 0).

Suppose $|a,1\rangle$ and $|a, 2\rangle$ be the two degenerate states. If we operate the $\mathcal{PT}$ operator on the states, then it will take the general form as 
\begin{equation}
    \mathcal{PT} |a,i\rangle = \sum\limits_{i'} |a,i'\rangle[w_a]_{i'i}, \quad \text{where} \quad i ,i'\in{\{1,2\}}.
\end{equation}
By utilizing the antiunitary property and Krammer's orthogonality condition, it can be observed that the two-fold degenerate manifold transforms via the unitary, antisymmetric sewing matrix $w_a = i\sigma_y$. Let us consider an operator $\hat{O}$ with a definite $\mathcal{PT}$ parity eigenvalue $\eta_o = \pm 1$, which means $(\mathcal{PT}) \hat{O}(\mathcal{PT}) ^{-1} = \eta_o \hat{O}$. Since $\mathcal{PT}$ is antiunitary, evaluating a matrix element involves taking a complex conjugate when we move onto the transformed basis as given below.
\begin{align}
    \langle a,i|\hat{O}|b,j\rangle &= \langle \mathcal{PT}(a,i)|(\mathcal{PT}) \hat{O}(\mathcal{PT})^{-1}|\mathcal{PT}(b,j)\rangle^*, \nonumber\\
    &=\eta_o  \langle \mathcal{PT}(a,i)| \hat{O}|\mathcal{PT}(b,j)\rangle^*.
\end{align}
Substituting our state expansion for the block matrix element $[O_{ab}]_{ij} =  \langle a,i|\hat{O}|b,j\rangle$, we get:
\begin{equation}
    [O_{ab}]_{ij} = \eta_o \sum\limits_{i'j'}[w_a]^*_{i'i} [O_{ab}]^*_{i'j'} [w_b]_{j'j}.
\end{equation}
In matrix form, we can write,
\begin{equation}
    O_{ab} = \eta_ow_a^\dagger O^*_{ab}w_b.
\end{equation}
We already discussed that the sewing matrix $w_a=w_b = i\sigma_y$. Therefore, plugging these sewing matrices results into the structural constraints on the $\hat{v}^\lambda$ ($\eta_v =+1$) and the spin matrices $\hat{S}^{s,\mu}$ ($\eta_S =-1$) become:
\begin{equation}
    v^\lambda_{ab} = +\sigma_y\ (v^\lambda_{ab})^*\ \sigma_y, \quad S^{s,\mu}_{ab} =  -\sigma_y\ (S^{s,\mu}_{ab})^*\ \sigma_y.
\end{equation}
Now, it is evident from Eq.~(\ref{eq:response}) that the shift conductivity do not depend on the isolated matrix elements, but they depend on gauge-invariant closed loops of matrix products, which appear inside the trace (represented by $\mathcal{M}_{ab}^{s,\mu;\nu\lambda}$).
From Eq.~(\ref{eq:shift2}) and (\ref{eq:shift3}), it is evident that $\mathcal{M}_{ab}^{s,\mu;\nu\lambda}$ involves $W^{s,\mu\nu}_{ab}$ which contains $S^{s,\mu}_{ab}$ and  $v^\lambda_{ab}$ resulting an odd $\mathcal{PT}$ parity in $W^{s,\mu\nu}_{ab}$ as expressed below
\begin{equation}
    W^{s,\mu\nu}_{ab} = -\sigma_y \ (W^{s,\mu\nu}_{ab})^* \ \sigma_y.
\end{equation}
Now, each term inside the loop trace of $\mathcal{M}_{ab}^{s,\mu;\nu\lambda}(\mathbf{k})$ contains one $\mathcal{PT}$-odd matrix $W^{s,\mu\nu}$ and one $\mathcal{PT}$-even velocity matrix $v$, which sets the number of $\mathcal{PT}$-odd operator to be 1 and it results,
\begin{equation}
    \mathcal{M}_{ab}^{s,\mu;\nu\lambda}(\mathbf{k}) = (-1)^1 (\mathcal{M}_{ab}^{s,\mu;\nu\lambda}(\mathbf{k}))^* = -(\mathcal{M}_{ab}^{s,\mu;\nu\lambda}(\mathbf{k}))^*.
\end{equation}
This is why we obtain Eq.~(\ref{eq:PT_M}), which forces the shift conductivity to be purely real.

\subsection{Mirror Symmetry Constraint}
Unlike $\mathcal{PT}$ symmetry, a mirror reflection $\mathsf{M}_\alpha$ (where $\alpha \in \{x, y\}$) is an unitary spatial symmetry. Its primary effect is to flip the momentum component perpendicular to the mirror plane. As from Eq.~(\ref{eq:mirror_x}) and (\ref{eq:mirror_y})
, it is evident that the symmetry connects the physics at the two different points of the Brillouin zone, $\mathbf{k} \  \text{and} \ \mathsf{M}_\alpha\mathbf{k} $. Since the energy eigenvalues are identical at these points, a state $|a,i\rangle$ at momentum $\mathbf{k}$ is mapped into the same degenerate band manifold $a$ at the reflected momentum. We track this internal state mixing using a unitary mirror sewing matrix $u_a$, such that
\begin{equation}
    \mathsf{M}_\alpha |a,i\rangle_{\mathbf{k}} = \sum\limits_{i'}  |a,i'\rangle_{\mathsf{M}_\alpha\mathbf{k}}[u_a]_{i'i}.
\end{equation}

Since $\mathsf{M}_\alpha$ is a unitary operator ($\mathsf{M}_\alpha^{-1} = \mathsf{M}_\alpha^\dagger$), evaluating the matrix elements does not involve complex conjugation. By inserting the identity $\mathsf{M}^{-1}_\alpha \mathsf{M}_\alpha = \mathbb{I}$, the matrix elements can easily maps over the reflected states as follows
\begin{align}
    \langle a,i|\hat{O}|b,j\rangle &=  \langle a,i|\mathsf{M}^{-1}_\alpha(\mathsf{M}_\alpha\hat{O}\,\mathsf{M}^{-1}_\alpha)\mathsf{M}_\alpha| b,j\rangle_{\mathbf{k}},\nonumber \\
    &=\sum\limits_{i'j'}[u_a]^*_{i'i}\langle a,i'|\mathsf{M}_\alpha\hat{O}\,\mathsf{M}^{-1}_\alpha|b,j'\rangle_{\mathsf{M}_\alpha\mathbf{k}}[u_b]_{j'j}.\nonumber
\end{align}
If the operator transforms under reflection with an intrinsic parity, $\eta_o^\alpha = \pm1$(meaning $\mathsf{M}_\alpha\hat{O}\,\mathsf{M}^{-1}_\alpha = \eta^\alpha_o \hat{O}$, we can express this block matrix element in a compact form as follows
\begin{equation}
    O_{ab}(\mathbf{k}) = \eta_o^\alpha u_a^\dagger O_{ab} (\mathsf{M}_\alpha\mathbf{k}) u_b.
\end{equation}
Now, we can determine the intrinsic spatial parity ($\eta_o^\alpha$) for our specific physical operators. The velocity operator $\hat{v}^\mu = \partial_{k_{\mu}}\hat{H}(\mathbf{k})$ picks up a negative sign if derivative is taken along the reflection axis ($\mu = \alpha$), because that specific component of the momentum is flipped. Therefore, it transforms as,
\begin{equation}
    v_{ab}^\mu(\mathbf{k}) = (-1)^{\delta_{\mu \alpha}} u_a^\dagger v^\mu_{ab}(\mathsf{M}_\alpha\mathbf{k})u_b.
\end{equation}
On the other hand, the spin operator $\hat{S}^s$ is a pseudovector. Under the spatial reflection, the component parallel to the mirror plane flips, while perpendicular components remain unchanged, which can be explained by introducing an intrinsic sign of $(-1)^{1+\delta_{s\alpha}}$. Now, the composite spin vertices ($S^{s,\mu}_{ab}\, \& \,S^{s,\mu;\nu,\lambda}_{ab}$) are built from the anti-commutation of the spin and velocity operator. Therefore, we can write
\begin{equation}
    S_{ab}^{s,\mu}(\mathbf{k}) = (-1)^{1+\delta_{s\alpha}+\delta_{\mu\alpha}} u_a^\dagger S_{ab}^{s,\mu} (\mathsf{M}_\alpha\mathbf{k}) u_b,
\end{equation}
\begin{equation}
    S_{ab}^{s,\mu;\nu}(\mathbf{k}) = (-1)^{1+\delta_{s\alpha}+\delta_{\mu\alpha}+\delta_{\nu\alpha}} u_a^\dagger S_{ab}^{s,\mu;\nu} (\mathsf{M}_\alpha\mathbf{k}) u_b.
\end{equation}

When we substitute these transformation rules into the trace definition for the response tensor $\mathcal{M}_{ab}^{s,\mu;\nu\lambda}(\mathbf{k})$ (followed by Eq.~(\ref{eq:shift2})), which consists of spin vertex operator $W\, \&$ velocity operator $v $. All the sewing matrices u cancel out completely utilizing their unitariness and cyclic property of the trace. Therefore, the transformation is then governed by the the total sign factor of $(-1)^{1+\delta_{s\alpha}+\delta_{\mu\alpha}+\delta_{\nu\alpha}+\delta_{\lambda\alpha}}$. Consequently, by defining the combined integer index as $n_\alpha = \delta_{s\alpha}+\delta_{\mu\alpha}+\delta_{\nu\alpha}+\delta_{\lambda\alpha}$, Eq.~(\ref{eq:mirror_on_M}) can be obtained.
\begin{equation}
\mathcal{M}^{s,\mu;\nu\lambda}_{ab}(\mathbf{k}) 
\xrightarrow{\;\mathsf{M}_\alpha\;} (-1)^{1+n_\alpha}\,
\mathcal{M}^{s,\mu;\nu\lambda}_{ab}(\mathsf{M_\alpha}\mathbf{k}).
\end{equation}

\subsection{Chiral Symmetry Constraint}
In the main text, we have mentioned that for N\'eel vector orientation along $x$, the chiral symmetry $\mathbb{S}_x= \kappa_x\sigma_y$. The chiral symmetry along with the $\mathcal{PT}$ symmetry and the Hermiticity property of the operators, strictly forces the $y$-polarized spin shift conductivity to vanish. In contrast, it leaves the $x$-(and $z$-) polarized components unconstrained. The chiral operator is Hermitian and $k$-independent. This implies $\mathbb{S}^2=\mathbb{I}$, and $\{\hat{\mathbb{S}}_x, \hat{H}\}=0$. Therefore, within a degenerate manifold $a$, the energy gets transformed as $\varepsilon_{\bar{a}}\to -\varepsilon_a$. The transformation corresponding to any generic operator can be given as $O_{\bar{a}\bar{b}}=\eta_o\ O_{ab}$ for any $\mathbb{S}\hat{O}\mathbb{S}^{-1} = \eta_o \hat{O}$.
Now the velocity operator is chiral-odd, and it can be given by,
\begin{equation}
   v^\beta_{\bar{a}\bar{b}}(\mathbf{k}) \xrightarrow{\mathbb{S}_x} - \,v^\beta_{ab}( \mathbf{k}),
\end{equation}

\begin{equation}
   v^{\alpha\beta}_{\bar{a}\bar{b}}(\mathbf{k}) \xrightarrow{\mathbb{S}_x} - \,v^{\alpha\beta}_{ab}( \mathbf{k}),
\end{equation}

where, $\alpha,\beta \in \{\mu, \nu, \lambda\}$ and  $\eta_v = -1$. For the spin operator, the transformation can be given as follows
\begin{equation}
S^{s,\mu}_{\bar{a}\bar{b}}(\mathbf{k}) \xrightarrow{\;\mathbb{S}_x\;} (-1)^{1+\delta_{sx}+\delta_{sz}} \, S^{s,\mu}_{ab}(\mathbf{k} ),
\end{equation}

\begin{equation}
S^{s,\mu;\nu}_{\bar{a}\bar{b}}(\mathbf{k}) \xrightarrow{\;\mathbb{S}_x\;} (-1)^{1+\delta_{sx}+\delta_{sz}} \, S^{s,\mu;\nu}_{ab}(\mathbf{k}).
\end{equation}
Here, we can consider $\eta_{S^s} = (-1)^{\delta_{sx}+\delta_{sz}}$, where, $s \in \{x,y,z\}$. In the expression of $W^{s,\mu;\nu}$ (followed by Eq.~(\ref{eq:shift3})) $\Delta^\nu_{ab}/\varepsilon_{ab}$ term is present which is chiral-even, therefore, $\eta_W = -\eta_{S^s}$.
From the transformation of these operators, we can get

\begin{equation}
    \mathcal{M}_{\bar{a}\bar{b}}^{s,\mu;\nu;\lambda} =\eta_{\mathcal{M}}^s \mathcal{M}_{ab}^{s,\mu;\nu;\lambda}, \quad \eta_{\mathcal{M}}^s = \eta_W \eta_v = \eta_{S^{s}}.
\end{equation}

Under the chiral symmetry operation, $f_{ab}$ transform as $f_{\bar{a}\bar{b}}\to - f_{ab}$
and $\delta(\varepsilon_{ab}-\Omega)$ transform as $\delta(\varepsilon_{\bar{a}\bar{b}}-\Omega) \to \delta(\varepsilon_{ba}-\Omega)$. Therefore, the spin shift conductivity tensor, given in Eq.~(\ref{eq:response}), with the $\mathcal{M}^{s,\mu;\nu\lambda} \,\text{and}\, W^{s,\mu;\nu}$ (followed by Eq.~(\ref{eq:shift2}) and (\ref{eq:shift3}), undergoes a chiral symmetry operation as shown below.

\begin{align}
\sigma^{s\mu;\nu\lambda}_{\text{shift}}(\Omega)
&=-\frac{i\pi q^3}{2}\int_k\dfrac{d^dk}{(2\pi)^d}\sum_{a\neq b}
\frac{-f_{ab}}{\varepsilon_{ab}^2}\,\eta_{\mathcal{M}}^s M^{s\mu;\nu\lambda}_{ab}\,
\delta(\varepsilon_{ab}+\Omega),\nonumber\\
&=-\eta_{\mathcal{M}}^s\,\sigma_{\text{shift}}^{s\mu:\nu\lambda}(-\Omega),
\end{align}

Since the operators are Hermitian, we have 
\begin{equation}
\mathcal{M}^{s,\mu;\nu\lambda}_{ba}=\bigl(\mathcal{M}^{s,\mu;\nu\lambda}_{ab}\bigr)^{*}.
\label{eq:herm}
\end{equation}
Again, under the  $\mathcal{PT}$ symmetry
\begin{equation}
\bigl(\mathcal{M}^{s,\mu;\nu\lambda}_{ab}\bigr)^{*}=-\,\mathcal{M}^{s,\mu;\nu\lambda}_{ab}.
\label{eq:imag}
\end{equation}
from Eq.~(\ref{eq:response}), we have for $(-\Omega)$ 
\begin{equation}
\sigma^{s,\mu;\nu\lambda}_{\text{shift}}(-\Omega)=-\frac{i\pi q^3}{2}\int_k\dfrac{d^dk}{(2\pi)^d}\sum_{a\neq b}
\frac{f_{ab}}{\varepsilon_{ab}^2}\mathcal{M}^{s\mu;\nu\lambda}_{ab}
\delta(\varepsilon_{ab}+\Omega),
\label{eq:B22}
\end{equation}
relabeling a,b, and using (Eqs.~(\ref{eq:imag}) and (\ref{eq:B22})) with even property of the delta function, we get  
\begin{equation}
\sigma^{s,\mu;\nu\lambda}_{\mathrm{shift}}(-\Omega)=+\,\sigma^{s,\mu;\nu\lambda}_{\mathrm{shift}}(\Omega).
\label{eq:B23}
\end{equation}
This indicates that the shift conductivity is an even function of frequency. From Eq.~(\ref{eq:herm}) we have 
\begin{equation}
\sigma^{s\mu;\nu\lambda}_{\mathrm{shift}}(-\Omega)=-\,\eta_{\mathcal{M}}^s\,\sigma^{s\mu;\nu\lambda}_{\mathrm{shift}}(\Omega),
\end{equation}

\begin{align}
\sigma^{s\mu;\nu\lambda}_{\mathrm{shift}}(-\Omega)&=-\,\eta_{S^s}\,\sigma^{s\mu;\nu\lambda}_{\mathrm{shift}}(\Omega),\nonumber\\
&=(-1)^{1+\delta_{sx}+\delta_{sz}} \,\sigma^{s\mu;\nu\lambda}_{\mathrm{shift}}(\Omega),
\end{align}

\begin{equation}
    \sigma^{s\mu;\nu\lambda}_{\mathrm{shift}}(\Omega)=(-1)^{1+\delta_{sx}+\delta_{sz}} \,\sigma^{s\mu;\nu\lambda}_{\mathrm{shift}}(\Omega).
\end{equation}

\subsection{$\mathcal{C}_{4z}\mathcal{T}$ Symmetry Constraint}
In this section, we derive the constraint imposed on the spin shift conductivity tensor by the combined four-fold rotation and time-reversal symmetry, $\Theta = \mathcal{C}_{4z}\mathcal{T}$. The combined symmetry operator $\Theta$ can be represented in the joint spin-orbital basis as follows
\begin{equation}
    \Theta = e^{-i\frac{\pi}{4}\kappa_z\sigma_z}(i\sigma_y)\mathcal{K}
\end{equation}
This has been shown in the main text. This symmetry maps the systems Hamiltonian as Eq.~(\ref{eq:C4zT_Ham}), i.e., in \textbf{k} space this operator transforms \textbf{k}(=($k_x,k_y$)) to R\textbf{k}(=($k_y,-k_x$)). Now here also we can track the mapping of the degenerate states of a given manifold $a$ with the help of a unitary antisymmetric sewing matrix $p_a$  as follows
\begin{equation}
    \Theta|a,i\rangle_{\mathbf{k}} = \sum\limits_{i'}|a,i'\rangle_{R\mathbf{k}} [p_a]_{i'i}
\end{equation}
Because of the antisymmetric nature, $p_a^T = -p_a$. Now the matrix element of the generic operator $\hat{O}$ is obtained by taking the complex conjugate when mapping onto the transformed basis as follows
\begin{equation}
    \langle a,i|\hat{O}|b,j\rangle = \langle \Theta(a,i)|\hat{O}|\Theta(b,j)\rangle^*.
\end{equation}
Substituting the state expansions, we can get
\begin{equation}
    O_{ab}(\mathbf{k}) = \left[p^\dagger_a (\Theta\hat{O}\Theta)_{ab}(R\mathbf{k})p_b\right]^*.
\end{equation}
Now, under $90^0$ rotation, the spatial coordinates transform via a rotational matrix $M$. We use the tilde notation to denote the spatial index transformation, such that $\tilde{x}=y \, \&\, \tilde{y}=-x$. The velocity operator, therefore, transforms as 
\begin{equation}
    \Theta v^\mu \Theta^{-1} = M_{\mu\nu} v^\nu(R\mathbf{k}), \quad \text{where} \quad M = \begin{pmatrix} 0 & 1 \\ -1 & 0 \end{pmatrix}. \nonumber
\end{equation}
Therefore, applying the matrix element rule the velocity elements transform as:
\begin{equation}
    v_{ab}^x(\mathbf{k}) = \left[p_a^\dagger v_{ab}^y(R\mathbf{k})\,p_b\right]^*,\, v_{ab}^y(\mathbf{k}) = -\left[p_a^\dagger v_{ab}^x(R\mathbf{k})\,p_b\right]^*.\nonumber
\end{equation}
If we extend this formulation, the second-order velocity derivative operator is defined as follows
\begin{equation}
\hat{v}^{\mu\nu}(\mathbf{k}) = \frac{\partial^2 \hat{H}(\mathbf{k})}{\partial k_\mu \partial k_\nu},\nonumber
\end{equation}
and, under the $\mathcal{C}_{4z}\mathcal{T}$ symmetry, this rank-2 tensor operator transforms with two rotation matrices which can be shown as 
\begin{equation}
\Theta \hat{v}^{\mu\nu}(\mathbf{k}) \Theta^{-1} = \sum_{\alpha, \beta} M_{\mu\alpha} M_{\nu\beta} \hat{v}^{\alpha\beta}(R\mathbf{k}).
\end{equation}
If we apply the anti-unitary transformation rule to the matrix elements, it yields
\begin{equation}
v_{ab}^{\mu\nu}(\mathbf{k}) = \sum_{\alpha, \beta} M_{\mu\alpha} M_{\nu\beta} \left[ p_a^\dagger v_{ab}^{\alpha\beta}(R\mathbf{k}) p_b \right]^*.
\end{equation}
When we use the compact tilde notation to absorb the signs and index swaps from the rotation matrix $M$ (mapping $\mu \to \tilde{\mu}$ and $\nu \to \tilde{\nu}$), this simplifies to:
\begin{equation}
v_{ab}^{\mu\nu}(\mathbf{k}) = \left[ p_a^\dagger v_{ab}^{\tilde{\mu}\tilde{\nu}}(R\mathbf{k}) p_b \right]^*.
\end{equation}

\begin{figure*}
    \centering
    \includegraphics[width=\textwidth]{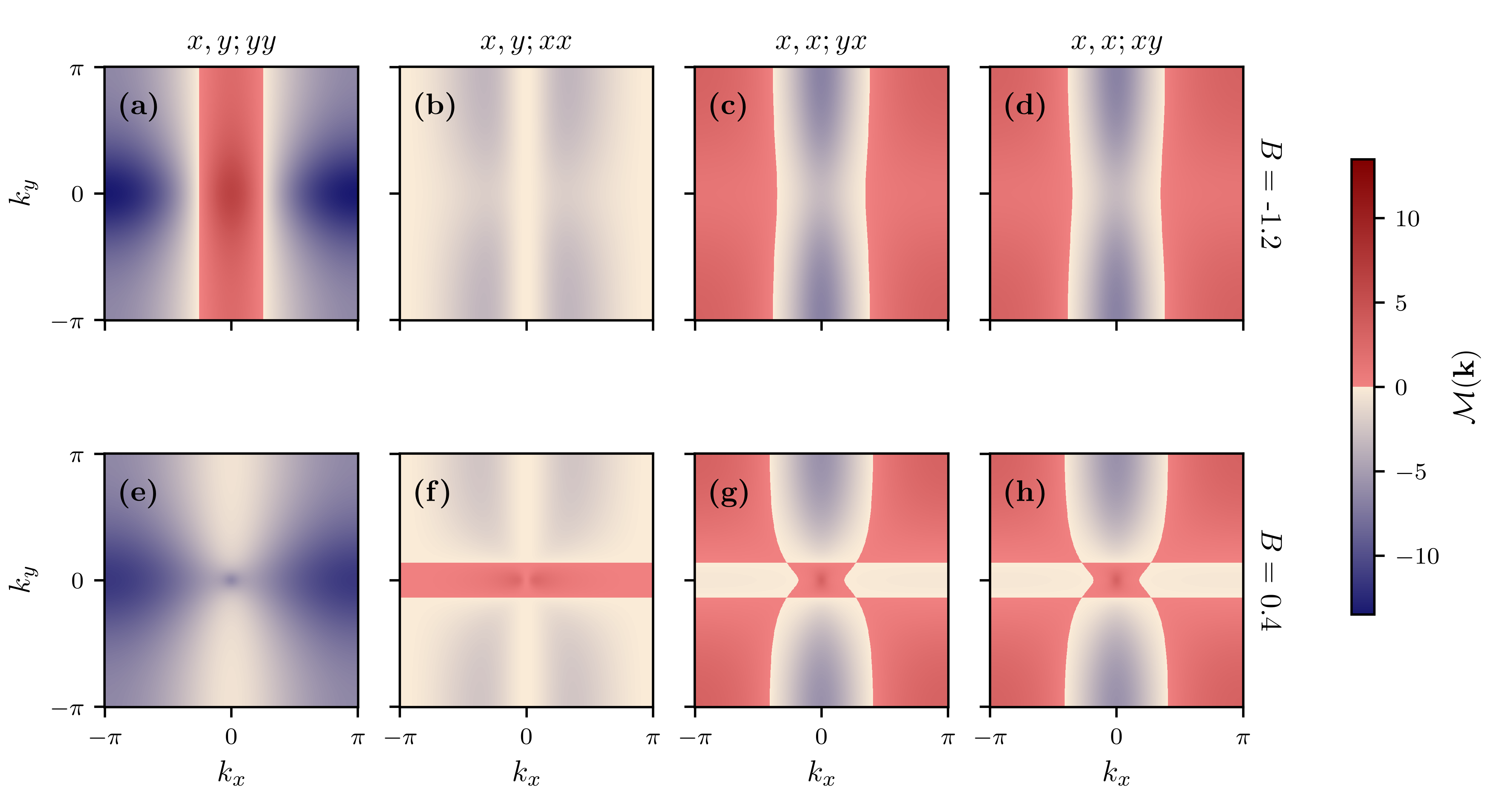}
    \caption{Contour plots of the response tensor corresponding to the spin shift conductivity near the $\Gamma$ point. The top row, (a)–(d), corresponds to the topological region ($B = -1.2t$), while the bottom row, (e)–(h), shows the trivial region ($B = 0.4t$). The rest of the parameters are fixed as in Fig.~\ref{fig2}. It depicts the contrast in the systems  behavior near $\Gamma$ point where local dirac mass is changing sign } 
    \label{fig9}
\end{figure*}

Consequently, when we trace over the band indices to calculate gauge-invariant observables, the unitary sewing matrices $p$ cancel out, which results the matrix elements of the second-order velocity derivative to effectively map as given below
\begin{equation}
v_{ab}^{\mu\nu}(\mathbf{k}) \xrightarrow{\mathcal{C}_{4z}\mathcal{T}} \left[ v_{ab}^{\tilde{\mu}\tilde{\nu}}(R\mathbf{k}) \right]^*.
\end{equation}

The spin operator on the other hand transforms as a pseudovector under rotation and it flips the sign after under time reversal operator. However, due to the spin-orbit coupling and the structure of the $\Theta$ operator, the transformation induces an explicit orbital dependence via the $\kappa_z$ matrix. If we investigate the transformation corresponding to each component of the spin matrices, then we obtain
\begin{align}
    \Theta\,(\kappa_0\otimes\sigma_x)\,\Theta^{-1}= -\kappa_z\otimes\sigma_y,\nonumber\\
    \Theta\,(\kappa_0\otimes\sigma_y)\,\Theta^{-1}= +\kappa_z\otimes\sigma_x,\nonumber\\
    \Theta\,(\kappa_0\otimes\sigma_z)\,\Theta^{-1}= -\kappa_0\otimes\sigma_z.
\end{align}

This defines the transformed spin index $s'$. If we combined these spin matrices with the velocity transformations, the conventional spin current and higher order spin vertices transform as follows
\begin{align}
    S_{ab}^{s,\mu}(\mathbf{k}) \xrightarrow{\mathcal{C}_{4z}\mathcal{T}} \left[ S_{ab}^{s',\tilde{\mu}}(R\mathbf{k}) \right]^*,\\
    S_{ab}^{s,\mu;\nu}(\mathbf{k}) \xrightarrow{\mathcal{C}_{4z}\mathcal{T}} \left[ S_{ab}^{s',\tilde{\mu};\tilde{\nu}}(R\mathbf{k}) \right]^*.
\end{align}
Now the response tensor $\mathcal{M}^{s,\mu;\nu\lambda}_{ab}$ contains a product of multiple operator inside the closed loop trace. When we apply the $\mathcal{C}_{4z}\mathcal{T}$ symmetry operation inside the trace, the $\mathcal{M}^{s,\mu;\nu\lambda}_{ab}(\mathbf{k})$ will transform as follows.
\begin{equation}
     \mathcal{M}^{s,\mu;\nu\lambda}_{ab}(\mathbf{k}) \xrightarrow{\mathcal{C}_{4z}\mathcal{T}} \left(\mathcal{M}^{s',\tilde{\mu};\tilde{\nu}\tilde{\lambda}}_{ab}(R\mathbf{k})\right).
\end{equation}
This constraint the spin conductivity tensor to
\begin{equation}
    \sigma^{s,\mu;\nu\lambda} = -\left(\sigma^{s',\tilde{\mu};\tilde{\nu}\tilde{\lambda}}\right)^*.
\end{equation}
Again $\mathcal{PT}$ symmetry already enforces the shift conductivty tensor to be real (from Eq.~(\ref{eq:shift_is_real})), which enforces
\begin{equation}
   \sigma^{s,\mu;\nu\lambda}  =-\sigma^{s',\tilde{\mu};\tilde{\nu}\tilde{\lambda}}.
\end{equation}

\section{Contour Plots of Response Tensor}
\label{AppC}
\renewcommand{\theequation}{C\arabic{equation}}
\setcounter{equation}{0}
In this section, we present the visualization of the response tensor $\mathcal{M}^{s,\mu;\nu\lambda}(\mathbf{k})$, focusing on the shift conductivity (Fig.~\ref{fig9}). In  this contour plot, the response tensor components are evaluated in the two-dimensional Brillouin zone, centered around the $\Gamma$ point ($k_x$, $k_y$ = 0). The color scale in the Fig~\ref{fig9} quantitatively maps the magnitude and sign of $\mathcal{M}(\mathbf{k})$, where the maroon color indicates a positive contributions and the dark blue color indicates a strong negative contributions.

\section{Conductivity Response across First Order Phase Transition}
\label{AppD}
\renewcommand{\theequation}{D\arabic{equation}}
\setcounter{equation}{0}
\begin{figure}[h]
\centering
\includegraphics[width=0.48\linewidth]{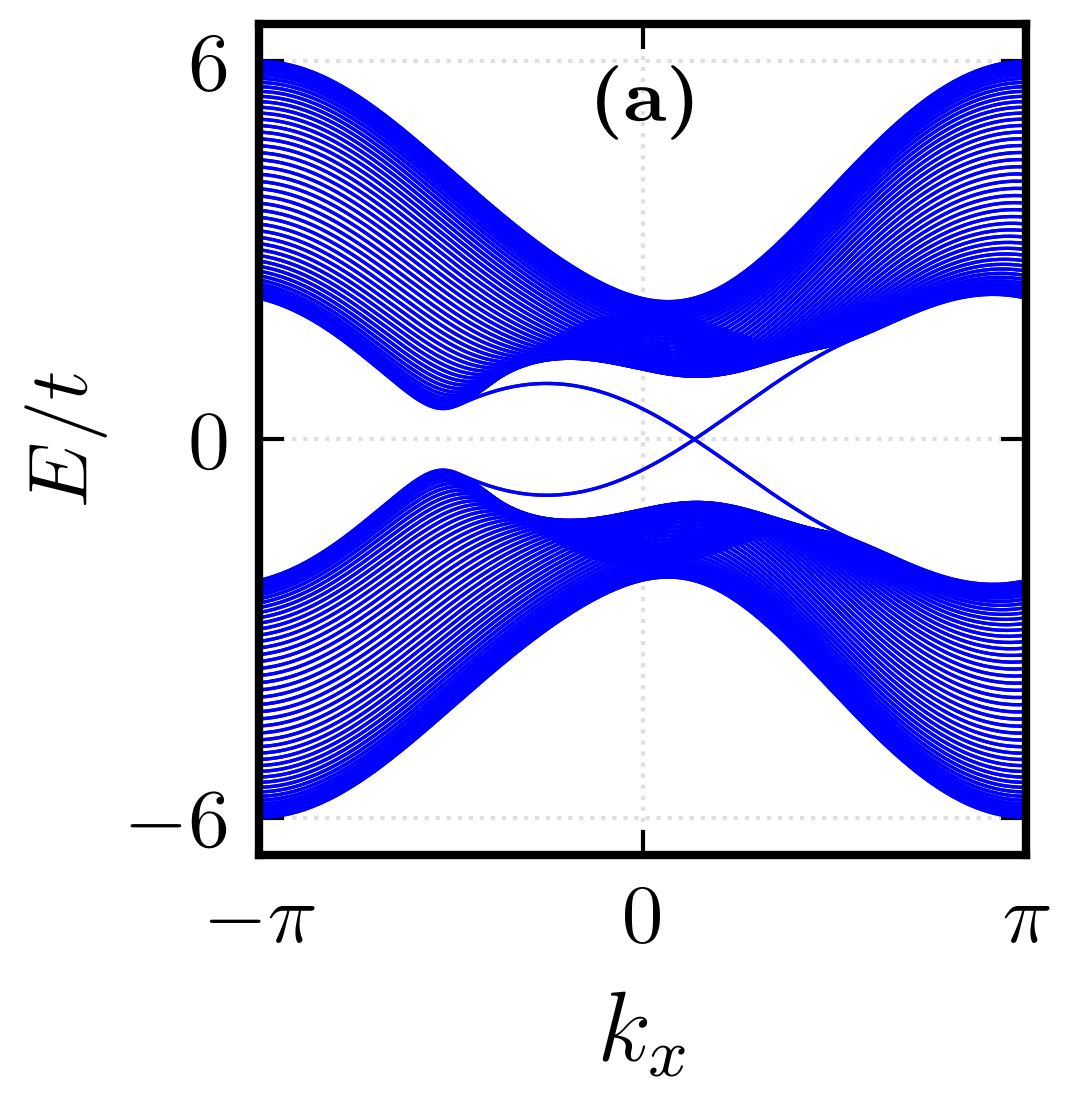}
\hfill
\includegraphics[width=0.48\linewidth]{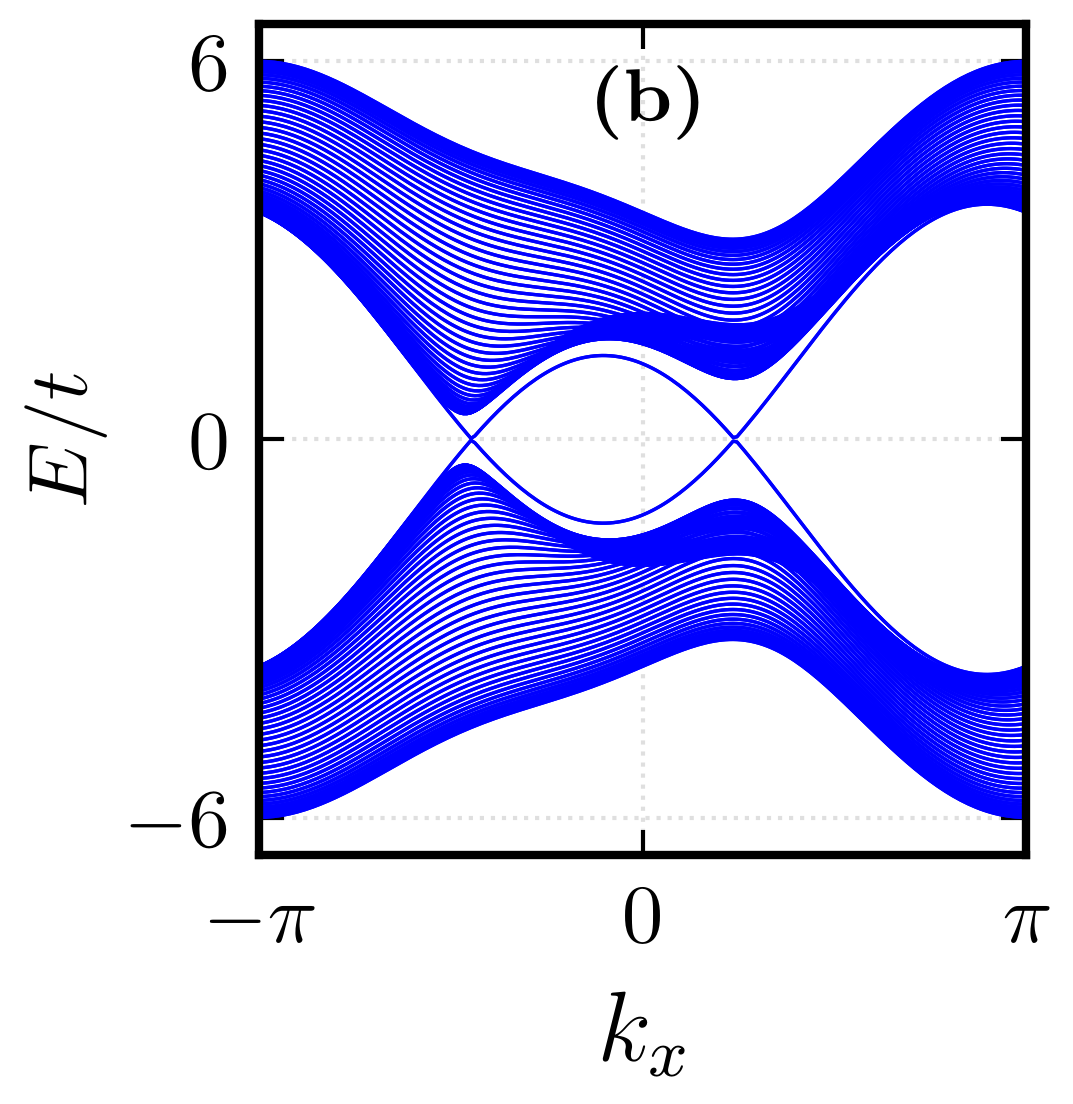}
\caption{Energy band diagrams illustrating the first-order topological phase transition. 
Parameters are fixed at $B = -2t$, $A = 1$, and $t = 1$. 
(a) Band structure for $J_a = 0.5$. 
(b) Band structure for $J_a = 1.5$.}
\label{fig10}
\end{figure}

When the N\'eel vector is along the $z$-axis, the energy of the Hamiltonian is 
\begin{equation}
    \varepsilon(\mathbf{k}) = \sqrt{ m(\mathbf{k})^2 + A^2 \sin^2(k_y) + \left[ A \sin(k_x) +J_a(\mathbf{k}) \right]^2 }.
\end{equation}
A topological phase transition occurs when the bulk gap closes, i.e., when $\varepsilon(\mathbf{k})=0$. Since the spectrum is a sum of positive-definite terms, this requires simultaneous vanishing of the mass and spin coupling terms:
\begin{equation}
    m(\mathbf{k}) = 0, \; \sin k_y=0, \quad \text{and} \quad A \sin k_x = J_a(\cos k_x - \cos k_y).
\end{equation}
At $k_y = 0$, this condition is satisfied when the altermagnetic coupling reaches the critical ratio:
\begin{equation}
    \left( \frac{J_a}{A} \right)_c = \sqrt{-\frac{4t}{B} - 1}.
\end{equation}

\begin{figure}[h]
\centering
\begin{tabular}{c c}
\includegraphics[width=0.49\linewidth]{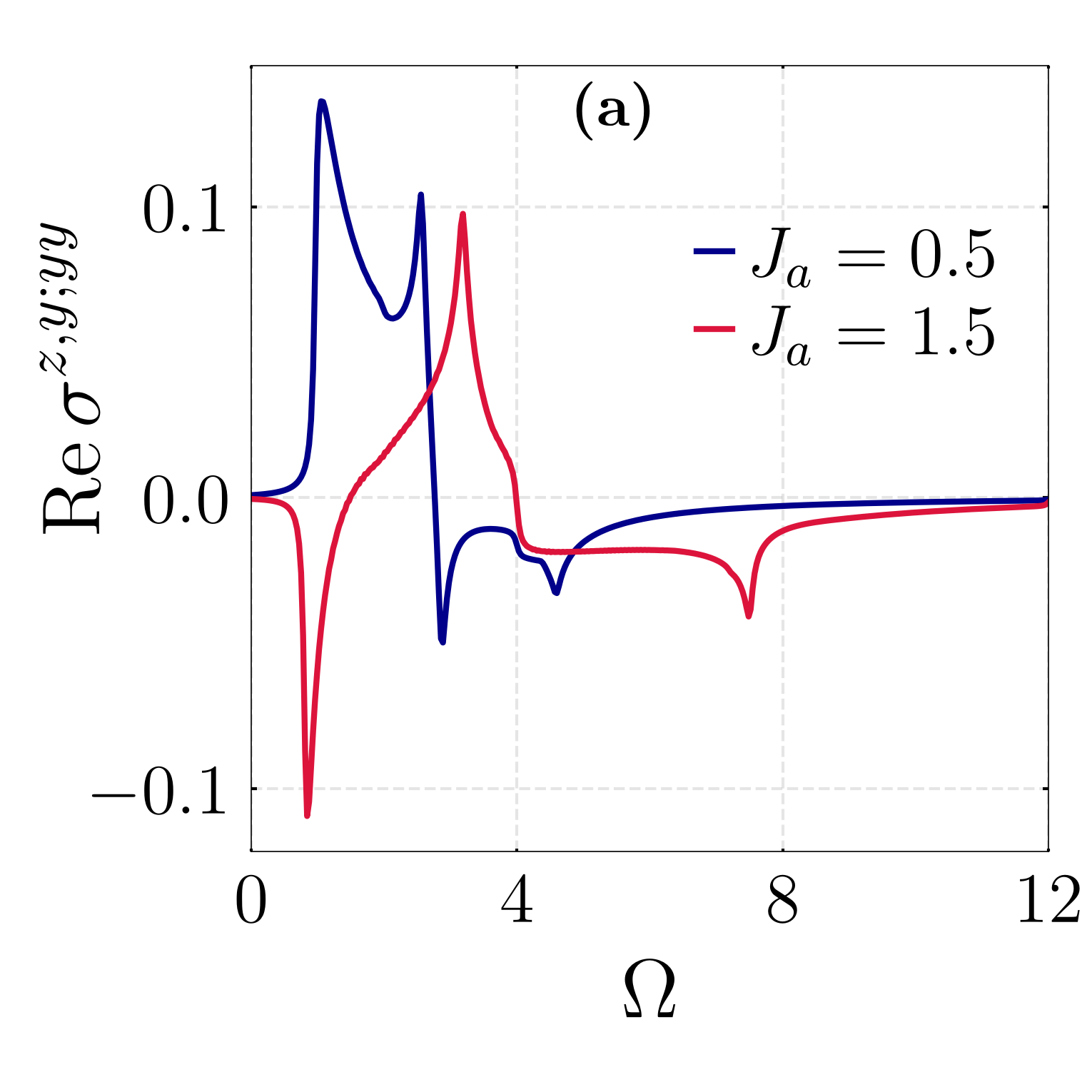}
& \includegraphics[width=0.49\linewidth]{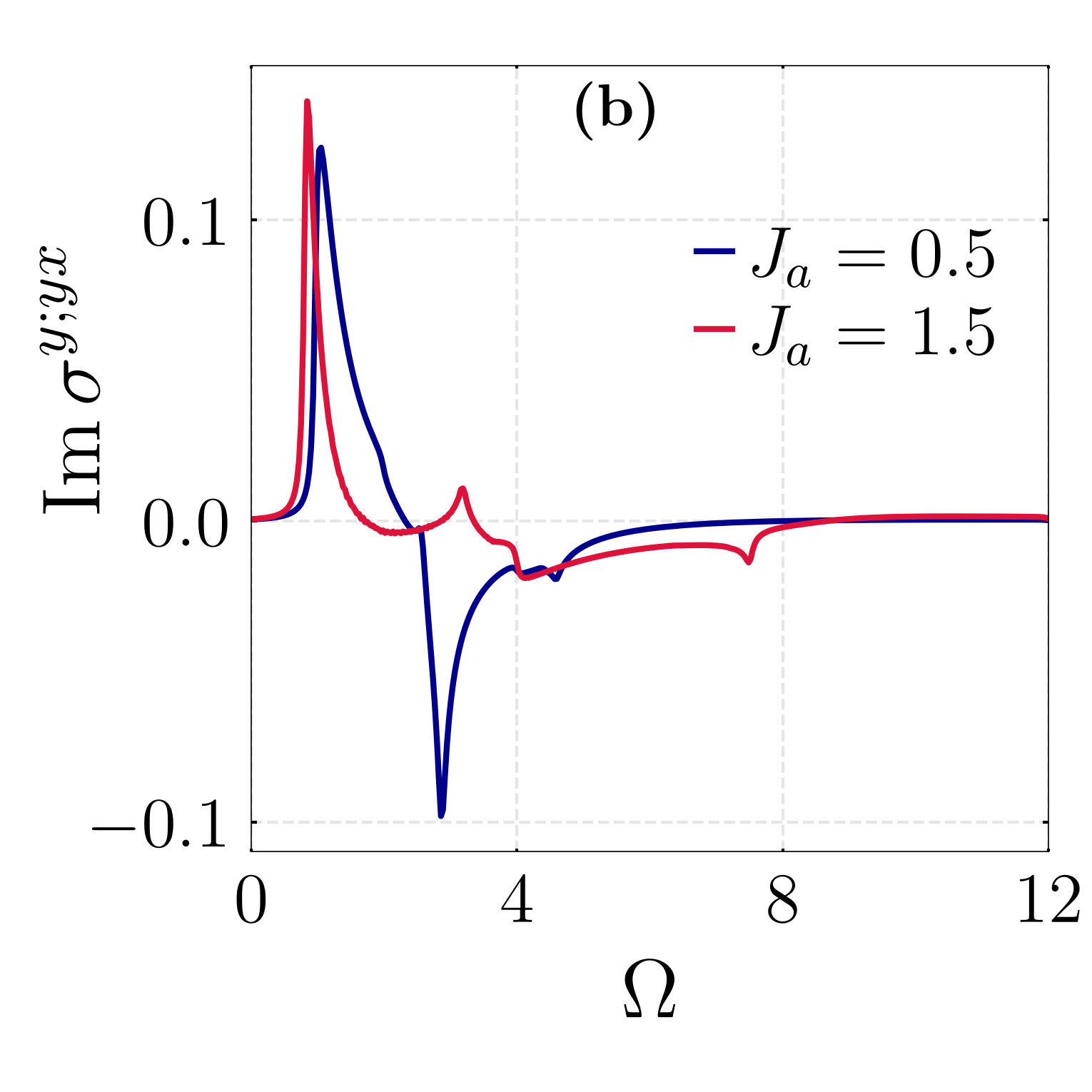}
    \end{tabular}
    \caption{Frequency dependence of the (a) spin shift conductivity $\text{Re}\,\sigma^{z,y;yy}$ and (b) charge shift conductivity $\text{Im}\,\sigma^{y;yx}$ across the phase transition. The plots are evaluated at parameters $J_a = 0.5$ (blue) and $J_a = 1.5$ (red), whereas the other parameters are fixed as in Fig.~\ref{fig10}.}
\label{fig11} 
\end{figure}
At critical coupling, the interplay between the parameter $A$ and the altermagnetic exchange $J_a$
 induces a band inversion that marks the boundary between distinct topological phases (Fig.~\ref{fig10}). A characteristic signature of this transition is the sign reversal of the linear spin shift conductivity, while the circular charge shift conductivity remains sign-preserving across the phase boundary, as demonstrated in Fig.~\ref{fig11}.

\newpage
\bibliography{referencess}
\end{document}